\documentclass[a4paper,11pt]{article}
\usepackage{jheppub} 
\usepackage{amsmath,amssymb,graphicx,hyperref}
\pdfoutput=1
\usepackage[table]{xcolor}
\usepackage{graphicx}
\usepackage{nicematrix}
\usepackage{tabularx}
\usepackage{xspace}
\usepackage{empheq}
\usepackage{verbatim}
\usepackage{amsmath}
\usepackage{amssymb}
\usepackage{url}
\usepackage{bbold}
\usepackage{slashed}
\usepackage{array}
\usepackage{aas_macros}
\usepackage{subcaption}
\usepackage{gensymb}
\usepackage{multirow}
\usepackage{threeparttable}
\usepackage{paralist}
\usepackage{xspace}
\usepackage{upgreek}
\usepackage{lipsum}
\usepackage[T1]{fontenc}
\usepackage{scalerel}
\usepackage{float}
\usepackage[framemethod=TikZ]{mdframed}
\usepackage{setspace}
\usepackage{mathtools}

\usepackage{slashed}

\usepackage[capitalise, english]{cleveref}

\allowdisplaybreaks
\renewcommand{\vec}[1]{{\mathbf{#1}}}

\newcommand{\di}{{\rm d}}

\makeatletter

    \def\CT@@do@color{%
      \global\let\CT@do@color\relax
            \@tempdima\wd\z@
            \advance\@tempdima\@tempdimb
            \advance\@tempdima\@tempdimc
    \advance\@tempdimb\tabcolsep
    \advance\@tempdimc\tabcolsep
    \advance\@tempdima2\tabcolsep
            \kern-\@tempdimb
            \leaders\vrule
                    \hskip\@tempdima\@plus  1fill
            \kern-\@tempdimc
            \hskip-\wd\z@ \@plus -1fill }
    \makeatother

\newcommand{\beq}{\begin{equation} }
\newcommand{\eeq}{\end{equation}} 
\newcommand{\bi}{\begin{itemize} }
\newcommand{\ei}{\end{itemize} }
\newcommand{\lp}{\left(}
\newcommand{\rp}{\right)}

\newcommand{\Q}{{\cal Q}}
\newcommand{\C}{{\cal C}}

\makeatletter
\providecommand*{\diff}%
	{\@ifnextchar^{\DIfF}{\DIfF^{}}}
\def\DIfF^#1{%
	\mathop{\mathrm{\mathstrut d}}%
		\nolimits^{#1}\gobblespace}
\def\gobblespace{%
	\futurelet\diffarg\opspace}
\def\opspace{%
	\let\DiffSpace\!%
	\ifx\diffarg(%
		\let\DiffSpace\relax
	\else
		\ifx\diffarg[%
			\let\DiffSpace\relax
		\else
  			\ifx\diffarg\{%
				\let\DiffSpace\relax
			\fi\fi\fi\DiffSpace}

\definecolor{Red}{rgb}{1.,0.,0.}

\usepackage{xspace}

\definecolor{cborange}{HTML}{e69f00}
\definecolor{cbgreen}{HTML}{009e73}
\definecolor{cbyellow}{HTML}{f1dd42}
\definecolor{cblblue}{HTML}{56b4e9}
\definecolor{cbblue}{HTML}{0072b2}
\definecolor{defgrey}{HTML}{9f9f9f}
\definecolor{defgreen}{HTML}{8eba42}

\newcommand{\On}{{\mathcal{O}_N}}
\newcommand{\Luv}{{\Lambda_\mathrm{UV}}}
\newcommand{\Lir}{{\Lambda_\mathrm{IR}}}

\begin{document}

\preprint{}

\title{Probing Neutrino Compositeness with Invisible and Displaced Signals}

\author[1]{\small Matteo Borrello,}
\emailAdd{matteo.borrello@unifi.it}
\author[2]{\small Marco Costa,}
\emailAdd{mcosta1@perimeterinstitute.ca}
\author[1]{\small Diego Redigolo,}
\emailAdd{diego.redigolo@fi.infn.it}
\author[3]{\small and Michele Tammaro}
\emailAdd{michele.tammaro@fi.infn.it}

\affiliation[1]{INFN, Sezione di Firenze Via G. Sansone 1, 50019 Sesto Fiorentino, Italy and\\ Department of Physics and Astronomy, University of Florence, Italy}
\affiliation[2]{Perimeter Institute for Theoretical Physics, 31 Caroline St N, Waterloo, ON N2L 2Y5, Canada}
\affiliation[3]{Galileo Galilei Institute for Theoretical Physics, Largo Enrico Fermi 2, I-50125 Firenze, Italy}

\abstract{We explore the possibility that neutrinos couple to an interacting sterile sector, providing a novel portal that generalizes the heavy neutral lepton portal to a composite setting. For a low confinement scale, high-energy neutrino beams can disintegrate into collimated sprays of hidden states, referred to as “dark jets”. This dynamics gives rise to two characteristic signatures in high energy neutrino beams. First, long-lived dark resonances can enhance the neutral-current to charged-current ratio. Second, shorter-lived dark states produced in neutrino neutral currents can produce single or multiple displaced vertices and even emerging jets, depending on the kinematics. These signals probe regions of parameter space beyond existing constraints from meson, electroweak, and Higgs decays, as well as from searches for displaced decays at beam dump experiments. We study these phenomena within broad classes of ultraviolet completions and identify scenarios in which high-energy neutrino beams provide leading sensitivity to neutrino compositeness. Such scenarios generically induce higher-dimensional contact interactions, which we classify and study alongside their complementary experimental signatures. Finally, we outline an experimental program spanning both the intensity and energy frontiers. Near-term neutrino facilities (DUNE, FPF) and running flavor experiments (LHCb, Belle II) can probe neutrino compositeness through neutrino disintegration into dark jets and displaced B-meson decays. Future colliders—particularly the Future Circular Collider (FCC-ee)—will ultimately provide the strongest sensitivity to the compositeness scale via displaced Z decays. 
}


\maketitle

\newpage

\section{Introduction}
\label{sec:Introduction}

How can we probe a new sector that couples predominantly to neutrinos? And how can we determine whether its dynamics is weakly or strongly coupled? These questions are well motivated theoretically. Many mechanisms explaining the smallness of neutrino masses predict new states and interactions coupled to neutrinos~\cite{Mohapatra:1986bd,Gonzalez-Garcia:1988okv}. Similarly, a broad class of dark-sector (DS) models that can evade cosmological constraints include interactions with the Standard Model (SM) primarily through neutrinos~\cite{Berlin:2019pbq,Aloni:2023tff,Giovanetti:2024orj}.

Experimentally, this program is particularly timely. Upcoming high-intensity neutrino facilities, as well as neutrino fluxes at the High-Luminosity (HL) LHC, provide access to previously unexplored kinematic regimes. In this work, we study signals of strongly interacting new physics in the neutrino sector and compare them to constraints from precision measurements of meson, electroweak, and Higgs decays. We also show that such scenarios constitute a compelling target for FCC-ee, where electroweak and Higgs observables will be measured with unprecedented precision.

We consider a strongly interacting sterile sector described by a fermionic operator $\mathcal{O}_N$ of scaling dimension $\Delta_N$, coupled to the SM via non-renormalizable portals suppressed by an ultraviolet (UV) scale $\Lambda_{\rm UV}$. The leading portal, $HL\mathcal{O}_N$, is the composite generalization of the heavy neutral lepton (HNL) portal and provides the main focus of this work~\cite{Chacko:2020zze,Ahmed:2023vdb,Ahmed:2025ldh,Borrello:2025hal}. The dynamics of the sterile sector is governed by a confinement scale $\Lambda_{\rm IR}$, which sets the mass scale of the infrared (IR) resonances. We focus on $1~\mathrm{MeV} \lesssim \Lambda_{\rm IR} \lesssim 100~\mathrm{GeV}$, where DS resonances evade Big Bang Nucleosynthesis constraints while remaining accessible at accelerators. In this regime, production leads to dark showers that hadronize on timescales $\tau \sim 1/\Lambda_{\rm IR}$ into multiple resonances~\cite{Strassler:2006im,Strassler:2008bv,Knapen:2016hky,Bernreuther:2019pfb,Albouy:2022cin}.

For processes with typical energy $E$ in the range $\Lambda_{\rm IR} < E < \Lambda_{\rm UV}$, inclusive production into DS jets can be computed using conformal methods~\cite{Georgi:2007si,Georgi:2007ek,Contino:2020tix}, with IR and UV corrections suppressed by ${\cal O}(\Lambda_{\rm IR}^2/s)$ and ${\cal O}(s/\Lambda_{\rm UV}^2)$, respectively. Requiring theoretical control motivates focusing on $\Delta_N \geq 5/2$. The resulting phenomenology is governed by the lifetimes of DS resonances, which depend on $\Lambda_{\rm UV}$, $\Lambda_{\rm IR}$, and $\Delta_N$, and can span microscopic to macroscopic scales.

Long-lived resonances generically dominate a large region of parameter space. In this regime, strongly coupled sterile sectors can enhance the neutral-current (NC) to charged-current (CC) ratio in neutrino deep inelastic scattering (DIS)~\cite{Borrello:2025hal}, a feature absent in weakly coupled scenarios. This ratio was measured with unprecedented precision at NuTeV~\cite{NuTeV:2001whx} and will be possibly probed with comparable precision at the future Forward Physics Facility (FPF) with the full luminosity of HL-LHC~\cite{FASER:2019dxq,FASER:2020gpr,Ismail:2020yqc,SNDLHC:2022ihg,Ariga:2025qup}\footnote{Throughout this paper we will refer to FPF has a future neutrino facility along the lines of FASER-$\nu$ and SND with the expected neutrino luminosity of HL-LHC. We will discuss the importance of the different possible detector configuration in \cref{sec:results}.}. The same long-lived states also contribute to invisible decays of mesons, the $Z$, and the Higgs.

For shorter lifetimes, DS resonances decay within the detector, producing displaced vertices (DV) with visible tracks and missing energy. Such signatures can arise both from SM decays and from neutrino NC interactions. In the former case, long-lived states originate from decays of SM particles, either at rest or in flight, while in the latter they are produced through neutrino NC scattering in the detector. In both cases, the signal depends on the available phase space and can interpolate between a single displaced vertex and a multiplicity of spatially distributed decays, forming emerging jet signatures always accompanied with some amount of missing energy in the form of SM neutrinos~\cite{Schwaller:2015gea,Linthorne:2021oiz,Cohen:2015toa}. 

The possibility to probe the DS parameter space can depend strongly on the UV completion at the cutoff scale $\Luv$. The most natural choice would be to have a weakly coupled QCD-like UV completion well above the EW. In this case the approximate conformality would be preserved all the way through the EW scale and the most sensitive probes are displaced $Z$ decays at LEP~\cite{DELPHI:1996qcc}, invisible Higgs decays at the LHC~\cite{Bernal:2023coo,ATLAS:2023tkt}, and a combination of invisible pion and kaon decays and CHARM beam dump searches~\cite{CHARM:1983ayi}. Remarkably, we will show that even after considering these constraints, the DUNE Near Detector~\cite{DUNE:2020lwj,DUNE:2020ypp} and SHIP~\cite{SHiP:2015vad} can still cover unexplored parts of the parameter space and can potentially observe displaced signatures with multiple vertices, motivating dedicated searches beyond those proposed for weakly coupled HNLs~\cite{Breitbach:2021gvv}.  

In a less generic scenario, the transition between the conformal regime and the weakly coupled UV regime occurs between the center-of-mass energy of neutrino experiments and the electroweak scale. It follows that searches for invisible decays and emerging jets at FPF and DUNE could probe regions left unconstrained by electroweak and Higgs decays. Scenarios with light Goldstone bosons from chiral symmetry breaking in the DS may reduce the displaced decay yield while increasing contributions to invisible decays.

Taken together, these prospects paint an exciting experimental landscape: neutrino experiments in the near-future can improve this sensitivity by implementing displaced vertex and emerging jets searches and reducing as much as possible the uncertainty on the NC/CC ratio. Further displaced signals are expected in $B$ meson decays at LHCb and Belle~II together as well as in future beam-dump experiments such as SHiP~\cite{SHiP:2015vad}. In the longer term a large portion of the parameter space can be covered by studies of invisible and displaced $Z$ decays at FCC-ee.
   
This paper is organized as follows. In \cref{sec:HLO} we introduce the composite heavy neutral lepton portal and discuss the assumptions underlying the conformal approximation in \cref{sub:conformal}, including an estimate of its theoretical uncertainties. We compute signal rates for the relevant processes in \cref{sec:CrossSecScaling}, and analyze the lifetimes and multiplicities of DS resonances in \cref{sec:WidthScaling}. Predictions for invisible and displaced signals are presented in \cref{sec:invisible} and \cref{sec:displaced}, respectively. In \cref{sec:CompositenessModel} we discuss explicit UV completions, identifying two benchmark scenarios: one in which the DS remains approximately conformal up to the electroweak scale, and one in which it becomes weakly coupled below it. We also examine the impact of Goldstone bosons. Our main results are summarized in \cref{sec:results}, with present and future experimental sensitivities collected in \cref{sec:present} and \cref{sec:future}. In \cref{sec:nusmeft} we classify higher-dimensional portals beyond the composite HNL portal. We conclude in \cref{sec:Conclusions}.

\section{The composite heavy neutral lepton portal}
\label{sec:HLO}
In this section we study in detail the phenomenological consequences of the HNL portal with the interacting DS. We assume for simplicity that the lowest lying resonances in the DS are fermionic (we will review this assumption in \cref{sec:UVcompletions}). As a consequence, the leading interaction of the SM can be described by the lowest-dimension operator connecting the SM to a fermionic operator in the DS, ${\cal O}_N$, with scaling dimension $\Delta_N$, which excites the fermionic multiparticle states. At dimension $\Delta_N+5/2$ the only portal we can write is
\begin{equation}\label{eq:compositeN}
\mathcal{Q}_{HL, \ell} = y_\ell\frac{\tilde{H} L_\ell \mathcal{O}_N}{\Lambda_{\rm UV}^{\Delta_N - 3/2}}\,,
\end{equation}
where $H$ is the Higgs doublet and $\tilde{H} = i\sigma_2 H^{\ast}$, $L_i$ is the left-handed lepton doublet of SM generation $\ell=1,2,3~(e,\mu,\tau)$. If the DS was simply composed of a single fermionic state $N$ with $\Delta_N = 3/2$, \cref{eq:compositeN} would reduce to the familiar renormalizable interaction of an HNL, whose phenomenology is well understood~\cite{Dasgupta:2021ies}. However, we are interested in the strongly interacting regime where ${\cal O}_N$ describes a multiparticle state, which we call DS jet, rather than a single weakly coupled fermion. We will see how this has important phenomenological consequences. 

\subsection{The conformal regime}\label{sub:conformal}
The near-conformal regime allows one to compute inclusive processes reliably as long as the typical energies involved are above the confinement scale $\Lambda_{\rm IR}$ and below the UV cutoff $\Lambda_\ast$. In general, the latter scale is different than the portal cutoff scale in \cref{eq:compositeN}, $\Lambda_{\rm UV}$; for simplicity here we take $\Lambda_\ast=\Luv$, while in \cref{sec:CompositenessModel} we relax this assumption.

Generic processes involving composite DS operators are naturally factorized between SM and DS dynamics. Consider creating a DS state $n$ together with an $m$-body SM final state, SM$_m$, from an initial SM $i$-body state, SM$_i$. The matrix element factorizes as
\begin{equation}
\mathcal{M}({\rm SM}_i \to n + {\rm SM}_m) = {\cal M}_{\rm DS} \times {\cal M}_{\rm SM}\,, 
\qquad 
{\cal M}_{\rm DS} = \langle n|\mathcal{O}_N|\Omega\rangle\,,
\end{equation}
where $\Omega$ is the vacuum of the theory and the Lorentz structures are suppressed for clarity. Squaring the amplitude gives
\begin{equation}
|\mathcal{M}({\rm SM}_i \to n + {\rm SM}_m)|^2 =
\langle \Omega|\mathcal{O}_N^\dagger|n\rangle\langle n|\mathcal{O}_N|\Omega\rangle \, |{\cal M}_{\rm SM}|^2\,,
\end{equation}
so that the DS dynamics factorizes from the SM one.

Focusing on scattering processes, the fully differential cross section, inclusive over all DS states $n$, reads
\begin{equation}
\mathrm{d}\sigma({\rm SM}_i \to {\rm DS}+{\rm SM}_m)
=
\frac{1}{\Pi_{\rm in}} \sum_n
|\mathcal{M}({\rm SM}_i \to n + {\rm SM}_m)|^2
\, \mathrm{d}\Pi_{\rm fin}\,,
\end{equation}
where $\mathrm{d}\Pi_{\rm in}$ and $\mathrm{d}\Pi_{\rm fin}$ are the initial and final state phase space measures, respectively. The final state phase space can be factorized as
\begin{equation}
\mathrm{d}\Pi_{\rm fin} = (2\pi)^4 \delta^4\Big(\sum_k p_k\Big)
\mathrm{d}\Phi_{\rm SM,fin}^m \, \mathrm{d}\Phi_{\rm DS}^n \, \frac{\mathrm{d}^4 p_D}{(2\pi)^4}\,,
\end{equation}
with $\mathrm{d}\Phi_{\rm SM,fin}^m$ the phase space of the $m$ SM states, $\mathrm{d}\Phi_{\rm DS}^n$ the $n$-body DS phase space at fixed momentum $p_D$, and the delta function enforcing momentum conservation.  

We can now exploit the optical theorem to relate the inclusive sum over DS states to the discontinuity of the two-point function of the interpolating operator as
\begin{equation}
\sum_n \int \mathrm{d}\Phi_{\rm DS}^n
\langle\Omega|\mathcal{O}_N^\dagger|n\rangle
\langle n|\mathcal{O}_N|\Omega\rangle
=
2\,\mathrm{Im}\!\Big(i \langle\Omega| T\, \mathcal{O}_N^\dagger \mathcal{O}_N |\Omega\rangle\Big),
\end{equation}
so that the fully inclusive cross section can be written as
\beq\label{eq:xsoptthm}
\begin{split}
\mathrm{d}\sigma({\rm SM}_i \to {\rm DS}+{\rm SM}_m)
&=
\frac{2}{\Pi_{\rm in}}
(2\pi)^4 \delta^4\Big(\sum_k p_k\Big)
\mathrm{d}\Phi_{\rm SM,fin}^m \frac{\mathrm{d}^4 p_D}{(2\pi)^4} \\
\quad &\times |{\cal M}_{\rm SM}|^2 \,
\mathrm{Im}\!\Big(i \langle\Omega| T\, \mathcal{O}_N^\dagger(p_D)\mathcal{O}_N(-p_D)|\Omega\rangle\Big)\,.
\end{split}
\eeq
\Cref{eq:xsoptthm} is completely general and does not rely on assumptions about the detailed structure of the DS spectrum. The complexity of the composite sector dynamic is entirely captured by the two-point function of $\On$ evaluated on the vacuum. In a generic strongly coupled theory, this correlator is incalculable, and for invariant masses close to the confinement scale $\Lambda_{\rm IR}$ it is expected to develop a complicated pattern of poles associated with light resonances.

However, if the invariant mass satisfies $4 \Lambda_{\rm IR}^2 \ll p_D^2 \ll \Lambda_{\rm UV}^2$, the dynamic is approximately conformal invariant. In this regime the two-point function admits an operator product expansion (OPE), with the leading term determined by the spin and scaling dimension $\Delta_N$ of $\mathcal{O}_N$, while deviations from conformality enter as power corrections in the confinement scale, $\mathcal{O}(\Lambda_{\rm IR}^2/p_D^2)$, as  well as inverse power corrections in the UV cutoff scale, $\mathcal{O}(p_D^2/\Lambda_{\rm UV}^2)$. Consequently, inclusive observables in the appropriate energy regime can be computed reliably in terms of the short-distance behavior of the two-point correlator, with controlled uncertainties. At the leading order the two-point correlator is conformal and its behavior is essentially fixed by naive dimensional analysis.\footnote{A standard example of the use of the conformal regime is the inclusive process $e^+ e^- \to \mathrm{hadrons}$. By the optical theorem, the total cross section is determined by the two-point function $\Pi(q^2)$ of the electromagnetic current, $\langle \Omega|T J^\mu(q)J^\nu(-q)|\Omega\rangle=(q^\mu q^\nu-q^2 g^{\mu\nu})\Pi(q^2)$, so that $\sigma = 4\pi \alpha^2\,\mathrm{Im}\,\Pi(s)/s$. Near the confinement scale, $\Pi(q^2)$ develops poles making differential information non-perturbative. Far above the confinement scale, the OPE applies and the leading conformal term dominates. The imaginary part $\mathrm{Im}\,\Pi(s) \propto N_C\sum_q Q_q^2$ is a constant, while power corrections are suppressed as $\mathcal{O}(\Lambda_{\rm QCD}^2/s)$~\cite{Ellis:1996mzs,Shifman:1978by}. Even near $\Lambda_{\rm QCD}$, coarse graining over large enough energy intervals effectively averages $\Pi(q^2)$ over the resonance structure, recovering the leading conformal short-distance behavior~\cite{Poggio:1975af}.
}

\begin{figure}[t]
    \centering
    \includegraphics[width=0.8\linewidth]{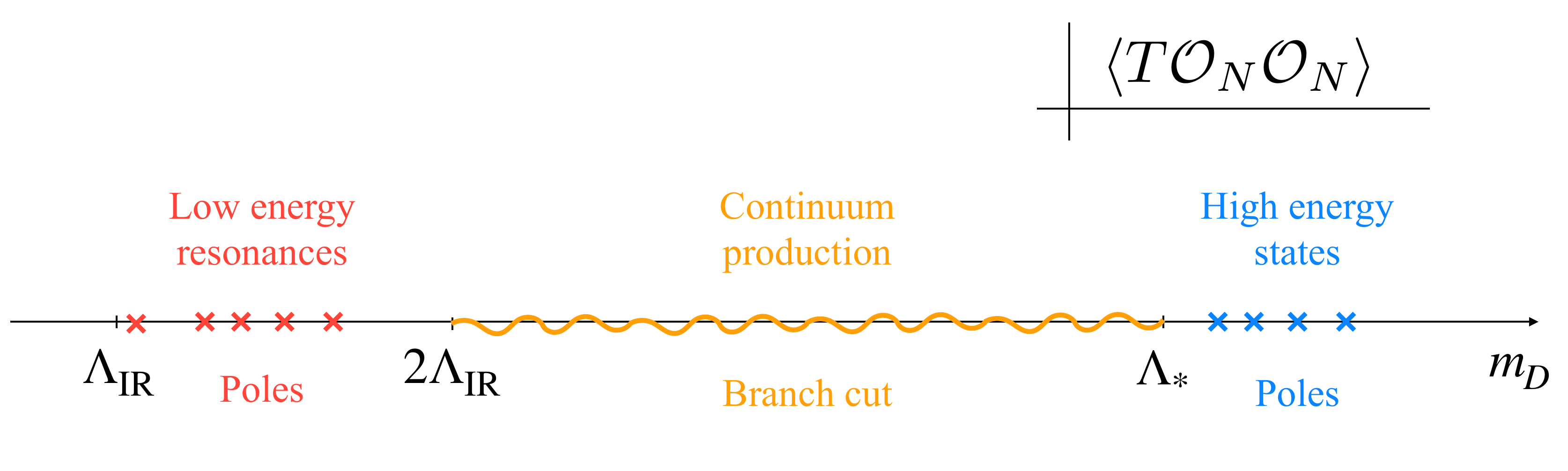}
    \caption{Sketch of the analytic structure of the DS operator two point function. The scale at which the conformal regime breaks is $\Lambda_\ast$, which in general can be different than the cutoff scale of the portal, $\Luv$. See \cref{sec:CompositenessModel} for details.}
    \label{fig:twopt}
\end{figure}

For fermionic operators, the imaginary part of the two-point function is well-defined for all $\Delta_N \geq 3/2$ and is given by~\cite{Borrello:2025hal}  
\begin{equation}\label{eq:twoptfunct}
    {\rm Im}\big(i\langle T \mathcal{O}_N(-p)\bar{\mathcal{O}}_N(p)\rangle\big) = A_N \, \slashed{p} \, (p^2)^{\Delta_N - 5/2}\,,
\end{equation}
where for convenience we define  
\begin{equation}
A_N  \equiv \frac{c_N \, 2^{3-2\Delta_N} \pi}{\Gamma(\Delta_N-3/2)\Gamma(\Delta_N+1/2)} \,,
\end{equation}
with $c_N > 0$ denoting the central charge of the DS.
This correlator captures the effect of the continuum production of DS jets, as schematically illustrated in \cref{fig:twopt}.

We want now to estimate corrections arising from both the IR and UV poles. In order to do so, we parametrize the production differential cross section as  
\begin{equation}
\frac{\di\sigma}{\di p_D^2} \propto \left(\frac{p_D^2}{s}\right)^{\alpha} \left(1 - \frac{p_D^2}{s}\right)^2 \,, \label{eq:scalingerror}
\end{equation}
where $s$ is the center of mass energy of the process, and require that it is not significantly modified by IR physics. This condition translates into the requirement that $\alpha > -1$, reflecting the fact that the conformal approximation is more reliable for UV-dominated cross sections and therefore less sensitive to strongly coupled dynamics at $p_D^2 \sim \Lambda_{\rm IR}^2$. For the limiting case $\alpha = -1$, one expects logarithmic corrections from IR physics to scale as $\sim \log(\Lambda_{\rm IR}^2 / s)$. Depending on the specific process, these IR contributions could produce non-negligible modifications of the signal yield; nevertheless, we retain this limiting case and impose the milder constraint $\alpha \geq -1$.

The requirement on $\alpha$ directly translates into a lower bound on the operator dimension, which controls the slope of the differential cross section. In practice, this enforces $\Delta_N \ge 5/2$ in order to maintain the calculability of all processes under consideration.\footnote{In a previous study, Ref.~\cite{Borrello:2025hal}, some of the authors focused on $\Delta_N = 7/2$, which imposes a stricter requirement on calculability. As a consequence, the phenomenology was dominated by invisible signatures from long-lived final states. As shown in \cref{fig:unfoldederror}, the case $\Delta_N = 5/2$ remains calculable, albeit with larger theoretical uncertainties compared to $\Delta_N = 7/2$. The smaller scaling dimension leads to shorter lifetimes, making displaced signatures more likely. Extending the conformal regime to $\Delta < 5/2$, as in previous studies~\cite{Chacko:2020zze,Ahmed:2023vdb,Ahmed:2025ldh}, introduces large theoretical uncertainties that can only be reduced through a more detailed modeling of the confining dynamics.} To quantify this statement, we estimate the theoretical uncertainty associated with IR corrections by evaluating the fraction of events arising from the production of DS states with low invariant mass, $p_D^2 \in [\Lambda_{\rm IR}^2, 4 \Lambda_{\rm IR}^2]$, which we interpret as the IR-sensitive component of the rate. The contribution that can be reliably calculated is instead obtained by integrating the production rate in \cref{eq:scalingerror} over the interval $p_D^2 \in [4 \Lambda_{\rm IR}^2, s]$.  

\begin{figure}[t!]
    \centering
    \includegraphics[width=0.75\linewidth]{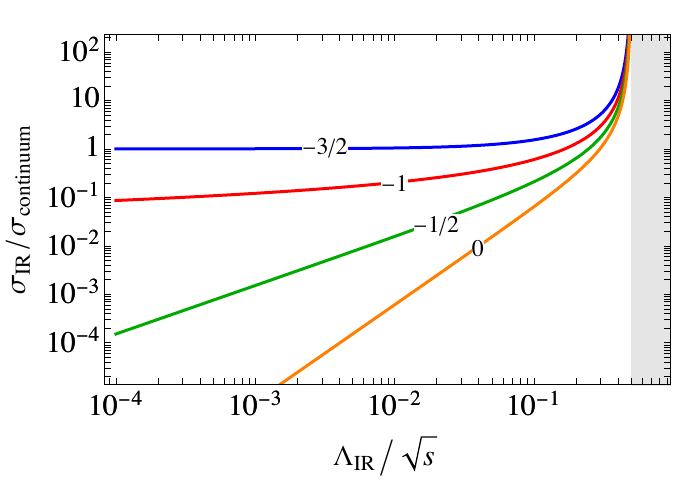}
  \caption{Estimate of the theoretical uncertainty on DS jet production arising from the breakdown of the conformal approximation, shown for different values of $\alpha$. The gray region is kinematically forbidden since $\Lambda_{\rm IR}^2 > s/4$. The more the production process probes the low-mass region, the larger the associated theoretical uncertainty.}
    \label{fig:confapprox}
\end{figure}

The resulting behavior is shown in \cref{fig:confapprox}. Two combined effects determine the overall theoretical uncertainty: the enhanced probability of radiating light dark states as $\alpha$ decreases, and the progressive reduction of the available phase space where the conformal approximation holds as $\Lambda_{\rm IR}$ approaches $\sqrt{s}$.

UV corrections are instead perturbative in our setup, where the theory is assumed to be asymptotically free in the UV and therefore does not dangerously affect our predictions, as long as $\sqrt{s}$ is sufficiently far from the cutoff scale.

After this discussion, one may ask whether the strength of the coupling in the DS plays any significant role. In principle, the conformal regime can be realized at both weak and strong coupling. Assuming a strongly coupled dark sector, however, tends to enhance the coefficient $A_N$ in \cref{eq:twoptfunct}, since the operator $\On$ interpolates a larger number of states compared to the weakly coupled case. Furthermore, strongly coupled quasi-conformal theories often generate sizable anomalous dimensions, which naturally motivates considering larger operator dimensions for the portal, ignoring lower dimensional ones. We come back to this last issues in \cref{sec:UVcompletions}, where we show that having large anomalous dimensions is a natural dynamical way of making the composite HNL portal the leading portal connecting the SM with the DS.

\subsection{Production mechanisms}
\label{sec:CrossSecScaling}
The portal in \cref{eq:compositeN} leads to the production of a DS jet in high-energy lepton scatterings as well as in SM particle decays. In general, the kinematics of this process depends on the invariant mass of the dark jet, $p_D^2$, which then hadronizes into DS resonances. The mass scale of the resonances is set by the confinement scale, $\Lir$. Thus the minimum invariant mass achievable can be defined as the production of a dark jet with only two DS resonances, $4\Lir^2$.

\subsubsection*{Scattering}
\begin{figure}[t!]
    \centering
    \includegraphics[width=0.45\linewidth]{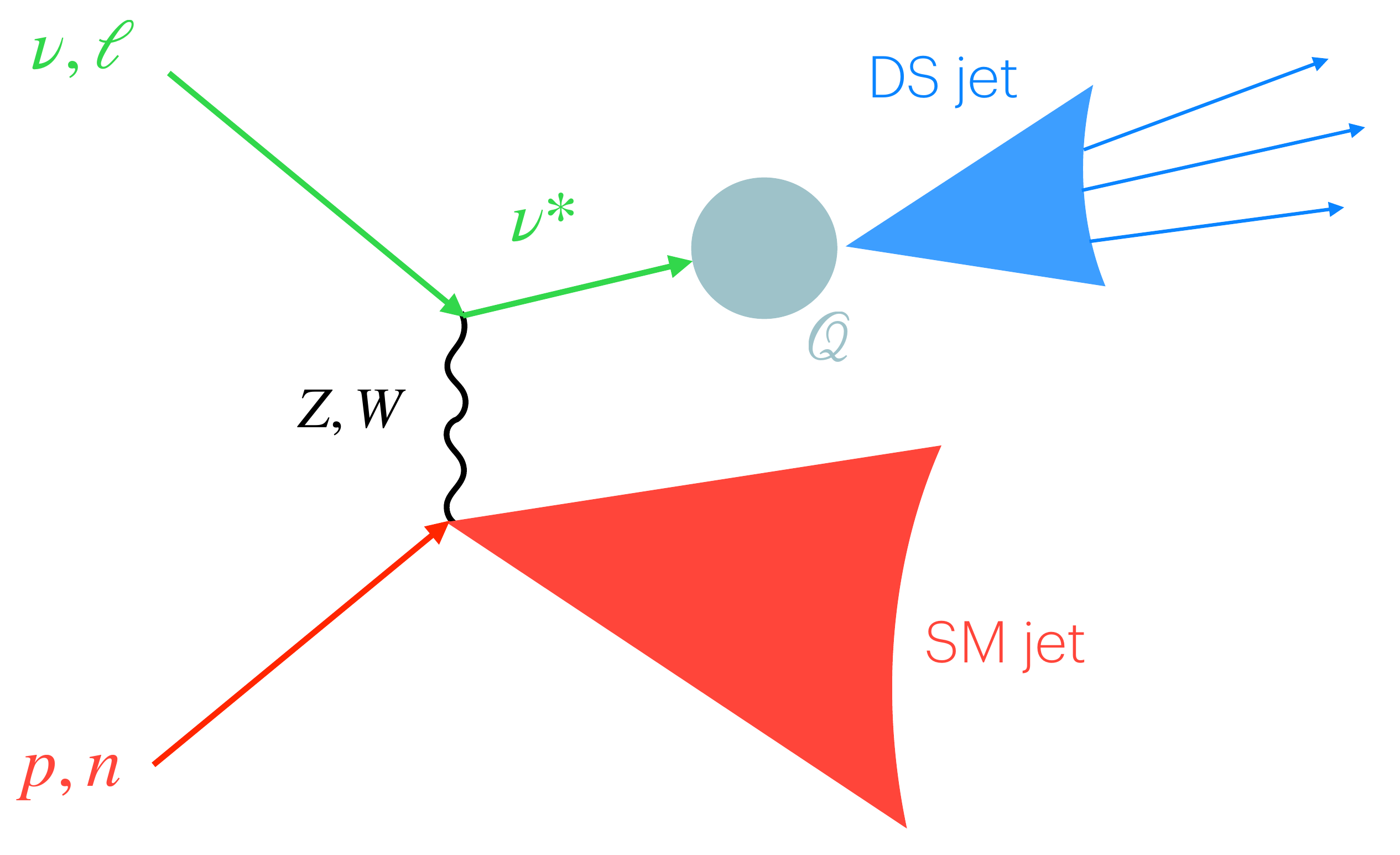}
    \caption{Sketch diagram for the production of DS jets from neutrino (electron) DIS against nucleons through NC (CC) weak interactions. The portal operator in \cref{eq:compositeN}, shown as a grey blob, is inserted on an \emph{off-shell neutrino propagator}. The differential cross section for this process is given in \cref{eq:diffscatt}. The DS jet then relaxes to light dark sector resonances whose dynamics will be illustrated in \cref{sec:WidthScaling}.}
    \label{fig:phenomixing}
\end{figure}
As illustrated in \cref{fig:phenomixing}, incoming neutrinos or charged leptons can undergo disintegration into DS jets through scattering off nucleons\footnote{In principle, neutrino-electron scattering could provide a clean and well-understood signal in instrumented detectors. However, its event rate is strongly suppressed compared to neutrino-nucleon scattering. For a neutrino of energy $E_\nu$ scattering off a stationary target of mass $m_T$, the center-of-mass energy satisfies $s \simeq 2 m_T E_\nu$. Consequently, the ratio of center-of-mass energies for electron and proton targets scales as $s_e/s_p \simeq m_e/m_p \approx 5 \times 10^{-4}$, leading to a correspondingly smaller phase space and cross section for neutrino--electron events~\cite{Borrello:2025hal}.}. From a phenomenological perspective, this mechanism is particularly relevant for experiments with intense neutrino beams. 

\begin{figure}[t!]
    \centering
     \includegraphics[width=0.5\linewidth]{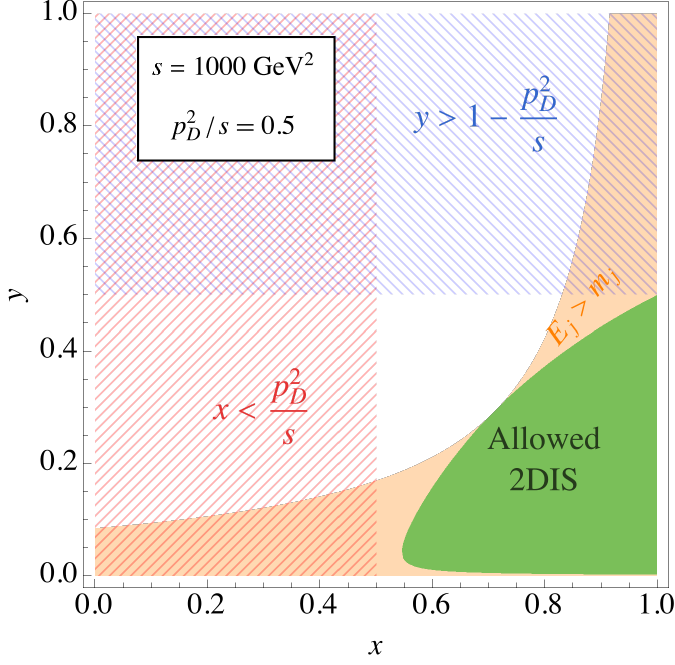}\hfill  
    \caption{Kinematics of double inelastic neutrino--proton scattering in the $(x,y)$ plane, fixing the center of mass energy and its ratio with the DS jet invariant mass squared to $s=1000~{\rm GeV}^2$ and $p_D^2/s = 0.5$, respectively. The \textbf{hatched red region} and the \textbf{hatched blue region} correspond to the kinematics boundaries of \cref{eq:formfactor}. The \textbf{orange shaded area} corresponds to the physical region where the SM jet energy satisfies $E_j^* > m_j$. The \textbf{green shaded region} defines the physically allowed domain for double DIS events, obtained by imposing $\cos^2\theta_j \leq 1$. A finite proton mass, $m_p$, slightly distorts the boundary of the allowed region near $y = 0$. For comparison, standard SM DIS occupies the full $[0,1]\times[0,1]$ square, up to corrections of order $m_p^2/s$.}
    \label{fig:kin2DIS}
\end{figure}

Defining the electroweak four-fermion interaction cross section as $\sigma_{\rm{EW}}=G_F^2 s/\pi$, the differential cross section in the dark jet invariant mass can be written as 
\begin{equation}\label{eq:diffscatt}
\frac{\di \sigma_{2\rm{DIS}}}{\di p^2_D}=\frac{A_Ny_\ell^2\sigma_{\rm{EW}}}{4\pi p_D^2}\left(\frac{v^2}{p^2_{D}}\right)\left(\frac{p^2_D}{\Lambda_{\rm{UV}}^2}\right)^{\Delta_N-3/2} \mathcal{F}(p^2_D/s)\ ,
\end{equation}
where we defined the form-factor as
\beq\label{eq:formfactor}
\begin{split}
\mathcal{F}(p^2_D/s,s)&\equiv\sum_q\int_{p^2_D/s}^1\!\!\!\! dx \\
&\times\int_0^{1-p^2_D/s}\!\!\!\!\!\!\!\!\!\! dy\left[\ell^2_qx\left(1-p^2_D/s\right)+r^2_q(1-y)\left(x(1-y)-p^2_D/s\right)\right] f_q(x,Q^2)\ ,
\end{split}
\eeq
where $x$ is the parton energy fraction, $y$ the inelasticity and $Q^2=xy s$ the momentum exchange. The term $f_q(x,Q^2)$ is the quark parton distribution function (PDF) inside the proton, which we take from Ref.~\cite{Clark:2016jgm}, and introduces an explicit dependence on the center of mass energy of the process. For a neutrino beam, the neutral current process is responsible for the production of DS jets, so both $\ell_q = I_{3,q} - Q \, \sin^2 \theta_W$ and $r_q = - Q \, \sin^2 \theta_W$ should be considered, where $I_{3,q}$ is the quark's weak isospin and $Q$ electric charge. For charged lepton beams, it is the charged current only that produces DS jets, so $r_q^2 = 0$ and $Q=0$ should be taken in \cref{eq:formfactor}. Notice that both the $x$ and $y$ integration ranges in \cref{eq:formfactor} are cut off by the mass of the produced DS jet relative to the center-of-mass energy. This behavior is typical of a doubly inelastic process, such as the one considered here. In \cref{fig:kin2DIS} we illustrate the kinematics of the scattering process, showing that the allowed phase space for the doubly inelastic scattering is always a subset of that allowed by the standard DIS in the SM. We provide details on the computation of the neutrino-nucleon DIS cross section in \cref{sec:NeutrinoNucleonScattering}.

In the massless limit, $p_D^2 \ll s$, the form factor in \cref{eq:formfactor} simplifies to $\mathcal{F}(0) \approx \sum_q x_q \left( \ell_q^2 + \frac{r_q^2}{3} \right) \approx 0.055$, with a residual weak dependence on $s$; here $x_q$ is the mean over the PDF of $x$. This number, plugged into \cref{eq:diffscatt}, reproduces the well-known feature of DIS: the energy scale of the process depends on $\hat{s} = x_q s$, reflecting the fraction of momentum carried by the struck quark inside the proton.

\begin{figure}[t!]
    \centering
    \includegraphics[width=0.49\linewidth]{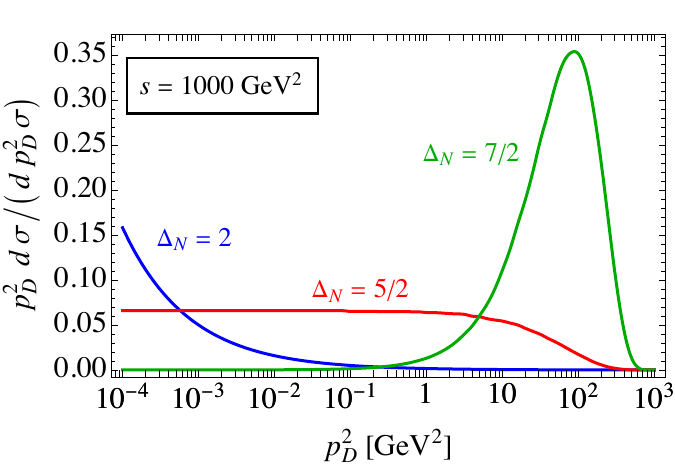}
      \includegraphics[width=0.49\linewidth]{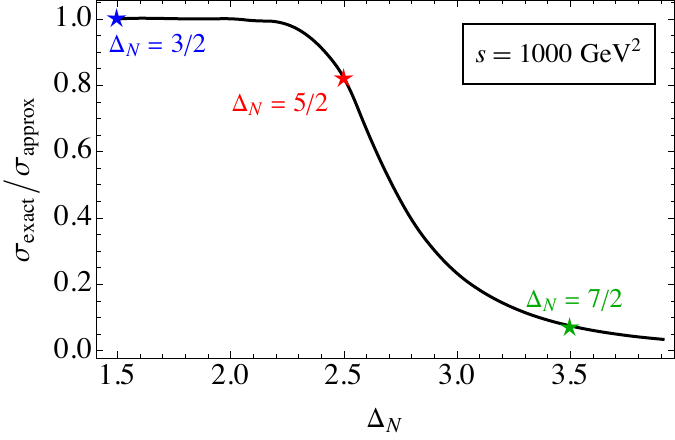}
    \caption{The importance of including the form factor in \cref{eq:formfactor} encoding the PDF and phase space constraint. {\bf Left:} Exact differential  cross sections in the jet invariant mass $p^2_D$. {\bf Right:} Ratio of the exact total cross section over the approximate one (ignoring the the form factor in \cref{eq:formfactor}) as a function of the operator dimension $\Delta_N$. }
    \label{fig:pdfeffect}
\end{figure}

We quantify the goodness of the massless limit approximation in \cref{fig:pdfeffect}. In the left plot, we compute the normalized differential cross section in the jet invariant mass, $p_D^2$, for a fixed incoming neutrino energy. For IR-dominated cross sections (smaller $\Delta_N$) the massless limit is an excellent approximation, while it becomes inaccurate for UV-dominated ones (larger $\Delta_N$), where the UV sensitivity of the hard cross section is suppressed by the kinematic fall-off of the PDFs. In the right plot of \cref{fig:pdfeffect} we explicitly report the error of the massless approximation on the total rate. We see that in the weakly coupled case, $\Delta_N = 3/2$, the approximation is essentially exact, whereas for $\Delta_N = 7/2$ it deviates from the true result by roughly an order of magnitude; the case of $\Delta_N=5/2$ is somehow in the middle, with a deviation of $\sim20$\%. 

\begin{figure}[t!]
    \centering
    \includegraphics[width=0.49\linewidth]{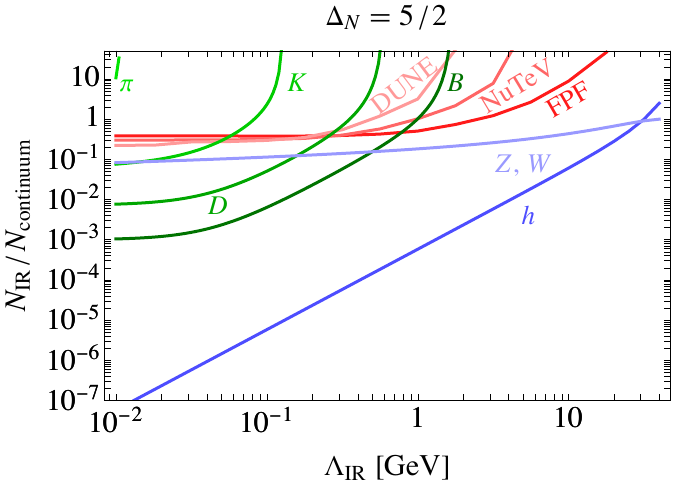}
    \includegraphics[width=0.49\linewidth]{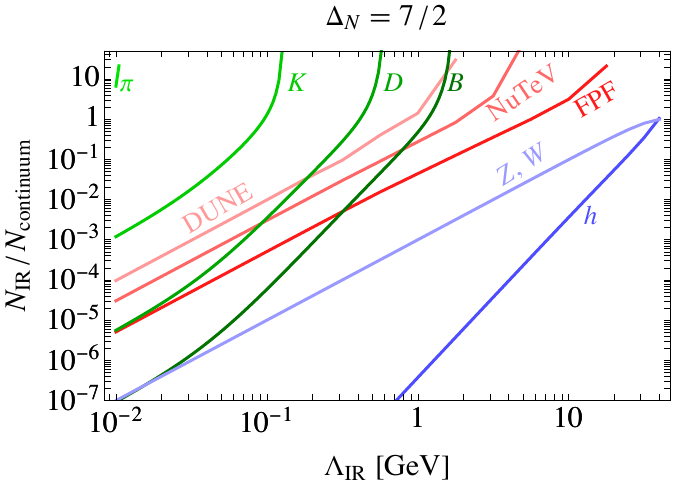}
    \caption{Estimate of the theoretical uncertainties for each probe considered in this work (see \cref{sec:results} for details), as function of $\Lir$, for $\Delta_N = 5/2$ ({\bf left}), and $\Delta_N = 7/2$ ({\bf right}). Following the discussion in \cref{sub:conformal}, we define the uncertainty as the ratio of signal events produced in the IR incalculable regime to those produced in the UV conformal regime.}
    \label{fig:unfoldederror}
\end{figure}

Given that the differential cross section \cref{eq:diffscatt} scales as $(p^2_D)^{\Delta_N-7/2}$, we can easily integrate it within the available phase space in order to get the inclusive total cross section. It reads

\begin{equation}\label{eq:totcatt}
\sigma_{2\rm{DIS}}\equiv\int_{4\Lambda_{IR}^2}^{s}\di p_D^2\frac{\di \sigma_{2\rm{DIS}}}{\di p^2_D}\approx\frac{A_Ny_\ell^2}{8\pi^2 v^2(\Delta_N-5/2)}\left(\frac{s}{\Lambda_{\rm{UV}}^2}\right)^{\Delta_N-3/2}\,. 
\end{equation}
With $\Delta_N=5/2$, the integral develops a logarithmic divergence, which is automatically cured by the minimal threshold required two produce two DS resonances. We then get 
\begin{equation}\label{eq:5/2log}
\sigma_{2\rm{DIS}}\vert_{\Delta_N=5/2}=\frac{A_Ny_\ell^2}{8\pi^2 v^2}\left(\frac{s}{\Lambda_{\rm{UV}}^2}\right)\log\left(\frac{s}{4\Lambda_{\rm{IR}}^2}\right)\,. 
\end{equation}
Following the discussion around \cref{eq:scalingerror}, we see that $\Delta_N=5/2$ is indeed the lowest operator dimension such that we can consider the cross section insensitive enough to IR effects to be reliably calculable in the conformal regime. Clearly, the theory error will depend on the specific process considered, the typical center-of mass energy of the process and its available phase space. We summarize the latter in \cref{fig:unfoldederror} for the phenomenological probes used in this work (see \cref{sec:results}).

\subsubsection*{Decay}
\begin{figure}[t!]
    \centering
    \includegraphics[width=0.45\linewidth]{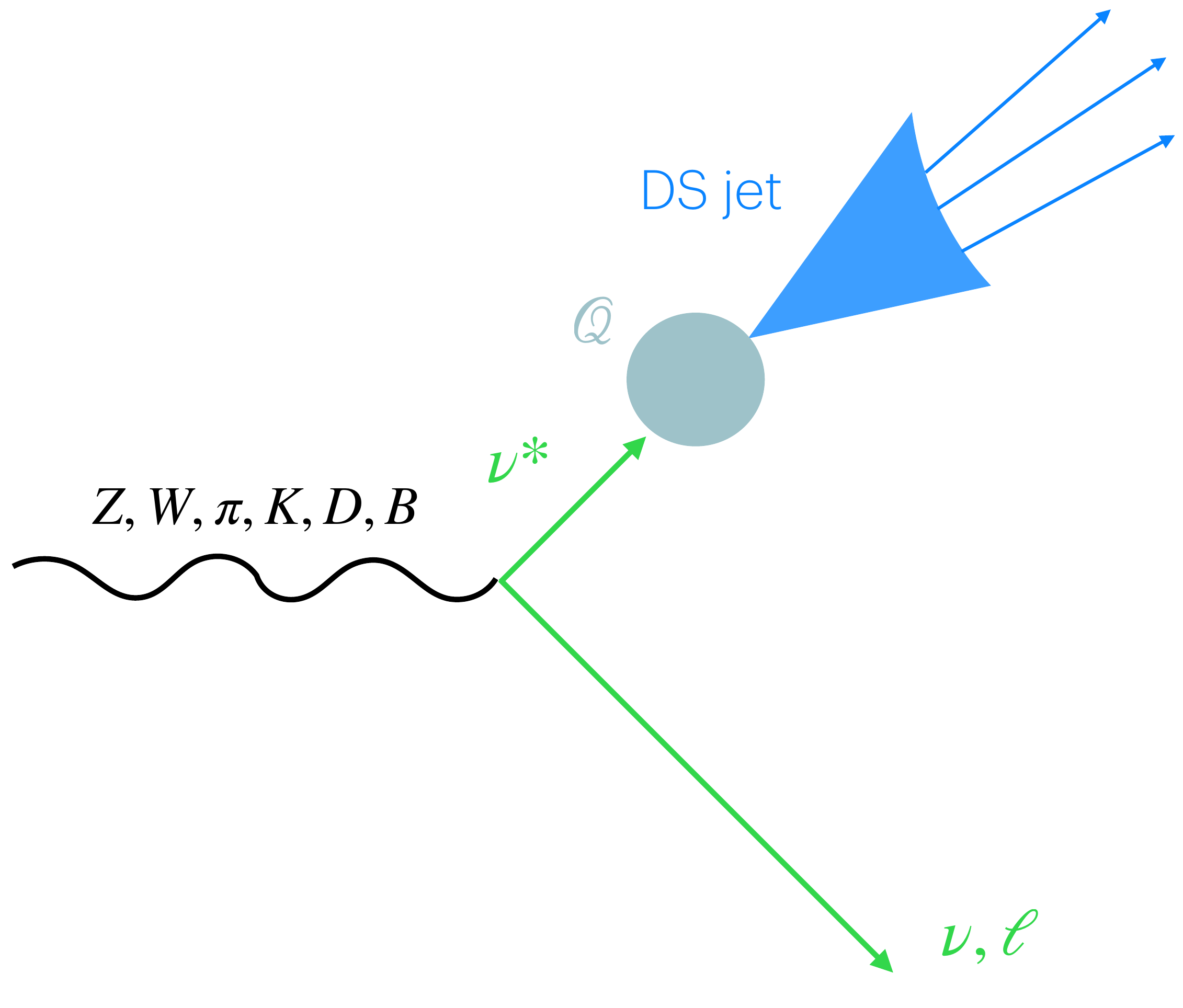}
      \includegraphics[width=0.45\linewidth]{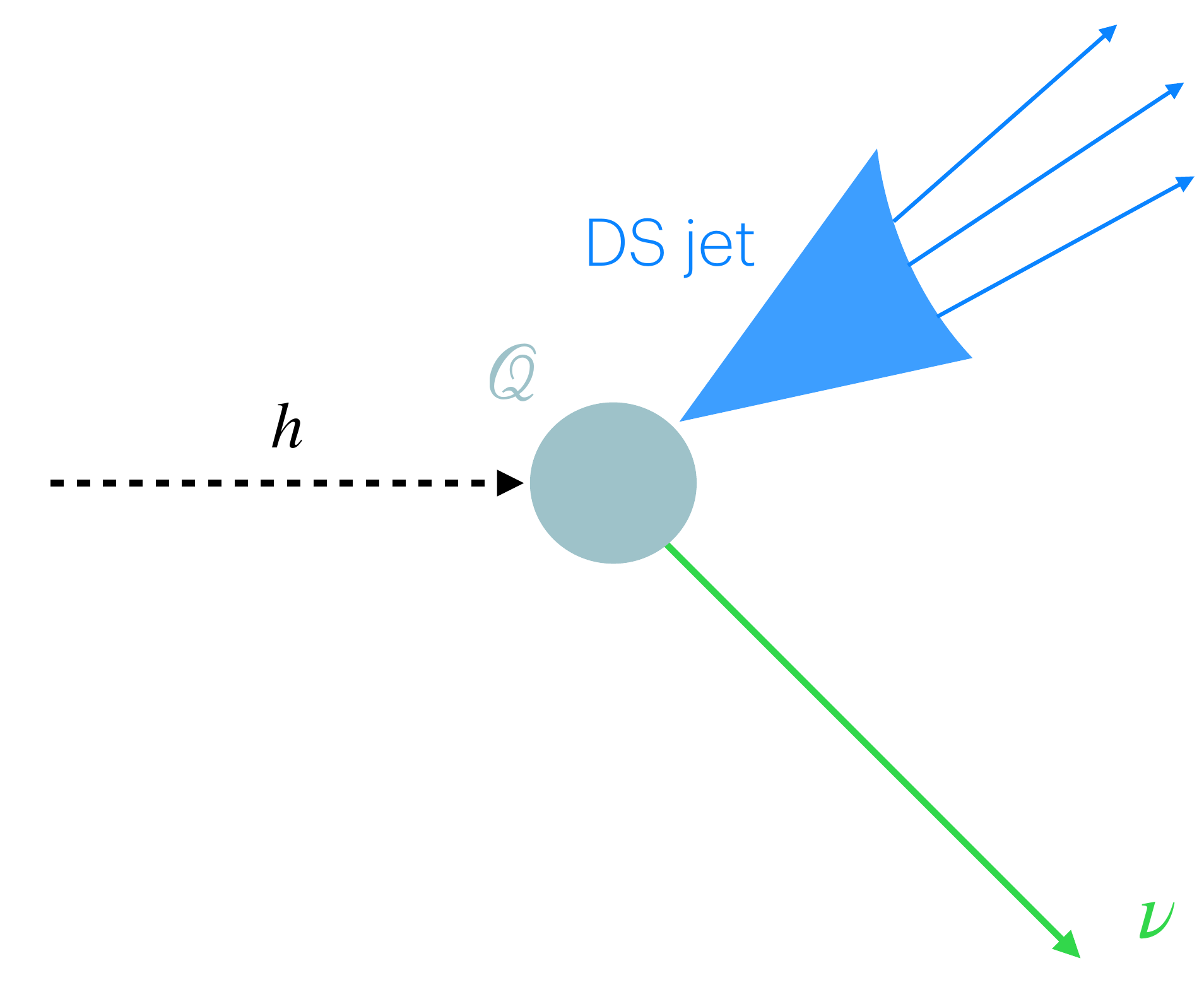}     
    \caption{Diagrams for the production of DS jets from decays of SM particles. The grey blob indicates the insertion of the $HL\On$ operator.
    {\bf Left}: gauge bosons and mesons decay, where the operator is inserted on the off-shell neutrino leg. {\bf Right}: Higgs decay, induced directly by the opeartor insertion.}
    \label{fig:phenocontact}
\end{figure}

SM particles can also decay into a dark jet, provided the decay is kinematically allowed. In this case, two distinct decay processes are possible, as illustrated in \cref{fig:phenocontact}:
\begin{enumerate}[(i)]
    \item The two-body\footnote{Additional probes of our model arise from three-body heavy meson decays. We discuss these in \cref{sec:present}, as the different Lorentz structure between SM and DS does not allow a simple analytical expression of the decay width.} decays of SM gauge bosons and mesons, shown on the left diagram of \cref{fig:phenocontact}. Analogously to the scattering process in \cref{fig:phenomixing}, below the EWSB scale these are mediated by the disintegration of an off-shell SM neutrino into a DS jet.
    \item The decay of the SM Higgs, shown in the right diagram of \cref{fig:phenocontact}. The latter is directly induced by the contact interaction in \cref{eq:compositeN}.
\end{enumerate}
The differential decay rates for meson, gauge boson and Higgs decays are 
\begin{align}
&\frac{\di \Gamma_{\bf m}}{\di p_D^2}=\frac{A_N y_\ell^2\Gamma_{\rm{SM,{\bf m}}}}{4\pi p^2_D}\left(\frac{m_{\bf m}^2}{m_\ell^2}\right)\left(\frac{v^2}{p^2_D}\right)\left(\frac{p^2_D}{\Lambda_{\rm{UV}}^2}\right)^{\Delta_M-3/2}\Pi_{\bf m}(p^2_D/m^2_{\bf m})\,,\qquad  &{\bf m}=\pi,K,D,B\label{eq:meson}\\
&\frac{\di \Gamma_X}{\di p_D^2}=\frac{A_N y_\ell^2\Gamma_{\rm{SM,b}}}{4\pi p^2_D}\left(\frac{v^2}{p^2_D}\right)\left(\frac{p^2_D}{\Lambda_{\rm{UV}}^2}\right)^{\Delta_N-3/2}\Pi_X(p^2_D/m^2_b)\,,\qquad  &X=W,Z\label{eq:boson}\\
&\frac{\di \Gamma_h}{\di p_D^2}=\frac{A_N y_\ell^2}{6 y_b^2}\frac{\Gamma_{\rm{SM},h}}{4\pi p_D^2}\left(\frac{p_D^2}{\Lambda_{\rm UV}^2}\right)^{\Delta_N-3/2}\Pi_h(p^2_D/m^2_h)\label{eq:higgs}
\end{align}
where $\Gamma_{\rm SM}$ indicates the SM total decay width of the relevant particle, and we approximated the total Higgs width with the $h\to b\bar b$ decay channel, where $y_b=0.024$. The $\prod_i(p_D^2/m_i^2)$ indicates the phase space factors, with the index $i$ indicating the different process considered; the full expressions of these terms are computed and collected in \cref{sec:SMdecay}. The latter are important to correctly predict the total number of event rate, especially for operators which are largely UV dominated, as the growth of the production rate for heavy invariant mass DS jets is compensated by the closure of the phase space. 

As a clear example, it can be immediately seen that for fixed values of $\Delta_N$ the production of DS jets from Higgs decays is more UV dominated than the gauge boson one. The reason is that the former does not involve the propagation of an off-shell neutrino. This would result in a substantial increase of the signal rate, compared to the $Z$ decay, for sufficiently irrelevant operator. 
Notice finally that the chiral suppression of SM decay rate in the meson decays in \cref{eq:meson} disappears since we are emitting a massive dark state.

The inclusive contribution to the SM branching ratios of EW and Higgs bosons can be estimated as
\begin{align}
&{\rm BR}({ Z/W}\to {\rm DS}+X_{\rm{SM}})\approx {\rm BR}_{\rm{SM}\to \nu X_{\rm{SM}}} \left(\frac{A_Ny_\ell^2}{4\pi(\Delta_N-5/2)}\right)\left(\frac{v^2}{m_{\rm{SM}}^2}\right)\left(\frac{m_{\rm{SM}}^2}{\Lambda_{\rm{UV}}^2}\right)^{\Delta_N-3/2}\, \!\!\!,\label{eq:BRmixing}\\  
&{\rm BR}(h\to {\rm DS}+\nu)\approx {\rm BR}_{h\to b\bar{b}}\left(\frac{A_Ny_\ell^2}{24\pi y_b^2(\Delta_N-3/2)}\right)\left(\frac{m_h^2}{\Lambda_{\rm{UV}}^2}\right)^{\Delta_N-3/2}\,. \label{eq:BRh}
\end{align}
We list all the SM processes considered in this work and their branching ratios in \cref{tab:SMdecay}. 

The two most relevant SM branching ratios for the DS phenomenology are $\mathrm{BR}(Z \to \nu \nu)=20\%$ and $\mathrm{BR}(h \to b\bar{b})=58\%$. For $\Delta_N=5/2$, the DS contribution to the $Z$ branching ratio is logarithmically enhanced by a factor $\log\!\left(m_Z^2/4\Lambda_{\rm IR}^2\right)$, analogously to what occurs for the neutrino disintegration cross section in \cref{eq:5/2log}. This logarithmic enhancement partially compensates for the overall suppression of the $Z$ branching ratio, rendering the DS contributions to the two branching ratios comparable. 
Conversely, for $\Delta_N=7/2$ we find $\mathrm{BR}(h \to \mathrm{DS} + \nu) \approx 8\,\mathrm{BR}(Z \to \mathrm{DS} + \nu)$. 

Meson decays become relevant only at low confinement scales and, up to differences in phase space and possible chiral suppression, their branching ratios can be obtained from \cref{eq:BRmixing} by replacing $m_Z^2 \to m_{\mathbf{m}}^2$. These branching ratios are also used to compute the expected signal yield at beam-dump experiments, as discussed in \cref{sec:present}.

\begin{table}[t]
    \centering
    \begin{tabular}{c| c c c c }
     SM &$m_{\rm SM}$ [GeV] &$X_{\rm SM}$& ${\rm Br}\lp {\rm SM}\to\nu X_{\rm SM} \rp$&Invisible bound\\
     \hline
    $h$ &125&$\nu$&$0.1\times 10^{-2}$$^*$&$10.7\times10^{-2}$$^\dagger$\\
    \hline
    $W^+$ &80&$\mu^+$& $11\times 10^{-2}$&$ 0.3\times 10^{-2}$\\
    $Z$ &91&$\nu$ & $20\times10^{-2}$&$0.1\times 10^{-2}$\\
    \hline
       $\pi^+$&0.140  & $\mu^+$&$99.99\times10^{-2}$&$ 0.8\times10^{-6}$\\
       &&$e^+$&$1.2\times 10^{-4}$&$0.8\times 10^{-6}$\\
       
       $K^+$ &  0.494&$\mu^+$&$64\times10^{-2}$&$0.2 \times10^{-2}$\\
 & &$e^+$&$1.6\times10^{-5}$&$1.4 \times10^{-7}$\\

         $D^+$ & 1.9&$\mu^+$& $3.7\times 10^{-4}$& $ 0.3\times 10^{-4}$\\
           & &$e^+$& & $ 9\times 10^{-6}$$^\dagger$\\
            & &$e^+X$& $16\times 10^{-2}$& $ 0.6\times 10^{-2}$\\
             & &$\mu^+X$& $18\times 10^{-2}$& $ 6\times 10^{-2}$\\

          $B^+$ & 5.3&$\mu^+$&&$9\times10^{-7}$$^\dagger$\\
            & &$e^+$&&$10\times10^{-7}$$^\dagger$\\
             & &$e^+X$&$11\times10^{-2}$&$5\times10^{-2}$\\
              & &$\mu^+X$&$11\times10^{-2}$&$7\times10^{-2}$\\
    \end{tabular}
    \caption{ Summary of decaying SM particles considered in this work. For each particle, we list the mass, the relevant SM final state in the decay, the inclusive branching ratio into DS jet (see \cref{eq:BRmixing,eq:BRh}), the invisible and displaced bounds. The former bound is given by twice the errror reported by PDG~\cite{ParticleDataGroup:2024cfk}.  $^\dagger$ and $^*$ refer to experimental upper bounds and theoretical prediction, respectively. Since the branching ratio of Higgs to two neutrinos is very small, we show the theoretical branching ratio to 4 neutrinos, ${\rm Br}(h\to 4\nu)\sim {\rm Br}(h\to ZZ)\times {\rm Br}(Z\to \nu\nu)^2$. 
    }
    \label{tab:SMdecay}
\end{table}

\subsection{Dark hadronization and resonance's lifetime}
\label{sec:WidthScaling}
After production, DS jets hadronize on a typical timescale $\tau \sim 1/\Lambda_{\rm IR}$ into DS resonances. In the strongly coupled regime, fragment evolution proceeds via intrinsically non-perturbative dynamics, making it difficult to sharply distinguish perturbative fragmentation from hadronization. This differs from weakly coupled scenarios, where partons are produced perturbatively and evolve through a calculable shower down to the confinement scale. As we will see this difference leaves precise imprints on the jet topology. 

The first consequence of strong coupling in the DS appears in the number of fragments per jet and in their momentum distribution. Assuming isotropic production in the jet rest frame, the Lorentz boost factor of an individual fragment can be approximated as
\begin{equation}
\gamma_i \sim \frac{E_D}{\Lambda_{\rm IR} \, \langle n_f(p_D) \rangle},
\end{equation}
where $E_D$ denotes the jet energy and $\langle n_f(p_D) \rangle$ is the average number of fragments.

We parametrize the jet multiplicity as $\langle n_f(p_D) \rangle \approx \left( \frac{p_D^2}{\Lambda_{\rm IR}^2} \right)^\xi$, assuming for simplicity that all fragments have roughly the same mass $\Lambda_{\rm IR}$. The exponent $\xi$ interpolates between different fragmentation regimes: values $\xi \ll 1/2$ roughly correspond to perturbative-like fragmentation, where the multiplicity grows slowly with energy, while $\xi \simeq 1/2$ describes the strongly coupled regime, which gives the maximal number of fragments $\langle n_f(p_D) \rangle_{\rm max} = \frac{p_D}{\Lambda_{\rm IR}}.$
The strong coupling limit reflects the democratic branching characteristic of maximally efficient energy splitting in a dense strongly interacting medium, in contrast to the collimated jets typical of perturbative dynamics (see  Ref.~\cite{Knapen:2016hky} for a discussion).

Following the original ideas of Fermi and Hagedorn~\cite{Fermi:1950frz,Hagedorn:1965st}, this scaling can also be interpreted as an average over a statistical ensemble in which, in each event, a jet of invariant mass $p_D^2$ fragments into $n_f \sim \sqrt{\frac{p_D^2}{\Lambda_{\rm IR}^2 + \mathbf{p}^2}}$ fragments, where $\mathbf{p}$ denotes the three-momentum of an individual fragment in the jet rest frame. Assuming local thermodynamical equilibrium in the interaction region, homogeneity and isotropy imply that all fragments carry the same magnitude of momentum while their directions are uniformly distributed. In this picture, each fragment follows a Maxwell-Boltzmann distribution with an effective temperature $T_{\rm IR}$, $f(\mathbf{p}) \sim \exp\Big[-\frac{\sqrt{\Lambda_{\rm IR}^2 + \mathbf{p}^2}}{T_{\rm IR}}\Big]$, and the mean number of fragments can be expressed as
\begin{equation}
\langle n_f(p_D^2) \rangle = 
\frac{1}{\mathcal{N}} \int_0^{\sqrt{p_D^2/4 - \Lambda_{\rm IR}^2}} 
\frac{p_D}{\sqrt{\Lambda_{\rm IR}^2 + \mathbf{p}^2}}
\exp\Big[-\frac{\sqrt{\Lambda_{\rm IR}^2 + \mathbf{p}^2}}{T_{\rm IR}}\Big] 
\mathbf{p}^2 \, {\rm d} |\mathbf{p}|, \label{eq:nframreal}
\end{equation}
where $\mathcal{N}$ is the normalization factor of the Maxwell--Boltzmann distribution.

\begin{figure}[t!]
    \centering
    \includegraphics[width=0.6\linewidth]{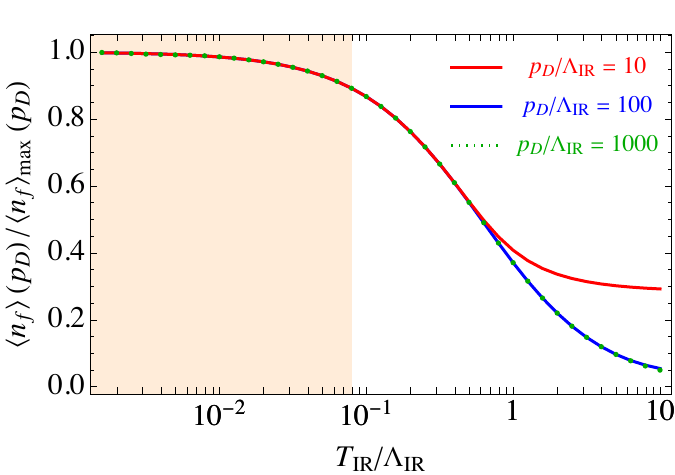}
    \caption{Ratio of the mean number of fragments, as defined in \cref{eq:nframreal}, with the maximal one, $\langle n_f \rangle_{\rm max}(p_D) = \frac{p_D}{\Lambda_{\rm IR}}$. Depending on the effective temperature of the thermal bath in the interaction region, $T_{\rm{IR}}$, the number of fragments decreases with a mild dependence on the jet invariant mass. The orange shaded region indicates the parameter region where the number of fragments is at least $90\%$ of its maximum.}
    \label{fig:nfragments}
\end{figure}

The result, shown in \cref{fig:nfragments}, compares the mean multiplicity with the maximal number of fragments allowed by kinematics. Producing fragments nearly at rest requires a sufficiently low bath temperature. For scales $\Lambda_{\rm IR} \sim T_{\rm IR}$, as suggested by AdS/CFT arguments~\cite{Hatta:2008qx} and phenomenological fits to light-meson forward fluxes~\cite{Becattini:2009ee}, the mean fragment multiplicity is reduced by roughly 40\% compared to the maximal number $\langle n_f(p_D) \rangle_{\rm max} = p_D / \Lambda_{\rm IR}$. Interestingly, this reduction exhibits only a mild dependence on the ratio $p_D / \Lambda_{\rm IR}$.
Keeping this reduction in mind, we will make use of the maximal number of fragments in the rest of the paper as a convenient reference for the kinematic upper bound. In this limit, each fragment effectively inherits the boost of the parent dark jet which greatly simplifies the signal kinematics. 

We estimate the proper lifetime of a single fragment by matching the operator in \cref{eq:compositeN} onto a one-particle resonance at energy \(E \sim \Lambda_{\rm IR}\). For fermionic operators, this matching can only be performed approximately as\footnote{This contrasts with quasi-conserved current operators, whose IR limit can be fully saturated by quasi-massless Goldstone bosons. A well-known example is the isovector axial quark current in the SM, which is matched to pions: \(\langle 0 | j_\mu^a | \pi^b(p)\rangle \approx \delta^{ab} i f_\pi p_\mu e^{-ipx}\).} 
\beq\label{eq:irmatching}
\langle 0 | \bar{\cal O}_N | \psi_s(p) \rangle \approx \kappa_{\rm IR}\Lambda_{\rm IR}^{\Delta_N - 3/2} \tilde u_s(p) e^{-ipx},
\eeq
where $p = (E_p, \vec p)$, with the energy being \(E_p = \sqrt{\vec p^2 + \Lambda_{\rm IR}^2}\), $\kappa_{\rm IR}$ is an unknown dimensionless matching coefficient, expected to be of order unity. We set $\kappa_{\rm IR}=1$ and discuss the phenomenological impact of this coefficient in \cref{sec:displaced}. The one-particle state is normalized as $\langle \psi_s(p) | \psi_{s'}(p') \rangle = 2 E_p (2\pi)^3 \delta^3(p - p') \delta_{ss'}$, and the spinors satisfy $\bar u_s(p) u_{s'}(p) = 2 E_p \delta_{ss'}$. 
In this framework, the same operator responsible for DS jet production also governs the decay of the light resonances. We assume that the DS contains no lighter states coupled to these fermionic resonances, so that the only decay channel is back to SM particles via the portal interaction. We will revisit these assumptions in \cref{sec:CompositenessModel}.

Under these assumptions the partial decay widths of the fermionic one-particle states to SM states can be expressed analogously to the HNL decays through charged- and neutral-current interactions. In the regime \(\Lambda_{\rm IR} < m_h\), these widths are identical, up to a normalization factor, to those derived in Refs.~\cite{Atre:2009rg,Bondarenko:2018ptm}. Consequently, the total width of $\psi$ is  
\beq\label{eq:GammaPsi}
\Gamma_\psi \approx \frac{G_F^2 \Lambda_{\rm IR}^5}{96 \pi^3} V_{\psi\nu_{\ell}}^2
= \frac{G_F^2 \Lambda_{\rm IR}^5}{96 \pi^3} \left(\frac{y_\ell v}{\Lambda_{\rm IR}}\right)^2 \left(\frac{\Lambda_{\rm IR}}{\Lambda_{\rm UV}}\right)^{2 \Delta_N - 3}\,,
\eeq
where we defined the mixing parameter
\begin{equation}
V_{\psi\nu_{\ell}} = \frac{y_\ell v}{\Lambda_{\rm IR}} \left(\frac{\Lambda_{\rm IR}}{\Lambda_{\rm UV}}\right)^{\Delta_N - 3/2}\,.    \label{eq:mixingangle}
\end{equation}
This width corresponds to a proper lifetime of
\beq
c \tau_\psi \approx 1.6\, y_\ell^2 \times 10^{-10 + 4 \Delta_N}~\mathrm{km} \, 
\left[\frac{10~\mathrm{MeV}}{\Lambda_{\rm IR}}\right]^{2\Delta_N} 
\left[\frac{\Lambda_{\rm UV}}{1~\mathrm{GeV}}\right]^{2\Delta_N - 3}\ ,
\eeq
which demonstrates that \(\tau_\psi\) can span several orders of magnitude depending on \(\Delta_N\), \(\Lambda_{\rm IR}\), and \(\Lambda_{\rm UV}\).

For the two cases of \(\Delta_N = 5/2\) and \(\Delta_N = 7/2\), proper displacements of the order of a typical detector size, $L \sim 1~\rm m$, can be achieved for mixing \(V_{\psi\nu_i} \sim 10^{-4}\) and \(\Lambda_{\rm IR} \sim 1~\rm GeV\):
\beq
\begin{aligned}
c \tau_\psi\vert_{5/2} &\approx y_\ell^2 \left(\frac{1.5~\rm GeV}{\Lambda_{\rm IR}}\right)^5 \left(\frac{\Lambda_{\rm UV}}{10~\rm TeV}\right)^2~\rm m, \\
c \tau_\psi\vert_{7/2} &\approx y_\ell^2 \left(\frac{5~\rm GeV}{\Lambda_{\rm IR}}\right)^7 \left(\frac{\Lambda_{\rm UV}}{1~\rm TeV}\right)^4~\rm m\, .
\end{aligned}
\eeq
As we discuss in the next section, such displacements, combined with a sizeable signal yield from neutrino disintegration, \cref{eq:totcatt}, enable emerging jets at future neutrino detectors like FPF and DUNE. Similarly, displaced decays of DS fragments back to the SM are expected in \(B\)-meson decays at LHCb and Belle-II, as well as in \(Z\) decays at FCC-ee, making composite neutrino models a powerful generator of new displaced signals at all these facilities. These signals would be inaccessible for weakly coupled HNLs, where the active-sterile mixing required to have the necessary displacement suppresses the signal yield too much. 

How the displaced jets appear in a detector can be inferred directly from the discussion above. Given the mean number of fragments and their proper lifetime, we can estimate the typical spatial separation of dark fragments in the detector as
\begin{equation}
\langle \Delta d_\psi \rangle \approx c\tau_\psi \, \gamma_i \beta_i \Delta \theta_i
\approx \frac{c\tau_\psi}{ n_{\rm f} }\, ,\label{eq:meandistance}
\end{equation}
where $\Delta\theta_i\approx1/(n_f\gamma_D)$ is the angular separation between the tracks in the lab frame, and in the last step we take the boost of each fragment to be equal to the boost of the dark excitation, \(\gamma_i = \gamma_D\), which is strictly valid only for very low effective temperatures in \cref{eq:nframreal}.

\begin{figure}[t!]
    \centering
    \includegraphics[width=0.6\linewidth]{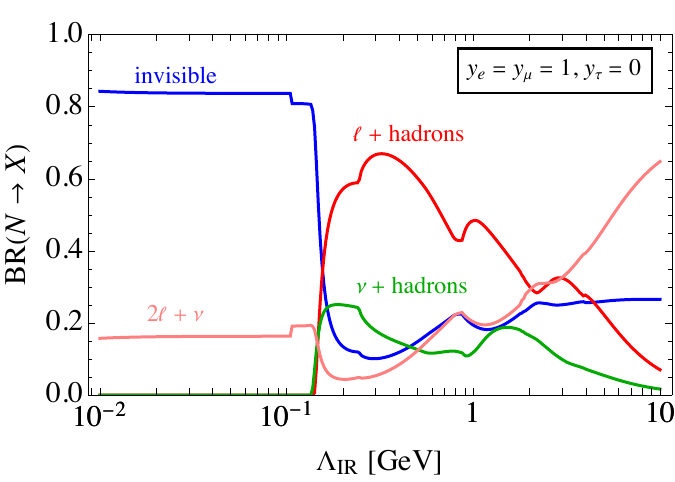}
    \caption{Branching ratios of HNL from Ref.~\cite{Atre:2009rg} for resonance mixing with $\nu_e$ and $\nu_\mu$ only.}
    \label{fig:HNL_br}
\end{figure}

The specific signatures of DS jets depend on the branching ratios of the fragments, which can be easily extracted from HNL decays~\cite{Plows:2022gxc,Liao:2017jiz,Atre:2009rg}, see \cref{fig:HNL_br}. As can be seen from \cref{fig:HNL_br}, although additional hadronic channels open up at larger fragment masses, the three-body leptonic widths scale more rapidly with $\Lir$. As a result, the fraction of purely hadronic decays decreases at high masses, making the probability of observing a displaced muon sizeable. 

For $\Lambda_{\rm IR} \ll 1~\rm GeV$, lifetimes become very long, producing missing-energy signatures at colliders and potential signals at beam-dump experiments such as CHARM and SHiP, where the decay volume is roughly 100 m. Prompt decays occur only for large \(\Lambda_{\rm IR}\), where the conformal approximation begins to break down (see \cref{fig:unfoldederror}). We therefore focus on long-lived and displaced signatures, which are both generic and calculable in this scenario.

\subsection{Invisible signals}
\label{sec:invisible}
If the fragment lifetimes are long compared to the size of the detector, the only available signal will be based on missing energy and momentum. In particular, the composite dynamics will contribute to the invisible branching ratio of the EW gauge and Higgs bosons, as computed in \cref{eq:BRmixing,eq:BRh}, respectively. 

At neutrino experiments the composite dynamic results in an enhancement of the NC to CC ratio, as first noticed in Ref.~\cite{Borrello:2025hal}. 
The SM NC events can be estimated as  
\begin{equation}
 B_{\rm NC}\approx N_\nu \sigma_{\rm t}^{-1}\sigma_{\rm{EW}}\sum_q x_q\lp\ell_q^2+\frac{r_q^2}3\rp\,,
\end{equation}
where $N_\nu$ is the total number of neutrinos and $\sigma_{\rm t}^{-1}$ is the surface density of the detector material, as summarized in \cref{tab:nuexp}. We approximate here the PDF average over the neutrino flux by assuming instead a monochromatic beam of neutrinos, with fixed energy $E_\nu=s/2m_p$. This seems a rough approximation, which however captures quite well the behavior of the full numerical results that we will show in \cref{sec:results}.  

\begin{figure}[t!]
    \centering
    \includegraphics[width=0.75\linewidth]{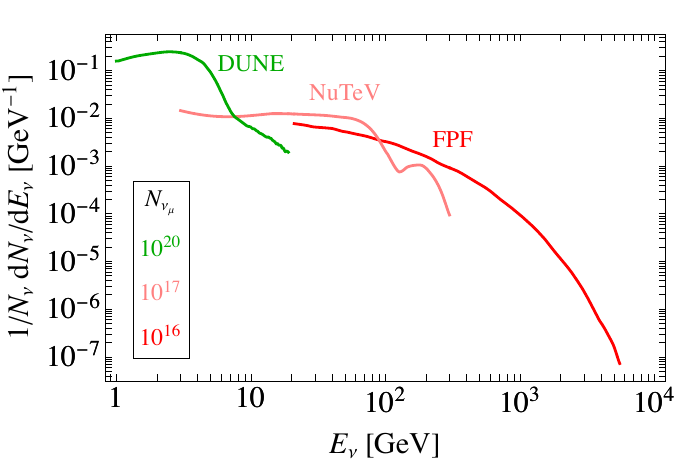}
    \caption{Summary plot of the normalized neutrino fluxes considered in this work. The green line indicates the \(\nu_\mu\) flux at DUNE Near-Detector~\cite{DUNE:2021cuw}, the pink is for NuTeV flux~\cite{NuTeV:2001whx} and the red for the FPF flux~\cite{MammenAbraham:2024gun,Feng:2022inv}.} 
    \label{fig:flux}
\end{figure}

The expected number of NC events produced by the DS portal can be written as 
\beq\label{eq:Signal:NC}
\begin{split}
   S_{\rm NC} &= N_\nu \sigma_{\rm t}^{-1}\sum_{q,\bar q}\int_{4\Lir^2}^{\hat s} \di p_D^2 \di x \di y f_q(x,x y s)\di E_\nu\frac{\di \Phi(E_\nu)}{\di E_\nu} \frac{\di^3\sigma_{2\rm DIS}(E_\nu)}{\di x \di y \di p_D^2}\,,\\
   &\approx B_{\rm{NC}}\frac{A_Ny^2}{4\pi(\Delta_N-5/2)}\left(\frac{v^2}{s}\right)\left(\frac{s}{\Lambda_{\rm{UV}}^2}\right)^{\Delta_N-3/2}\,,
\end{split}
\eeq
where $\hat s=2xm_pE_\nu$ is the center of mass energy for the scattering on a parton $q$, with proton PDF $f_q$; the flux of incoming neutrinos crossing the detector is $\di\Phi(E_\nu)/\di E_\nu$. For completeness we provide the explicit neutrino fluxes in \cref{fig:flux}. In the second line of \cref{eq:Signal:NC} we use the same approximation on the PDF dependence as for the SM cross section (see \cref{fig:pdfeffect} for an assessment of the goodness of this approximation).

\begin{figure}[t!]
    \centering
    \includegraphics[width=0.49\linewidth]{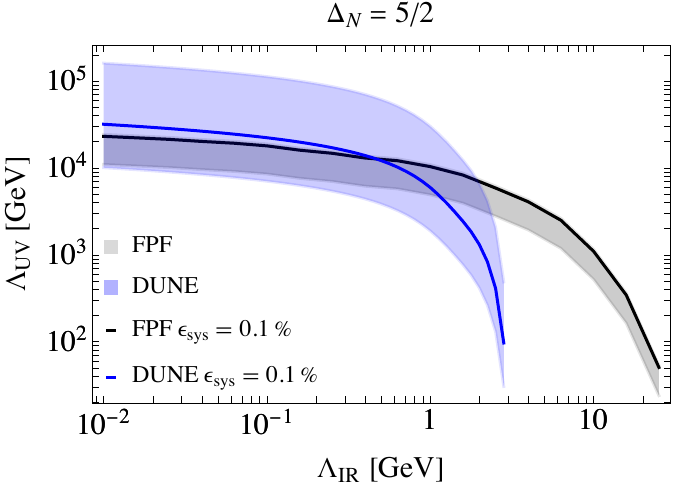}
     \includegraphics[width=0.49\linewidth]{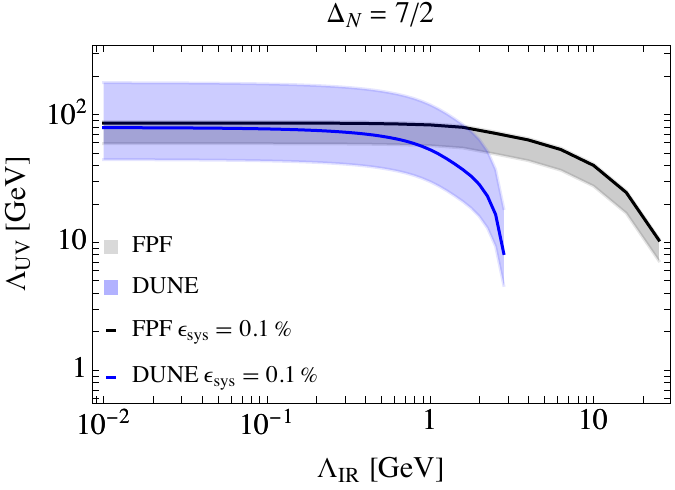}
    \caption{Effect of systematics uncertainties at DUNE ({\bf blue}) and FPF ({\bf black}). The colored region covers parameter space as $\epsilon_{\rm sys}$ ranges from 0 to 0.01. Solid lines represent the level of systematic uncertainties obtained at NuTeV~\cite{NuTeV:2001whx}.}
    \label{fig:systplt}
\end{figure}

The expected sensitivity can be estimated by assuming that the measured data is compatible with the background predictions, $B_{\rm{NC}}$. We define the test statistic $\lambda=S_{\rm NC}^2/(B_{\rm{NC}}+\epsilon_{\rm{sys}}^2B_{\rm{NC}}^2)$, which approximates the Poisson likelihood in the high statistics limit. The one-sided 95\% exclusion is obtained solving $\lambda= 2.71$, and can be written as an expected lower bound on the portal cut-off $\Lambda_{\rm{UV}}$,
\beq
\Lambda_{\rm{UV}}\gtrsim \sqrt{s}\left[\frac{y_\ell^2A_N}{6.4\pi(\Delta_N-5/2)}\left(\frac{v^2}{s}\right)\sqrt{\frac{B_{\rm{NC}}}{1+\epsilon_{\rm{sys}}^2B_{\rm{NC}}}}\right]^{1/2\Delta_N-3}\,.
\eeq
For future neutrino experiments as FPF (DUNE), the typical expected number of neutral currents is $B_{\rm{NC}}\approx 10^6\, (10^8)$ for a center of mass energy of the neutrino-proton collision of $\sqrt{s}\approx 100\,\rm{GeV}\, (3\, GeV)$. Then the expected reach on the cut-off is $\Lambda_{\rm{UV}}\gtrsim 1\, (20)\,\rm{TeV}$ for $\Delta_N=5/2$ and  $\Lambda_{\rm{UV}}\gtrsim 100\,  (250)\,\rm{GeV}$ for $\Delta_N=7/2$. The latter result is based on the assumption that the systematic uncertainties on this measurement can be brought to a negligible level compared to the background statistics. This implies  $\epsilon_{\rm{sys}}\lesssim 10^{-3}\, (10^{-4})$ for FPF (DUNE). In \cref{fig:systplt} we show the expected reach varying the systematic uncertainties. We will discuss more about the challenges of these measurements in \cref{sec:results}.

In principle, one could attempt to bin the new physics contribution to neutral-current (NC) events in terms of jet energy and angular distributions. However, the broad energy spectrum of the neutrino fluxes shown in \cref{fig:flux} precludes any meaningful discrimination between signal and background, as extensively discussed in Ref.~\cite{Borrello:2025hal}.

The expected reach of neutrino experiments should be compared with the present constraints from SM invisible branching ratios. For $\Delta_N=5/2,7/2$, the leading constraints comes from $Z$ and $h$ decays to invisible final states across the majority of the parameter space. It is instructive to impose the current upper bounds on $Z$ and $h$ invisible branching ratios (see \cref{tab:SMdecay}) and compute the rate of signal events at neutrino experiments. Taking $\Delta{\rm Br}_{\rm inv}$ as the limit on invisible $Z$ decays, we get
\begin{align}
    N_{\rm events}\lesssim B_{\rm NC} &\lp\frac{s}{m_Z^2}\rp^{\Delta_N-5/2}\frac{\Delta{\rm Br}_{\rm inv}}{\rm Br_{\rm inv}}\approx 5\times 10^{3} \lp\frac{B_{\rm NC}}{10^6}\rp\lp\frac{s}{m_Z^2}\rp^{\Delta_N-5/2}\,,\label{eq:upperinv}
\end{align}
which corresponds to $5\times 10^{3}, (6\times 10^{2})$ events for $\Delta_N=5/2, (\Delta_N=7/2)$ and $s=1000$ GeV$^2$. This estimate suggests that the constraint on $\Luv$ derived from enhancement in NC events becomes comparable to that obtained from electroweak boson decays only for $\Delta_N = 5/2$, where the signal event yield is of the order of the statistical error, $N_{\rm events} \sim \sqrt{B_{\rm NC}}$.

As a final remark, it is important to notice that the increase of the NC yield is a peculiar feature of an interacting DS that is not present in standard HNL scenarios. In the absence of interaction in the sterile sector, \cref{eq:compositeN} reduces to a rotation between the interaction and mass eigenstates. Consequently, inclusive processes experience no enhancement in this scenario. In fact, they do experience a suppression due to the reduced available phase space if the HNL is massive (see \cref{app:strongISS} for more details).

\subsection{Displaced signals}\label{sec:displaced}

\begin{figure}[t!]
        \centering

 \begin{minipage}{.75\linewidth}
        \centering
        {\small Displaced event at proton beam dump}
        \includegraphics[width=\linewidth]{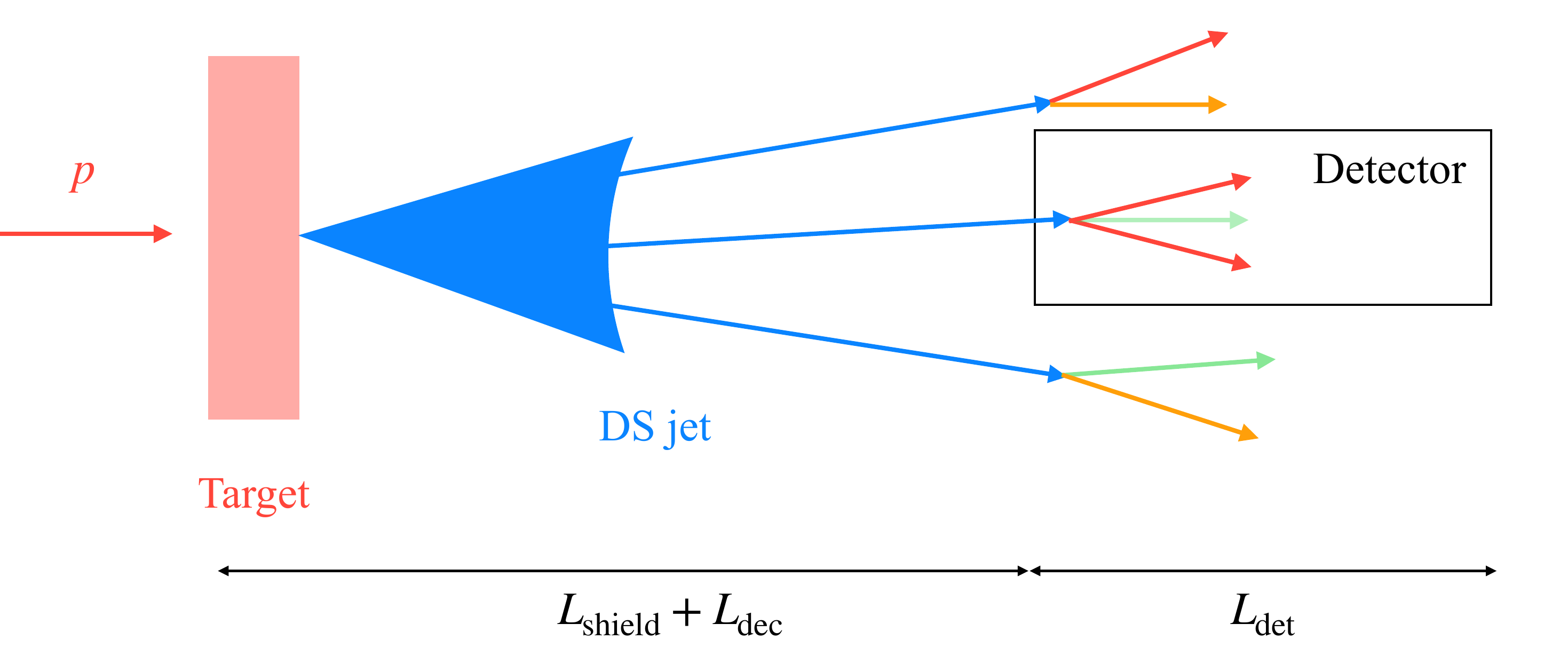}

    \end{minipage}
    \begin{minipage}{0.6\linewidth}
        \centering
        
        {\small Emerging jet at neutrino beam}
        \includegraphics[width=\linewidth]{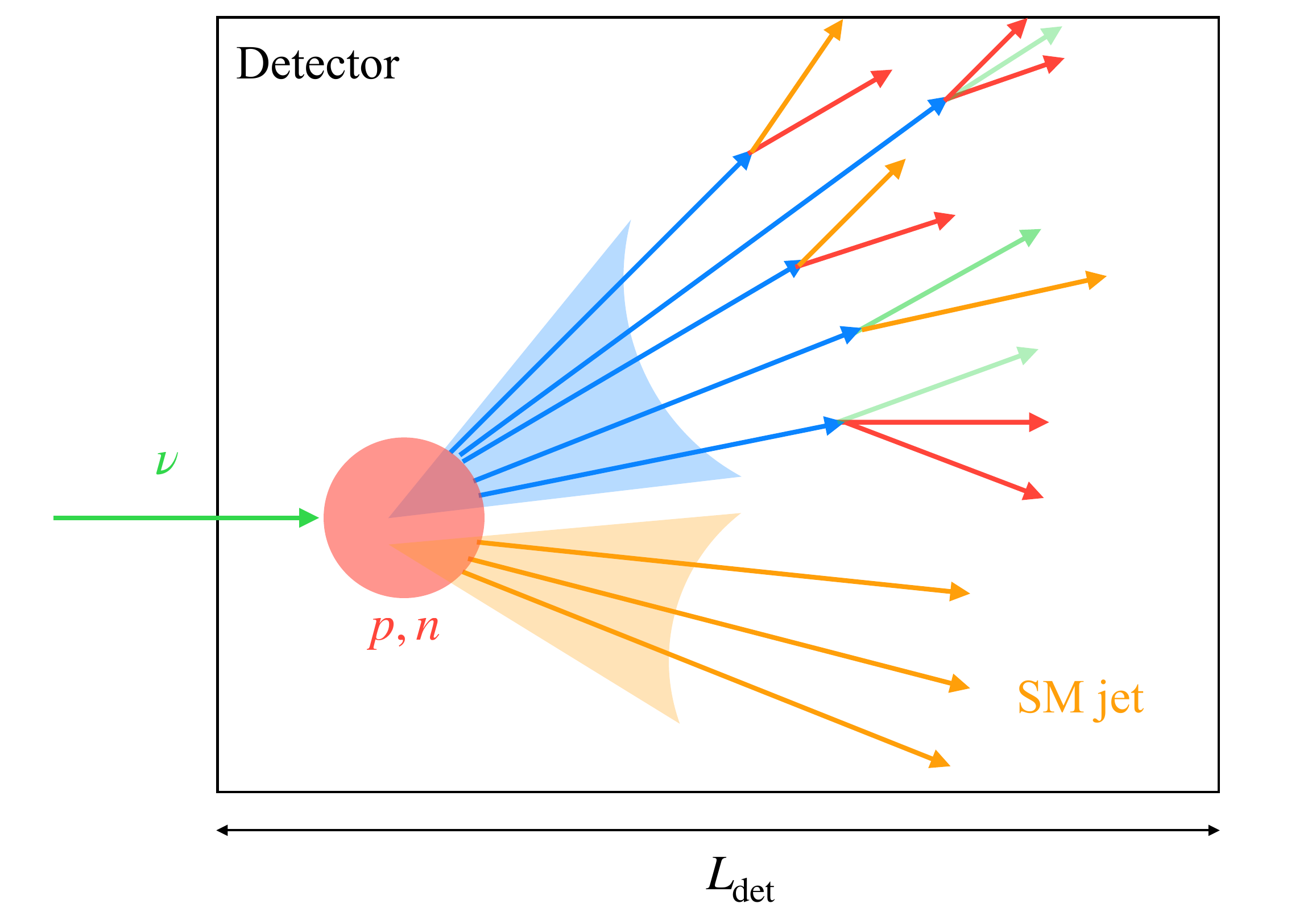}
        
    \end{minipage}

    \vspace{0.5cm}

    \caption{
    Sketch of DS production at proton beam-dump facilities, where a single displaced vertex is observed in the detector ({\bf upper panel}), and at neutrino experiments, featuring an emerging jet signature with many visible tracks appearing displaced from the primary vertex ({\bf lower panel}). Blue (green) arrows indicate dark sector particles (neutrinos) which do not leave tracks in the detector, red (orange) arrows are visible leptons (hadrons). The branching ratios of the decay channels are given in \cref{fig:HNL_br}. 
    }
    \label{fig:emvsdispl}
\end{figure}

If the fragments lifetimes are shorter than or comparable to the detector size, the possibility to observe a displaced vertex opens up. We consider three main classes of displaced signals. First, in beam-dump experiments or neutrino experiments used in dump mode, SM resonances are produced at the target, and their decays into DS jets generate dark resonances that propagate through the decay volume, with length $L_{\rm dec}$, and subsequently decay back into SM particles within a detector of size $L_{\rm det}$ (\cref{fig:emvsdispl}, upper panel).

A second possibility arises in collider experiments, where EW gauge bosons and the Higgs are produced nearly at rest and decay into DS jets, which hadronize into dark-sector resonances. These subsequently decay back into SM states with a measurable displacement inside the detector modules.

Third, in neutrino experiments, an incoming neutrino beam interacts within the detector and produces DS jets. These may decay back into SM particles within the same detector of size $L_{\rm det}$, yielding an observable displaced signature (\cref{fig:emvsdispl}, lower panel). In this case, the signal consists of the primary interaction point, where the NC hadronic activity is observed, together with displaced vertices in the same detector. In some neutrino experiments, a hybrid scenario is also possible: dark resonances are produced in a detector of size $L_{\rm det}$ through NC interactions, propagate over a distance $L_{\rm dec}$, and are then detected in a downstream module of size $L_{\rm det}'$.

\paragraph{Signal strength}
We are now ready to estimate the signal strength in displaced events and characterize their nature. 
Assume that in a given scattering event, a DS object of invariant mass $p_D^2$ is produced and hadronizes into $n_f$ light fragments, each carrying the boost of the parent jet, $\gamma_D$. As discussed above, we assume here that the dark hadronization produces the maximal number of fragments, $n_f\equiv\langle n_f(p_D) \rangle_{\rm max}$.

The probability of having a single fragment $i\in\{1,n_f\}$, with proper lifetime $\tau_\psi$, decaying between the minimum observable displacement in the detector, $\Delta L$, and the detector length, $L_{\rm det}$, is 
\begin{equation}
    P_{\mathrm{dec}}= \exp\left(-\frac{\Delta L}{c\tau_\psi (\gamma \beta)_D}\right)-\exp\left(-\frac{L_{\rm det}}{c\tau_\psi (\gamma \beta)_D}\right)\approx \frac{L_{\rm det}}{c(\gamma\beta)_D\tau_\psi}\ ,
\end{equation}
where in the last equation we expanded for $c\tau_\psi(\gamma\beta)_D\gg L_{\rm det}$ and ignored a small correction proportional to $\Delta L\ll L_{\rm det}$. In order to capture detector configuration with a non-instrumented decay volume we can just shift the detector length in the formula above: $L_{\rm det}\to L_{\rm det}+L_{\rm dec}$. Since the single fragment has an irreducible invisible branching ratio, as shown in \cref{fig:HNL_br}, this has to be subtracted in order to select the probability of having visible displaced signatures; namely, $P_{\mathrm{dec}}\rightarrow P_{\mathrm{dec}}(1-\rm{BR}_{\rm{inv}})$.

For a DS object fragmenting into $n_f$ fragments, each with an individual decay probability $P_{\rm decay}$, the probability to observe at least $n$ displaced decays is naturally given by the binomial distribution
\begin{equation}
P(\ge n \text{ decays}) = \sum_{k=n}^{n_f} \binom{n_f}{k} \, P_{\rm dec}^k \, (1-P_{\rm dec})^{n_f-k} \,.
\end{equation}
The probability of seeing \emph{exactly} one fragment decaying inside the detector is then given by  
\begin{equation}
P(1 \text{ decay}) = n_f \, P_{\rm dec} \, (1-P_{\rm dec})^{n_f-1} \,.
\end{equation}
These particular events with only a single fragment decaying will be indistinguishable from a weakly coupled HNL decaying displaced. A more distinctive signature arises when at least 2 fragments decay inside the detector volume into visible final states. This multiple displaced vertices signature is a hallmark of strongly coupled dark sectors, which allows a clear distinction from a weakly coupled HNL. Moreover having more than one DV inside the detector is extremely beneficial in terms of SM background suppression. 

How likely a given experiment is to detect multiple DV depends on the lifetime of the fragments compared to the decay volume and the detector size. If the lifetime is longer than the sum of the typical decay volume length and the detector size, then the probability of having two tracks will be suppressed. Conversely, if the lifetime matches the decay volume size, multiple DVs can be reconstructed in the detector without any penalty and emerging jets signatures with multiple DVs of different displacement can appear. 

One might be interested in selecting a particular decay channel of the dark resonances to suppress SM background. For example, requiring at least two leptons in a DV, effectively suppresses SM backgrounds from SM neutrino interactions, which produce at most one lepton. The probability to observe $n$ displaced decays with at least two leptons in the final state is 
\begin{equation}
P(\ge n \text{ decays},2\ell) = \sum_{k=n}^{n_f} \binom{n_f}{k} \, P_{\rm dec}^k \, (1-P_{\rm dec})^{n_f-k}\left(1-(1-\rm{BR}_{2\ell+\nu})^k\right) \,,\label{eq:dilep}
\end{equation}
where $(1-\rm{BR}_{2\ell+\nu})^k$ is the probability of having zero di-leptons pair in a vertex for $k$ fragments, which we take the complement of. $\mathrm{BR}_{2\ell+\nu}$ is typically $\sim 20\%$ at low $\Lambda_{\rm IR}$ (see \cref{fig:HNL_br}).

\paragraph{Beam dump experiments}
A single displaced vertex accompanied by a dilepton pair constitutes a generic and essentially background-free signature, yielding a sizable event rate in the kinematic regime relevant for proton beam-dump experiments. This signature was, in fact, the primary target of the CHARM and CHARM~II searches for heavy neutral leptons (HNLs)~\cite{CHARM:1985nku,CHARMII:1994jjr}. The corresponding signal yield can be expressed as
\begin{equation}\label{eq:1DV}
N_{\rm 1DV(2\ell)} = \sum_{M} \int \di p \, \di \Omega \, \di p_D^2 \, \Phi_{\rm{M}}(p,\Omega)\,\,\frac{\di \rm{BR}(M \to \mathcal{O}_N+X)}{\di p_D^2}\,P(\ge 1 \text{ decay}, 2\ell)\,,
\end{equation}
where $\Phi_M(p,\Omega) = \di^2N_M/\di\Omega \di p$ denotes the flux of mesons $M$ with momentum $p$, which dominate HNL production~\cite{Bondarenko:2018ptm}. The parametrization of this flux is detailed in \cref{app:mesonfluxes}. The differential branching ratio with respect to the DS jet invariant mass is convoluted with the probability that a jet produces at least one DV with a dilepton pair, as given in \cref{eq:dilep}. This probability exhibits a nontrivial dependence on the energy distribution of the parent meson flux, which determines the boost of the DS jets and, consequently, that of their decay products.

\begin{figure}[H] 
    \centering

    \begin{subfigure}[b]{0.47\linewidth}
        \centering
        \includegraphics[width=\linewidth]{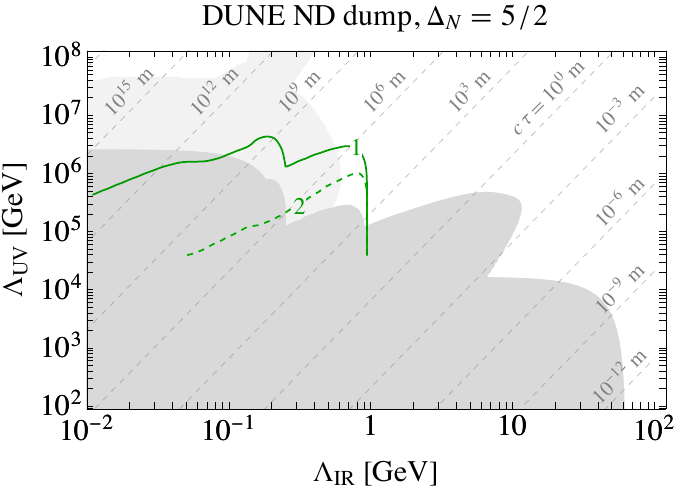}
    \end{subfigure}
    \hfill
    \begin{subfigure}[b]{0.47\linewidth}
        \centering
        \includegraphics[width=\linewidth]{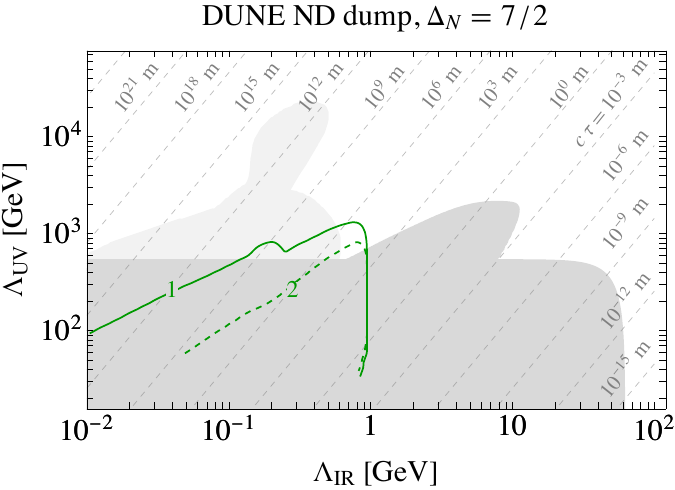}
    \end{subfigure}

    \vspace{0.1cm} 

    \begin{subfigure}[b]{0.47\linewidth}
        \centering
        \includegraphics[width=\linewidth]{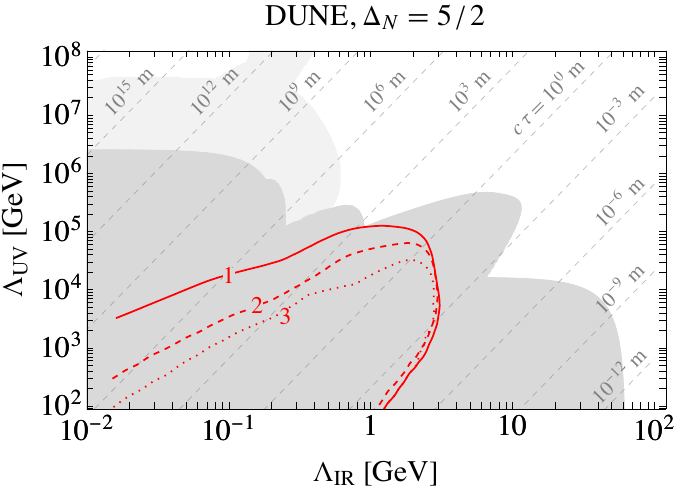}
    \end{subfigure}
    \hfill
    \begin{subfigure}[b]{0.47\linewidth}
        \centering
        \includegraphics[width=\linewidth]{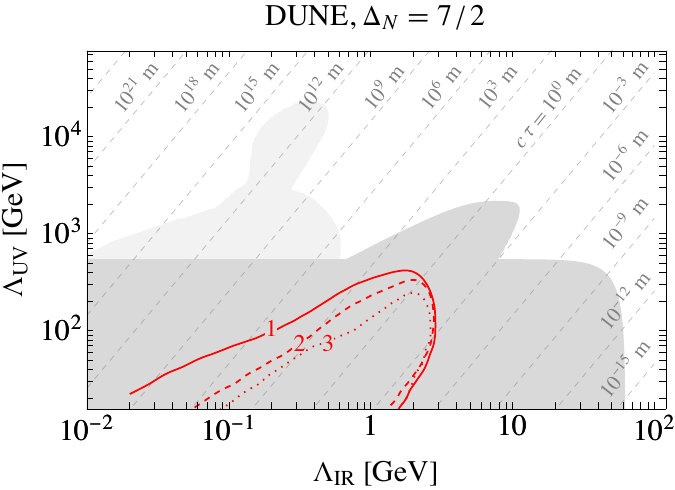}
    \end{subfigure}
    \caption{Expected reach for increasing track multiplicity, for $\Delta_N = 5/2$ (left column) and $\Delta_N = 7/2$ (right column). {\bf Top row:} reach for DUNE ND in dump mode (production at target). {\bf Bottom row:} reach for displaced vertex searches at DUNE ND (production in the detector). In each plot, the dotted gray lines indicate isocontours of the fragment decay length. For details on the production mechanism and the experimental setups see \cref{sec:CrossSecScaling} and \cref{sec:future}, respectively. The grey shaded region is excluded by current constraints, assuming that the conformal regime persists up to the electroweak scale (see \cref{sec:CompositenessModel}).}
    \label{fig:jets_LE}
\end{figure}

In the context of future beam-dump experiments, the observation of multiple displaced vertices provides a particularly distinctive signature. In particular, requiring at least two DVs would unambiguously discriminate the composite HNL portal from minimal scenarios. However, in the kinematically accessible region, the probability of observing more than one DV is suppressed due to the long lifetime of the fragments compared to the size of the decay volume. As shown in the top row of \cref{fig:jets_LE}, the only region of parameter space where the rate for two DVs is not strongly suppressed relative to the single-DV case corresponds to $\Lir \approx 1\,\mathrm{GeV}$, i.e.\ close to the kinematic threshold of planned beam-dump experiments, such as DUNE ND (in dump mode) and SHIP, which have comparable luminosities and center-of-mass energies as reported in \cref{tab:dumpexp}.

\paragraph{Collider experiments}
Colliders, both hadronic and leptonic, produce large numbers of EW gauge and Higgs bosons. As we show in \cref{sec:results}, the leading observable is a DV from a DS jet produced by $Z$ bosons decaying at rest. Using the decay width in \cref{eq:boson}, we can write the branching ratio as 
\begin{align}\label{eq:displbr}
    {\rm Br}(Z\to {\rm displ}&)\approx \frac{1}{\Gamma_Z}\int_{4\Lir^2}^{m_Z^2}\frac{\di\Gamma(Z\to \On)}{\di p_D^2}P(\ge n \text{ decays})\notag\\&\approx \left[V_{\psi\nu}^2L_{\rm det}\Gamma_\psi\right]\times\left[\frac{A_N}{4\pi}\lp\frac{m_Z^2}{\Lir^2}\rp^{\Delta_N-2}\hat\Pi_{\Delta_N}(m_Z,\Lir)\right]\,,
\end{align}
where we defined a new phase space factor
\begin{equation}\label{eq:PiHat}
\hat\Pi_{\Delta_N}(m_Z,\Lir)=\int_{4\Lir^2}^{m_Z^2}\frac{\di p_D^2}{m_Z^2(\gamma\beta)_D}\Pi_{Z}\left(\frac{p_D^2}{m_Z^2}\right)^{\Delta_N-3}\,,
\end{equation}
and the boost of the DS excitation is $(\gamma\beta)_D=\sqrt{(p_D^2 + m_Z^2)^2/(4p_D^2  m_Z^2)-1}$. 

\begin{figure}[t!] 
    \centering

    \begin{subfigure}[b]{0.47\linewidth}
        \centering
        \includegraphics[width=\linewidth]{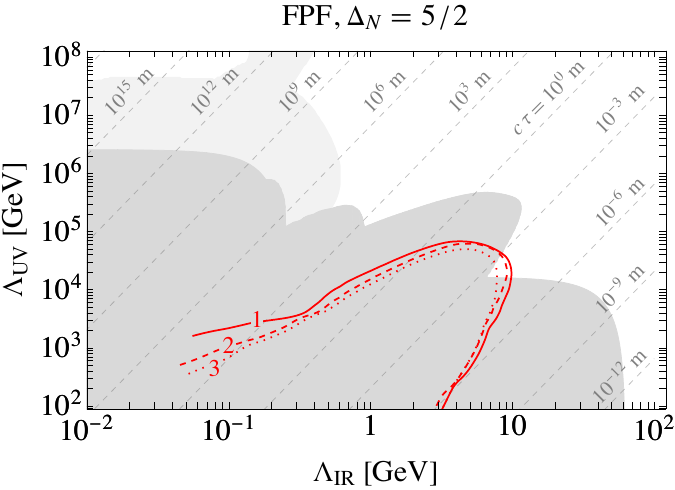}
    \end{subfigure}
    \hfill
    \begin{subfigure}[b]{0.47\linewidth}
        \centering
        \includegraphics[width=\linewidth]{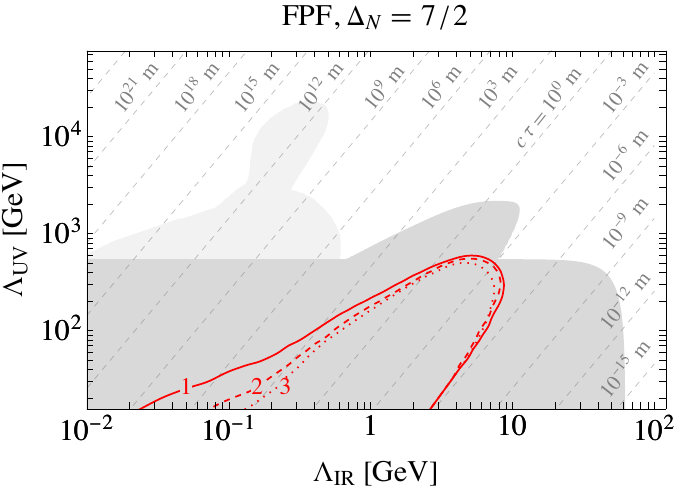}
    \end{subfigure}
    
    \vspace{0.1cm} 

    \begin{subfigure}[b]{0.47\linewidth}
        \centering
        \includegraphics[width=\linewidth]{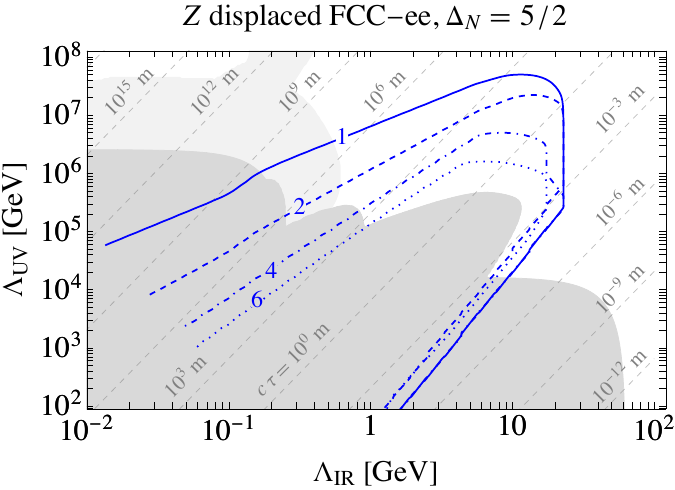}
    \end{subfigure}
    \hfill
    \begin{subfigure}[b]{0.47\linewidth}
        \centering
        \includegraphics[width=\linewidth]{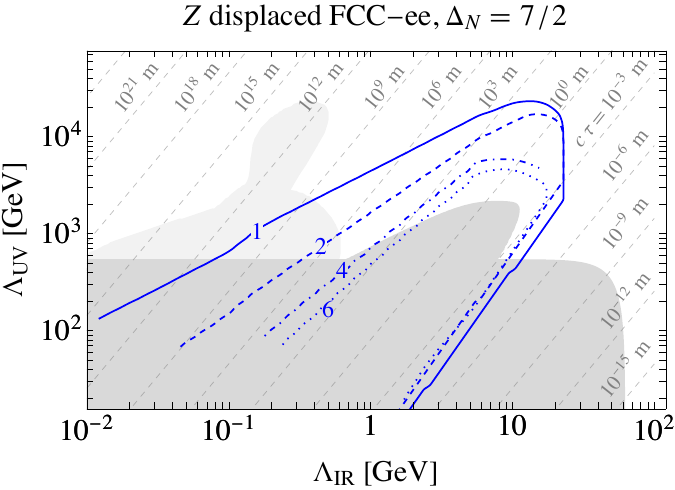}
    \end{subfigure}
    \caption{Same as \cref{fig:jets_LE}, for FPF (top row) and FCC-ee (bottom row).}
    \label{fig:jets_HE}
\end{figure}

The clean environment of lepton colliders does not require focusing on a specific visible branching ratio, although requiring a dilepton pair could be advantageous at the LHC. In considering current experiments, we set $n=1$ in order to recast LEP HNL searches~\cite{DELPHI:1996qcc}. For future high-energy colliders such as FCC-ee, we instead investigate whether composite signatures with multiple DVs can be observed in light of existing constraints. The results shown in the last row of \cref{fig:jets_HE} are encouraging in this respect, indicating that searches requiring at least two DVs—or even higher DV multiplicities, as expected in emerging-jet signatures—could probe a large fraction of the currently unconstrained parameter space.

\paragraph{Neutrino experiments} 
In general, computing the number of events at neutrino experiments requires the fully differential distribution in the DS jet invariant mass, $p_D^2$, and energy, $E_D$. We can arrive at a simpler expression by considering that in the center of mass frame the visible and the DS jets are produced back-to-back, and that in the DS jet rest frame the fragments are produced almost at rest. Thus the average boost of the fragments can be approximated with the boost between the center of mass and the lab frame, $(\gamma\beta)_D\approx\sqrt{s}/m_p$.

With this approximation, the expected number of displaced decays at neutrino experiments can be written as
\begin{align}\label{eq:displN}
S_{\rm{displaced}} &\approx N_\nu \, \sigma_{\rm t}^{-1} \int_{4\Lambda_{\rm{IR}}^2}^{s} \di p_D^2\frac{d\sigma_{2\rm{DIS}}}{dp_D^2} \, P(\ge n \text{ decays}, 2\ell) \notag \\
&\approx \left[ B_{\rm{NC}} \, V_{\psi\nu}^2 \, L_{\rm{det}} \, \Gamma_\psi \right] \times \left[ \left(\frac{A_N m_p}{4\pi \Lambda_{\rm{IR}}}\right)\left(\frac{s}{\Lambda_{\rm{IR}}^2}\right)^{\Delta_N - 5/2} \tilde{\mathcal{F}}_{\Delta_N}(\Lambda_{\rm{IR}}^2, s) \right] \, ,
\end{align}
where $\Gamma_\psi$ is given in \cref{eq:GammaPsi}. For simplicity, we fix $\mathrm{BR}_{2\ell+\nu} = 1$. The form factor $\tilde{\mathcal{F}}_{\Delta_N}$ is defined as
\begin{equation}\label{eq:dispff}
\tilde{\mathcal{F}}_{\Delta_N}(\Lambda_{\rm{IR}}^2,s)\equiv\frac{1}{\sum_q x_q(\ell_q^2+r_q^2/3)}\int_{4\Lir^2}^{s}\frac{dp_D^2}{s}\mathcal{F}(p^2_D/s,s) \lp \frac{p_D^2}{s}\rp^{\Delta_N-3}\,,
\end{equation}
where the PDF form factor $\mathcal{F}(p^2_D/s,s)$ is taken from \cref{eq:formfactor}. The form factor for displaced decays exhibits a modified power-law dependence, scaling as $\Delta_N - 3$ rather than $\Delta_N - 7/2$, due to the contribution from the fragment multiplicity in the decay probability.

\Cref{eq:displN} shows explicitly that at fixed mixing angle between the single particle states, $V_{\psi\nu}$, the displaced events for an interacting DS exhibit an energy-dependent enhancement compared to the usual weakly coupled HNL: the term in the squared parenthesis scales as $m_p/
\sqrt{s}$, which is related to the boost between the center of mass and the lab frame. We observe that for $\Delta_N = 5/2$ this enhancement is moderate, whereas the $\Delta_N = 7/2$ scenario maximally benefits from this effect. 

We consider both final states with $n=1$ (at least one DV) and $n=2$ (at least two DVs). The latter provide a clear discriminator with respect to minimal HNL scenarios, where a neutral-current interaction is followed by the displaced decay of a single HNL into dileptons or hadrons~\cite{CHARM:1985nku, CHARMII:1994jjr, Mishra:1987xh}. In contrast to the beam-dump case, displaced vertices produced in neutral-current events are more likely to yield multiple displaced decays, particularly in experiments probing large center-of-mass energies $\sqrt{s}$, such as the FPF. This is illustrated in \cref{fig:jets_LE,fig:jets_HE}: while DUNE can produce more than one DV only near its kinematic threshold, the expected yields at the FPF for single and multiple DVs are comparable, so that even emerging-jet signatures could become observable (see \cref{tab:nuexp} for a summary of the key features of the planned neutrino experiments).

Using the constraint from $Z$ decays we can bound from above the number of expected displaced events at future neutrino experiments as
\begin{equation}
S_{\rm{displaced}}\lesssim \left(\frac{B_{\rm{NC}}}{10^6}\right)\left(\frac{s}{m_Z^2}\right)^{\Delta_N-2}\frac{m_p}{\sqrt{s}}\frac{\tilde{\mathcal{F}}_{\Delta_N}(\Lambda_{\rm{IR}}^2,s)}{\hat\Pi_{\Delta_N}(m_Z,\Lir)}\,,  \label{eq:upperdis}
\end{equation}
where $\hat\Pi_{\Delta_N}$ is defined in \cref{eq:PiHat}.
From this equation, it is clear that the expected number of displaced events at the FPF, assuming $B_{\rm{NC}} = 10^6$, is below unity. Displaced searches at the FPF may nevertheless have a slight advantage over existing constraints due to their favorable kinematics at large $\Lir$, as shown in \cref{fig:jets_HE}.  Conversely, DUNE, with its projected full luminosity corresponding to $B_{\rm{NC}} = 10^8$, can in principle probe new regions of parameter space with a non-zero displaced signal for $\Delta_N = 5/2$. However, much of this region is challenged by the CHARM beam-dump constraints discussed above.

The upper bounds in \cref{eq:upperdis,eq:upperinv} strongly motivate the discussion in \cref{sec:breakingCFT}, where we show how the constraints from EW physics can be circumvented by changing the UV behavior of the theory, leading to larger available signal regions for future neutrino experiments.

\begin{figure}
    \centering
    \includegraphics[width=0.49\linewidth]{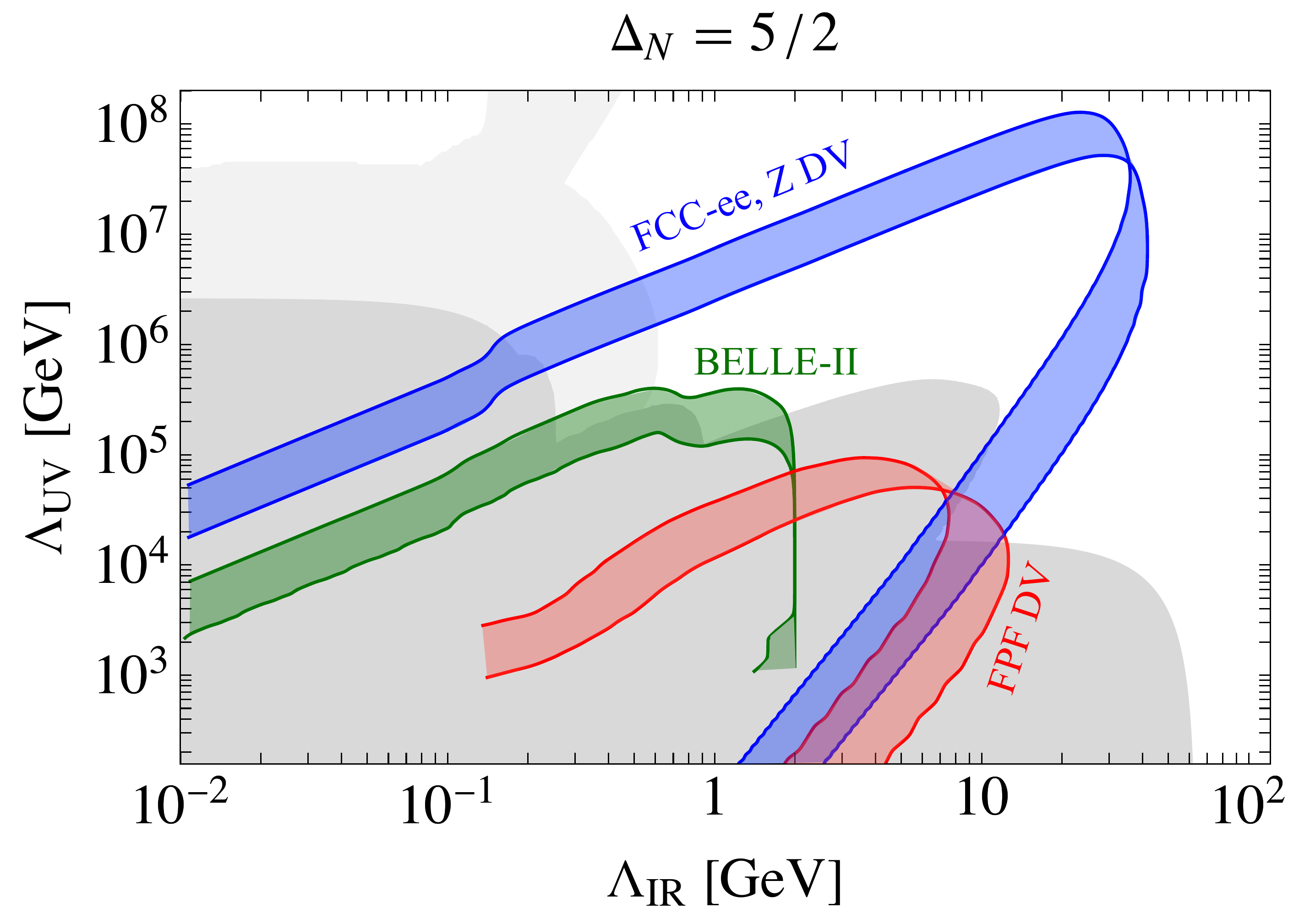}
    \includegraphics[width=0.49\linewidth]{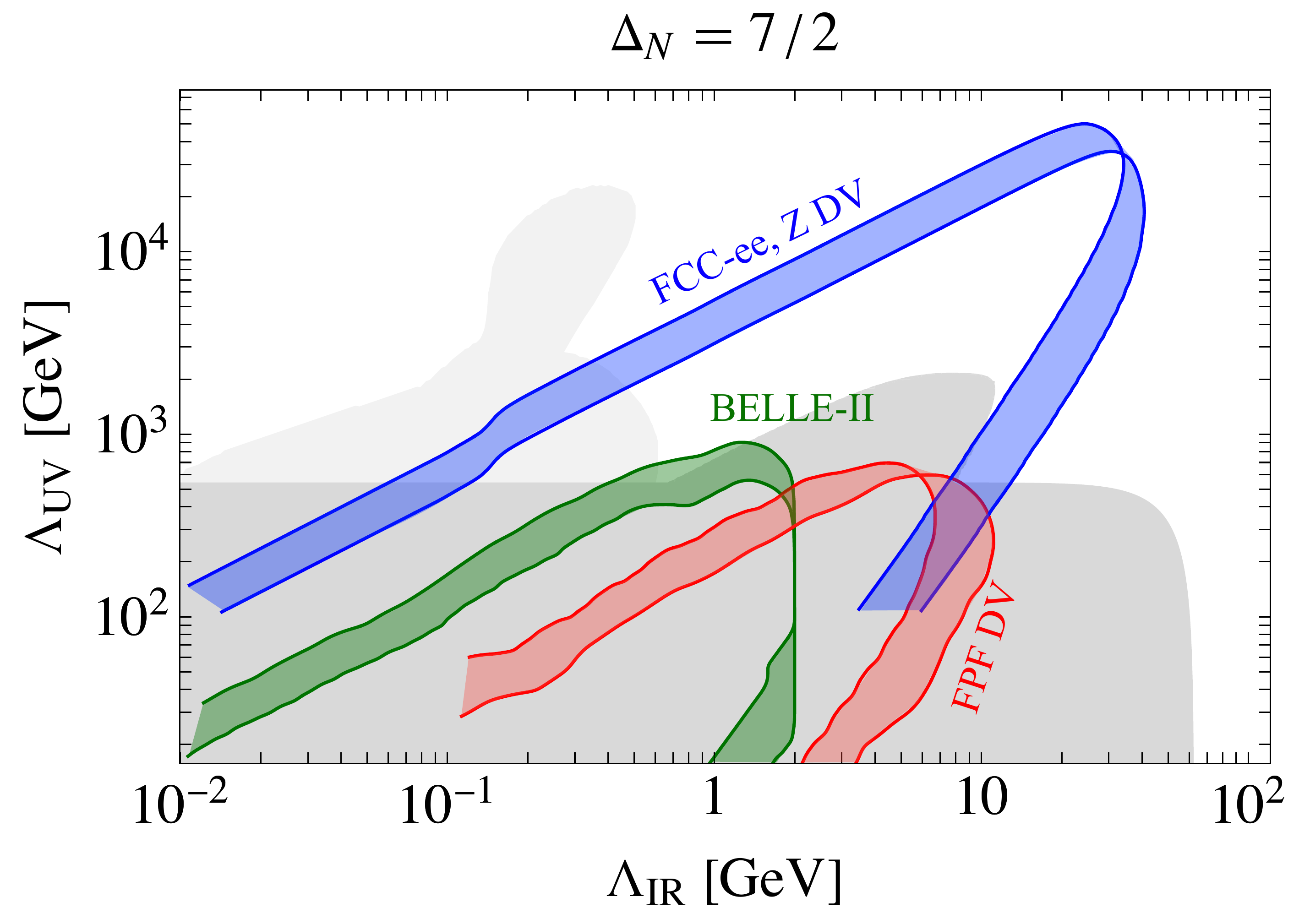}
    \caption{Uncertainty associated with the unknown dimensionless matching coefficient $\kappa_{\rm IR}$ in \cref{eq:irmatching}, for $\Delta_N=5/2$ (left) and $\Delta_N=7/2$ (right). The colored bands show the variation of the projected sensitivity to displaced-vertex searches as $\kappa_{\rm IR}$ is varied from $1/3$ to $3$. We show the excluded region for $\kappa_{\rm IR}=1$ as a reference.}
    \label{fig:lifetimeeffect}
\end{figure}

As a final remark, the lifetime of the dark-sector resonances carries an uncertainty associated with the dimensionless coefficient $\kappa_{\rm IR}$ in the matching of \cref{eq:irmatching}. This coefficient rescales the decay width of the light states according to
\begin{equation}
\Gamma_\psi\to\kappa_{\rm IR}^2\Gamma_\psi\,.
\end{equation}
In the long-lived regime, the probability for a fragment to decay inside the detector is linear in $\Gamma_\psi$, as shown in \cref{eq:displbr,eq:displN} and the production rate scales as
\begin{equation}
S_{\rm displaced}\propto\kappa_{\rm IR}^2\frac{1}{\Lambda_{\rm UV}^{4\Delta_N-6}}\, ,
\end{equation}
and therefore, at fixed signal yield,
\begin{equation}
\Lambda_{\rm UV}^{\rm reach}(\kappa_{\rm IR})=\Lambda_{\rm UV}^{\rm reach}(\kappa_{\rm IR}=1)\kappa_{\rm IR}^{1/(2\Delta_N-3)}\, .
\end{equation}
Consequently, increasing the matching coefficient by a factor of three increases the asymptotic reach in $\Lambda_{\rm UV}$ by a factor $\sqrt{3}\simeq1.73$ for $\Delta_N=5/2$ and by a factor $3^{1/4}\simeq1.32$ for $\Delta_N=7/2$. A factor-three decrease gives the inverse shifts. Away from the long-lived regime the dependence is non-linear because the full exponential decay probability must be retained. We therefore also evaluate the effect numerically in \cref{fig:lifetimeeffect}, where we compare the projected displaced sensitivities of FPF, Belle-II and FCC-ee for $\kappa_{\rm IR}=1/3,3$.

\section{UV completions and benchmark scenarios}
\label{sec:CompositenessModel}
So far, the discussion has assumed that the DS is characterized by three parameters: the scaling dimension $\Delta_N$ of the fermionic composite operator $\mathcal{O}_N$, the cutoff scale $\Lambda_{\mathrm{UV}}$, and the IR scale $\Lambda_{\mathrm{IR}}$. 

\begin{figure}[htbp]
    \centering
    
    \begin{subfigure}[b]{0.45\textwidth}
        \centering
        \caption*{CFT until the UV} 
        \includegraphics[width=.8\textwidth]{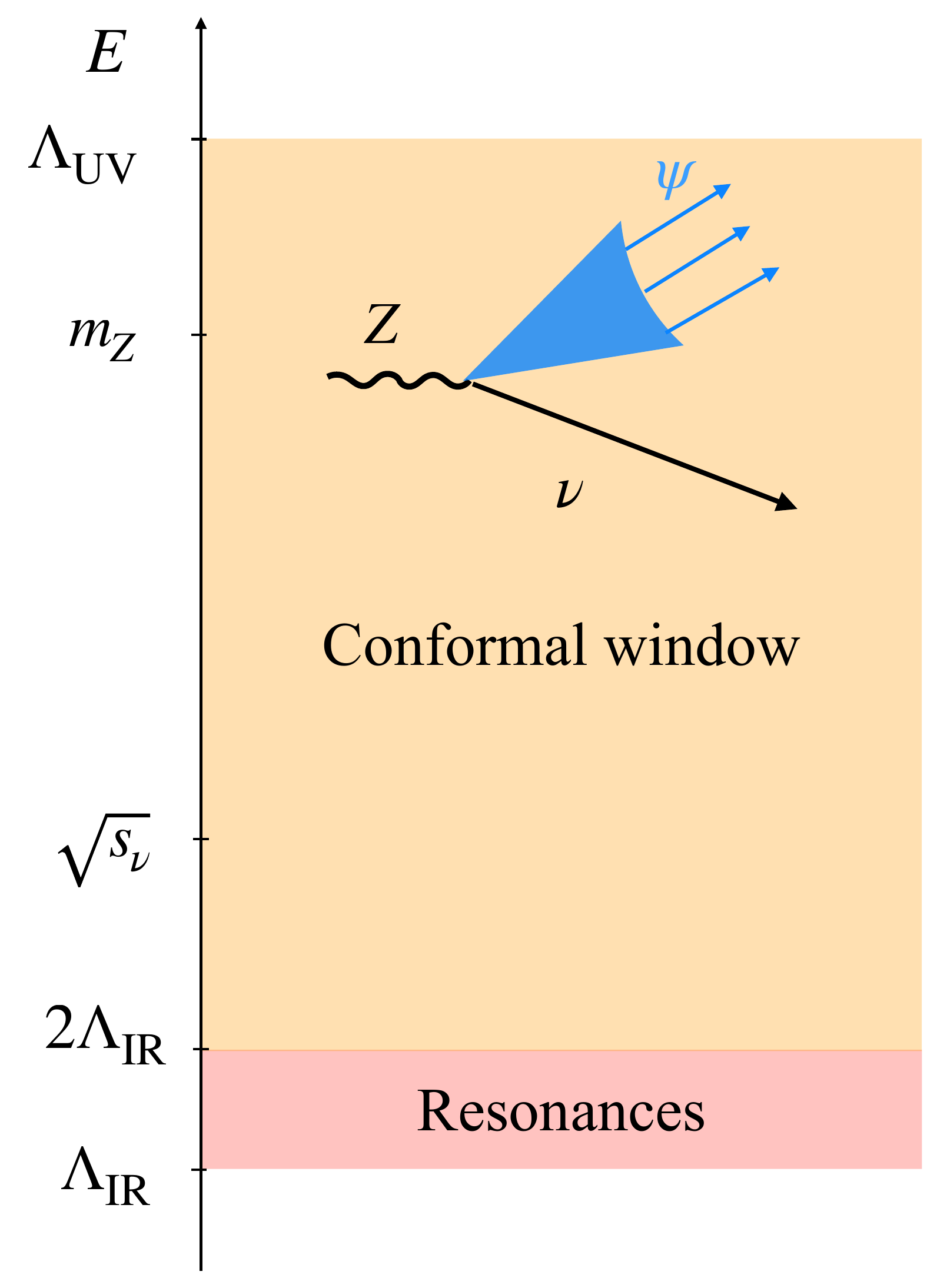}
    \end{subfigure}
    \hfill
    \begin{subfigure}[b]{0.45\textwidth}
        \centering
        \caption*{Broken CFT}
        \includegraphics[width=.8\textwidth]{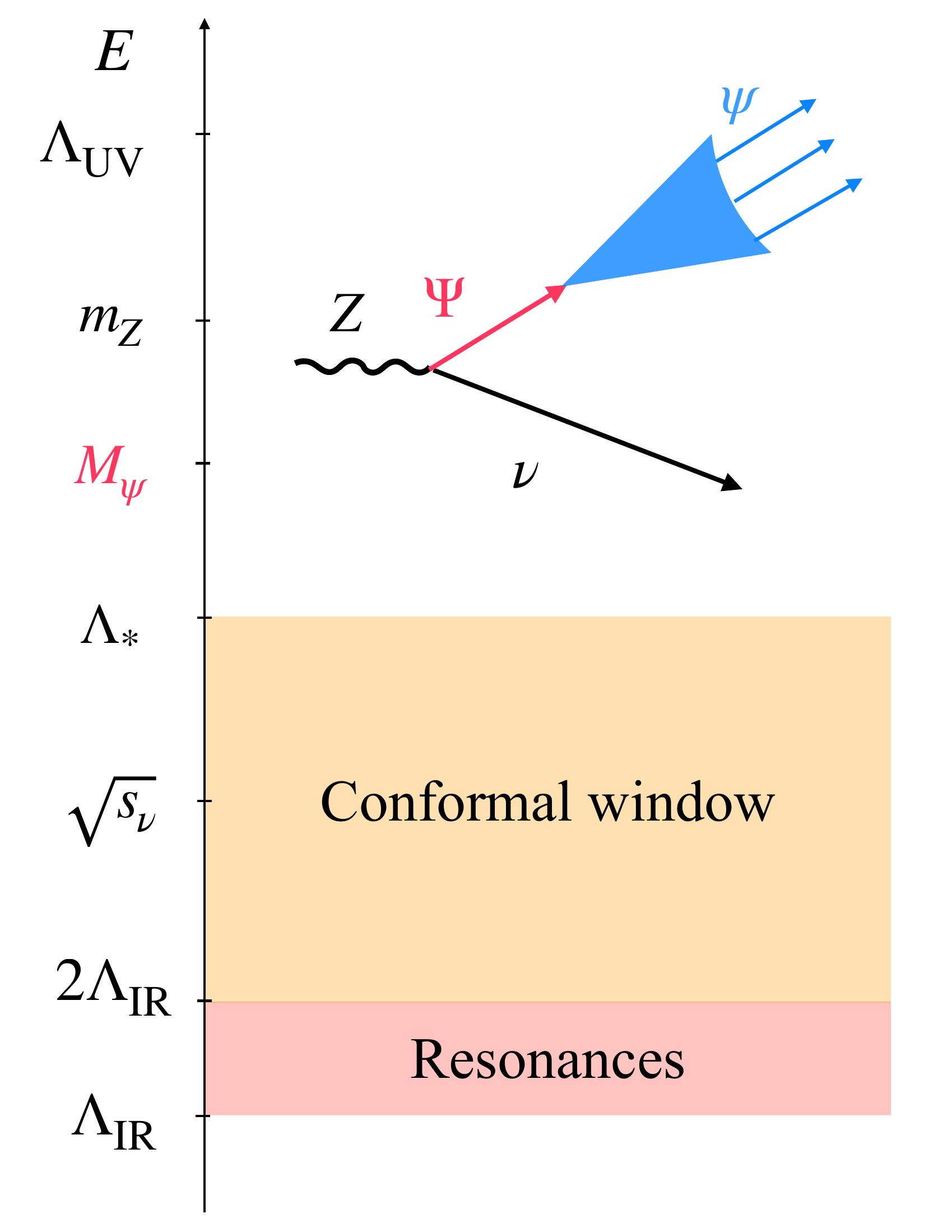}
    \end{subfigure}
    
    \vspace{0.5cm} 
    
    \begin{subfigure}[b]{0.45\textwidth}
        \centering
        \caption*{CFT until the UV with pNGBs}
        \includegraphics[width=.8\textwidth]{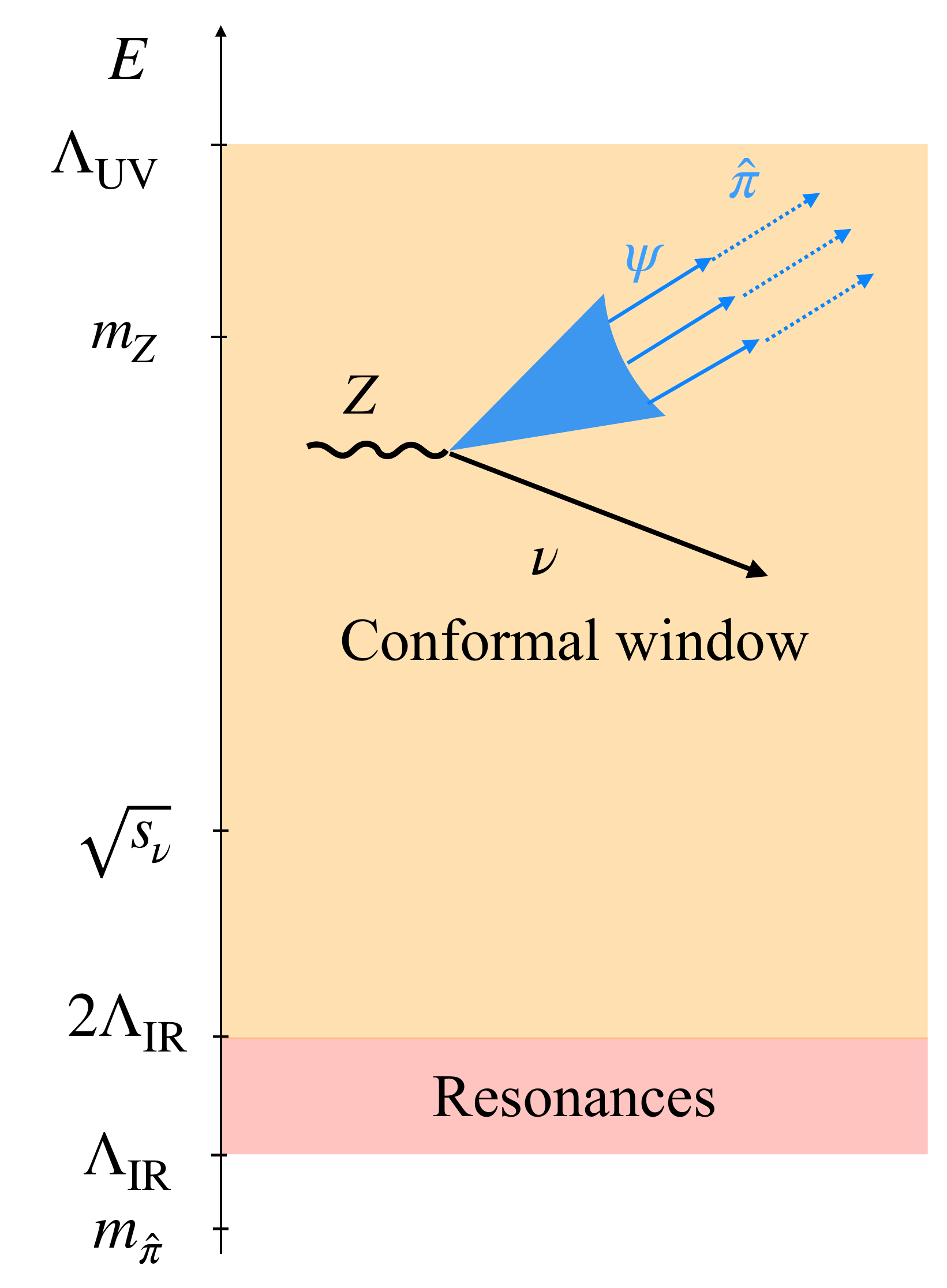}
    \end{subfigure}
    \hfill
    \begin{subfigure}[b]{0.45\textwidth}
        \centering
        \caption*{Broken CFT, with pNGBs}
        \includegraphics[width=.8\textwidth]{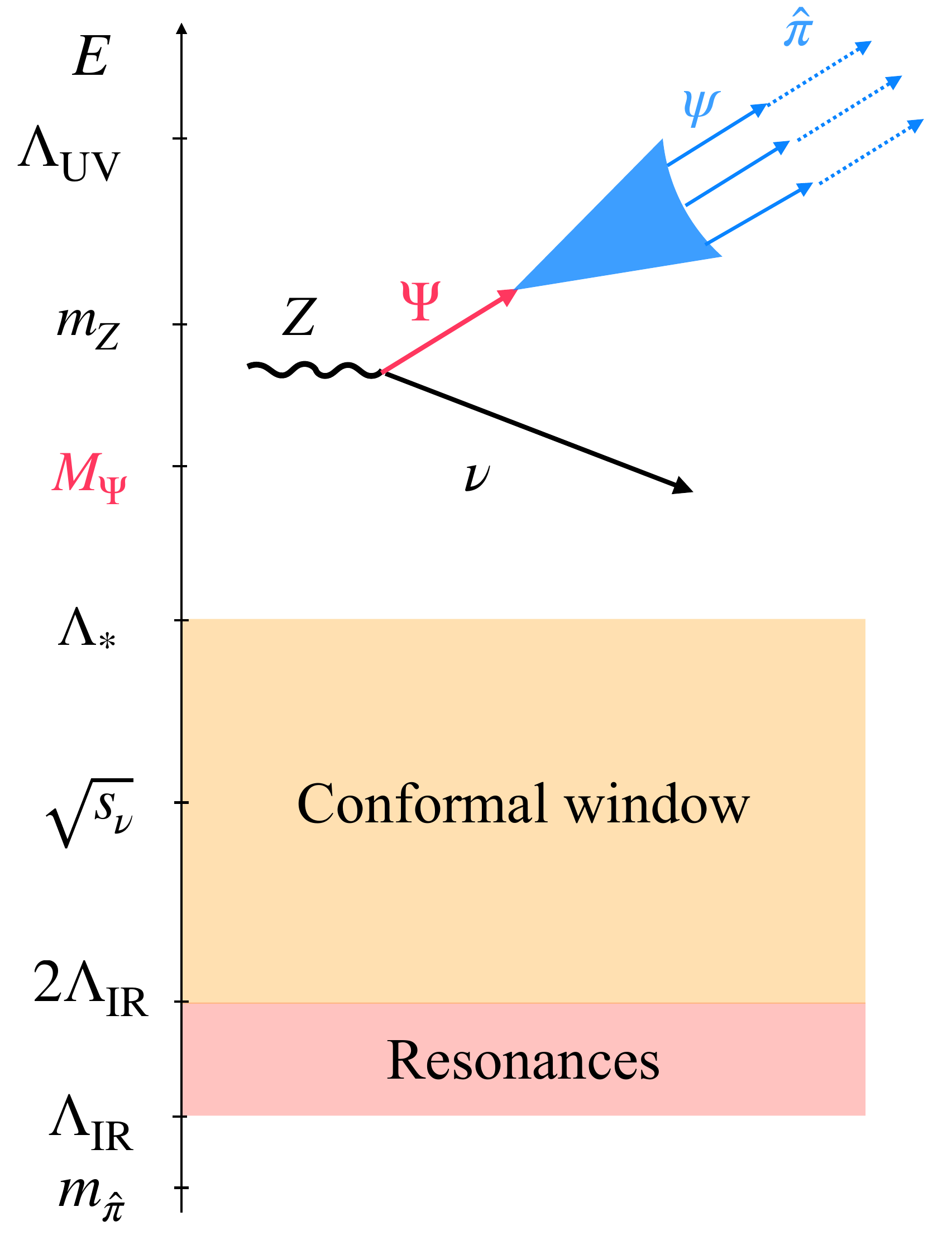}
    \end{subfigure}
    
       \caption{
       Sketch of the dark sector mass spectrum and the resulting $Z$ decay phenomenology across different realizations, described in \cref{sec:CompositenessModel}. The orange (red) bands indicate the conformal window (IR resonance) region, while $\sqrt{s_\nu}<m_Z$ indicates the typical center of mass energy at a neutrino facility. {\bf Top left:} The conformal window extends up to the $\Luv$, leading to direct $Z$ decay into a conformal shower (blue cone). {\bf Top right:} The conformal symmetry is broken at $\Lambda_*$ by a heavy state $\Psi$, with mass $\Lambda_*<M_\Psi<\Luv$. The $Z$ boson decays to a $\Psi$, shown as a red line, prior to the development of the dark shower. {\bf Bottom:} Same as the top panel, with inclusion of light pNGBs, $\hat{\pi}$, emerging in the IR regime, $m_{\hat{\pi}}< \Lir$.} 
    \label{fig:conformal_windows}
\end{figure}

In this section, we discuss how deviations from this minimal setup affect our predictions. The hierarchy of energy scales in these non-minimal cases is illustrated in \cref{fig:conformal_windows}. First, we show that the presence of an intermediate scale $\Lambda_{\mathrm{IR}} < \Lambda_* < \Lambda_{\mathrm{UV}}$, which terminates the conformal regime, can relax EW constraints. We illustrate this possibility with a concrete scenario motivated by high-energy neutrino facilities.

We then examine the robustness of the resonance lifetime and branching ratios derived in \cref{sec:WidthScaling}. In particular, if confinement is accompanied by the spontaneous breaking of internal symmetries in the dark sector, light pseudo-Nambu--Goldstone bosons (pNGBs) generically arise. Their presence enhances the invisible branching fraction of fermionic resonances, thereby reducing the displaced signal yield in favor of invisible signatures across the parameter space.

Moreover, we sketch several UV completions that realize these deviations from the vanilla scenario. This allows us to study the interplay between the $HL\mathcal{O}_N$ portal and other higher-dimensional portals involving the fermionic operator $\mathcal{O}_N$. The phenomenology of these operators will be outlined in \cref{sec:nusmeft}, where they may lead to novel signatures at the HL-LHC.

We also show that weakly coupled QCD-like UV completions generically predict the existence of lower-dimensional scalar operators $O_S$ and vector currents $J_V$, inducing lower-dimensional portals to the Standard Model that tend to dominate the phenomenology. Avoiding this feature requires strong Dark Sector dynamics that generate large anomalous dimensions, lifting the scalar and vector operators relative to the fermionic one.

\subsection{Breaking the conformal regime below the EW scale}\label{sec:breakingCFT}

As discussed in \cref{sec:invisible,sec:displaced}, the parameter space of composite neutrino scenarios is strongly constrained by modifications of the EW gauge bosons and Higgs branching ratios. This leaves very little room for invisible signals at neutrino experiments and essentially no room for displaced signals. 

These results rely on the assumption that the conformal regime extends up to and beyond the masses of the Higgs and EW bosons, $m_{h,Z}$. This case is shown in the top left plot of \cref{fig:conformal_windows}. However, the bounds can be relaxed if the conformal window terminates at a scale $\Lambda_* < m_Z$, after which the theory becomes UV-free. This scenario was outlined in Ref.~\cite{Borrello:2025hal}, and its relevant scales are schematically illustrated in the top right panel of \cref{fig:conformal_windows}.

A possible realization involves coupling the Higgs boson coupling to a HNL, $\Psi$, which is a singlet under both the SM and DS gauge groups, with mass satisfying $\Lambda_*<M_\Psi<m_Z$, and with a large decay width into DS states, controlled by standard dark showering dynamics. The light DS states can eventually decay to SM particles inside the detector, producing displaced signals. However, the Higgs and $Z$ bosons can interact only with the weakly coupled $\Psi$, rather than the full conformal spectrum of the dark theory. At energies below $M_\Psi$, integrating out the HNL while the DS is still perturbative generates the portal of \cref{eq:compositeN}, reproducing the neutrino phenomenology discussed earlier.  

As an explicit example, consider the UV Lagrangian:  
\begin{equation}\label{eq:uvmodel}
    \mathcal{L}_{\rm UV} = \tilde{y}\, H L \Psi + M_\Psi \bar{\Psi} \Psi + y_D \frac{\bar{\Psi} \chi \Phi^n}{\Lambda_D^{\,n-1}},\quad \frac{3}{2}+n \equiv \Delta_N\,,
\end{equation}
where $\chi$ and $\Phi$ are fermionic and scalar DS fields charged under the DS gauge group, and $\Lambda_D$ is generally different from the UV cutoff $\Lambda_{\rm UV}$ defined in \cref{eq:compositeN}.  

For $n>1$, ensuring perturbativity up to the cutoff scale $\Lambda_{\rm UV}$ requires estimating radiative corrections from the $\bar{\Psi} \chi \Phi^n$ interaction to the mass and wavefunction of $\Psi$. By closing $n$ loops on the $\Psi$ leg gives
\begin{align}
    \frac{\delta M_\Psi}{M_\Psi} &\sim C\frac{y_D^2}{16\pi^2} \left( \frac{\Lambda_{\rm UV}^2}{16\pi^2 \Lambda_D^2} \right)^{n-1}\,,
\end{align}
where $C$ is a multiplicity factor that depends on details of the DS structure; we keep it a free parameter here, while we compute it explicitly in specific UV-complete models later on in the discussion. Typical values of $C$ can reach up to $\mathcal{O}(10^2)$ relative to the $A_N$ values used in our results.
Requiring $\delta M_\Psi / M_\Psi \lesssim \mathcal{O}(1)$, we obtain
\begin{equation}\label{eq:lowerboundpert}
C^{-1/2}y_D^{-1}\Lambda_D^{\,n-1} \gtrsim \left( \frac{1}{4\pi} \right)^{n} \Lambda_{\rm UV}^{\,n-1}.
\end{equation}
A similar, though less stringent, bound can be obtained from the perturbativity requirement $\Gamma_\Psi < M_\Psi$.
%
Integrating out $\Psi$ we obtain the IR Lagrangian  
\begin{equation}
\mathcal{L}_{\rm IR} = \tilde y y_D \frac{H L \chi \Phi^n}{M_\Psi \Lambda_D^{\,n-1}},
\end{equation}
which can be matched to the portal interaction of \cref{eq:compositeN} by the relation  
$y_D^{-1}\Lambda_D^{\,n-1} = \tilde y \Lambda_{\rm UV}^{\,n}/M_\Psi$.
For $\Delta_N = 5/2$, $\Lambda_D$ cancels out as the interaction in \cref{eq:uvmodel} becomes renormalizable. To evade Higgs bounds, we take $\tilde y$ as small as possible and saturate the perturbativity bound on $\Lambda_D$ from \cref{eq:lowerboundpert}, giving  
\begin{equation}\label{eq:pertplusmatch}
C^{-1/2} \tilde y (4\pi)^n \Lambda_{\rm UV} = M_\Psi\,.
\end{equation}
We also require $M_\Psi$ to exceed the typical center-of-mass energy of neutrino collisions, $ \sqrt{s_\nu} \sim 40~(3)~\rm{GeV}$ for FPF (DUNE). Combining the latter with \cref{eq:pertplusmatch} gives  
\begin{equation}\label{eq:lowercut}
\Lambda_{\rm UV} \gtrsim \frac{C^{1/2}\sqrt{s_\nu}}{\tilde y (4\pi)^n}\,,
\end{equation}
which sets the minimum cutoff for a given neutrino beam energy. The two cases primarily discussed in this work are $n=1$ $(\Delta_N=5/2)$ and $n=2$ $(\Delta_N=7/2)$.
For simplicity we set in the following $C=1$, since the scaling of the minimum $\Luv$ is straightforward with higher $C$. For more realistic values of $C \sim 10^2$, the minimum value allowed by $\Luv$ increases by a factor of $10$.  

Lower bounds on $\tilde y$ arise from constraints on the Higgs invisible width. For long-lived $\Psi$, this leads to $\tilde y \lesssim 10^{-2}$. Avoiding the latter bound requires $\Lambda_{\rm UV} \gtrsim 320~{\rm GeV}$ with $y_D = 4\pi$ for $n=1$, and $\Lambda_{\rm UV} \gtrsim 25~{\rm GeV}$ with $\Lambda_D = 2~\rm{GeV}$ for $n=2$. In the latter case, a stronger lower bound on the cutoff comes from requiring that the mixing angle defined in \cref{eq:mixingangle} is perturbative, $V_{\psi\nu}<1$, which gives $\Lambda_{\rm UV} > 130~\rm{GeV}.$ 

For short-lived $\Psi$, stronger bounds arise from the $Z$ boson branching ratios into displaced SM particles, see \cref{tab:SMdecay}. Assuming light fragments decay inside the detector with ${\cal O}(1)$ probability, we have
\begin{equation}
{\rm Br}(Z \to \nu + (\Psi \to {\rm displaced})) \sim \theta_{\Psi\nu}^2\frac{ \Gamma(Z \to \nu\nu)}{\Gamma_Z} = \theta_{\Psi\nu}^2 \, {\rm Br}(Z \to \nu\nu) < 10^{-6},
\end{equation}
where $\theta_{\Psi\nu} = y v /(\sqrt{2} M_\Psi)$, which requires  
\begin{equation}
\tilde y \lesssim 2 \times 10^{-3} \frac{M_\Psi}{v} \approx 4 \times 10^{-4} \left( \frac{M_\Psi}{40~{\rm GeV}} \right).
\end{equation}
Using \cref{eq:lowercut}, this translates to $\Lambda_{\rm UV} > 8~{\rm TeV}$ with $y_D = 4\pi$ for $n=1$, and $\Lambda_{\rm UV} > 635~{\rm GeV}$ with $y_D = 0.5$ and $\Lambda_D \approx 2~{\rm GeV}$ for $n=2$.  

This discussion demonstrates that these models can satisfy EW and Higgs constraints while still allowing interesting signals at neutrino beam experiments. The parameter space and the potential reach of future high-energy neutrino experiments are discussed in \cref{sec:results}.  
Further constraints on this construction arise from precision measurements. If active neutrinos mix with a heavy neutral lepton $\Psi$ with mixing angle $\theta_{\Psi\nu}$, then at energies below $M_\Psi$ this mixing induces non-unitarity of the neutrino mixing matrix, thereby modifying the neutrino couplings to gauge bosons~\cite{Antusch:2006vwa, Antusch:2014woa, Blennow:2023mqx}. These effects provide the dominant constraints for $M_\Psi > v$.

In the UV completion of \cref{eq:uvmodel}, the state $\Psi$ is lighter than the electroweak scale but heavier than a few tens of GeV, as required by $M_\Psi \gtrsim \sqrt{s_\nu}$. Its decays are predominantly invisible, due to its dominant coupling to the DS states. Constraints on this scenario were studied in~\cite{deGouvea:2015euy}, where the strongest bound was found on the muon mixing angle, $\theta_\mu < 0.05$. For $\tilde{y} = 10^{-2}$, this translates into $M_\Psi \gtrsim 30$~GeV. Additional constraints, for instance from CKM unitarity tests~\cite{Blennow:2023mqx}, could further tighten the bound on $\theta_\mu$, thereby putting additional pressure on this construction.

\subsection{Effects of light resonances}\label{sec:goldstone}
In this section we want to study the role of light resonances in the phenomenology of the $HL\mathcal{O}_N$ portal. In absence of them, the branching ratios of the fermionic DS resonances (DS baryons) are inherited from the HNL case as discussed in \cref{sec:WidthScaling}. By coupling the DS baryons $\psi$ to light dark particles, it is possible to enhance the branching ratio of $\psi$ into invisible states.

As a concrete example, we consider QCD-like UV completions, where spontaneous chiral symmetry breaking gives rise to light pNGBs~\cite{Coleman:1969sm, Callan:1969sn}. Following the CCWZ construction, IR baryons are ``dressed'' with the pNGB fields to implement the broken axial symmetries in the effective Lagrangian describing the strongly coupled sector. In particular, the baryon $\psi$ interacts with the pNGBs $\hat{\pi}$ schematically via
\begin{equation}
{\cal L}_{\rm IR} \supset v\left(\frac{\Lir}{\Luv}\right)^{\Delta_N-3/2}\frac{\hat{\pi}}{f}   \bar{\psi}' \nu' 
\end{equation}
with $4\pi f \sim \Lir$. The leading effect is that $\psi$ can decay into $\nu + \hat{\pi}$, rendering the decay largely invisible. The pNGB itself can decay visibly via a 4-body process~\cite{Ahmed:2025ldh} with width
\begin{equation}
\Gamma_{\hat{\pi}} \sim \frac{1}{8\pi} \frac{1}{16\pi^2} \left(\frac{\Lir}{\Luv}\right)^{2\Delta_N-3} G_F^2 v^4 \frac{m_{\hat{\pi}}^5}{\Lir^4}.
\end{equation}
Comparing boosted lifetimes between the DS baryon and the pNGBs, the ratio of visible decay widths is
\begin{equation}
\frac{(\Gamma/\gamma\beta)_{\hat{\pi}}}{(\Gamma/\gamma\beta)_\psi} \sim r^6 \frac{v^2}{\Lir^2}, \quad r \equiv \frac{m_{\hat{\pi}}}{\Lir},
\end{equation}
which governs the relative number of visible signal events. While heavier pNGBs can possibly enhance the visible signal, in the most likely scenario the pNGBs are light enough that the DS baryon decays almost entirely invisibly. This is the assumption adopted in \cref{sec:results}, where we set $\mathrm{BR}(\psi \to \text{inv}) \sim 1$. 

\subsection{Sketches of UV completions}
\label{sec:UVcompletions}

In this section, we outline possible UV completions at energy scales $E > \Lambda_{\rm UV}$, that generate the mixing operator introduced in \cref{eq:compositeN} upon integrating out heavy degrees of freedom below $\Lambda_{\rm UV}$. A detailed understanding of such UV completions is essential for several reasons.  First, they clarify the possible existence of further portal structures involving $\mathcal{O}_N$, together with the expected magnitude of their corresponding Wilson coefficients. Second, they determine whether additional scalar or vector operators in the dark sector are generically present, potentially giving rise to complementary portal interactions, as well as the natural size of their associated couplings.
Finally, they determine the size of higher dimensional operators involving SM-only fields that can be constrained using precision data.
Ultimately, the construction of explicit UV completions provides a framework to identify which classes of dark sector scenarios give rise uniquely to the phenomenology associated with the portal discussed in \cref{sec:HLO}.

The simplest UV completions arise from theories with gauge group $G_{\rm SM} \times SU(N_c)$, where the dark color group confines at $\Lambda_{\rm IR} \lesssim 10\,\mathrm{GeV}$. We focus on scenarios in which the light dark sector states are neutral under the SM. 
We further assume SM-singlet mediators at $\Luv$.
Adding heavy fields with non-trivial SM charges can generate extra higher-dimensional operators that we list in \cref{sec:nusmeft}.

The remaining freedom lies in the choice of the dark color representations.  We first consider the minimal case of fields in the fundamental representation, analogous to QCD. We then present a non-weakly coupled UV completion, illustrating how the dimension of $\mathcal{O}_N$ can be parametrically separated from that of scalar and vector operators.

The connection between the composite neutrino portal and neutrino Majorana masses in the context of the inverse see-saw has been discussed in Ref.~\cite{Chacko:2020zze}. Since it does not impact the phenomenology considered here, we defer a brief summary to \cref{app:strongISS}, where we also compare with other composite neutrino scenarios~\cite{Arkani-Hamed:1998wff,Grossman:2010iq}.

\paragraph{Fundamental dark color representations}\label{sec:fundamental_dc}
In the minimal setup, there are three types of fields: $N$ (fermion, SM and dark singlet), $\chi$ (fermion, SM singlet, fundamental of $SU(N_c)$), and $\phi$ (scalar, SM singlet, fundamental of $SU(N_c)$). These may come with flavor indices, and we do not assume yet which fields are heavy (i.e., with mass $\sim \Lambda_\mathrm{UV}$).

The most general Lagrangian including these fields is
\begin{equation}\label{eq:uv_5/2}
\mathcal{L}_\mathrm{DS} \supset y \tilde{H} \bar{L} N + y_\mathrm{DS} \bar{\chi}_i N \phi^i + \lambda_{\phi H} H^\dagger H \phi^\dagger \phi + V(\phi),
\end{equation}
where $i$ is an $SU(N_c)$ index and $V(\phi)$ contains quadratic and quartic scalar terms and we suppressed SM flaor indices for simplicity. To avoid spontaneous breaking of the dark gauge symmetry, the quadratic term is taken positive.

For specific values of $N_c$, additional interactions appear. For $N_c=3$:
\begin{equation}\label{eq:su_3_break}
\mathcal{L}_{N_c=3} = \lambda_2 \epsilon^{ijk} \phi^i \chi^j \chi^k + \frac{\lambda_3}{6} \epsilon^{ijk} \phi^i \phi^j \phi^k,
\end{equation}
provided there are enough scalar flavors for these structures to be non-vanishing. Similar terms can exist for other $N_c$ if $\phi, \chi$ transform in representations other than the fundamental.

Focusing on $N_c=3$ and assuming $N$ is heavy ($m_N \sim \Lambda_\mathrm{UV}$), if $\lambda_2 = \lambda_3 = 0$, the theory has two unbroken $U(1)$ symmetries: $U(1)_d = U(1)_{\chi+\phi}$ and a generalized lepton number under which $N$ and $\chi$ are charged.\footnote{Strictly speaking, this is broken by the Majorana operator $\mathcal{O}_{2N}$ required in the inverse seesaw mechanism. Since the Majorana scale $\mu$ is small, we neglect it here.} These forbid operators such as $\chi \phi \phi$, $\chi \chi \phi$, $\phi^3$, and $\chi^3$ to appear in a portal with the SM, since the SM fields are not charged under the $U(1)_d$ symmetry. In this scenario, the composite HNL operator is $\mathcal{O}_N \propto \bar{\chi} \phi$ with $\Delta_N = 5/2$. Integrating out $N$ at lowest order generates the IR portals:
\begin{equation}\label{eq:sym_term_singelt}
\mathcal{L}_{\Delta_N=5/2} \supset \frac{\lambda_{H\phi} y_\mathrm{DS}}{\Lambda_\mathrm{UV}} (\tilde{H} \bar{L}) (\bar{\chi} \phi) + \frac{\lambda_{H\phi} y_\mathrm{DS}^2}{16\pi^2 \Lambda_\mathrm{UV}} (H^\dagger H) (\bar{\chi} \chi),
\end{equation}
where the first term realizes the composite HNL portal (\cref{eq:compositeN}) and the second induces an Higgs portal-like phenomenology.

If both $\lambda_2$ and $\lambda_3$ are nonzero, the accidental $U(1)$ symmetries are broken, and all gauge-singlet operators are allowed. Assuming only $\chi$ is light, the composite HNL operator becomes $\mathcal{O}_N \propto \chi^3$ with $\Delta_N = 9/2$. Integrating out the heavy fields generates
\begin{equation}\label{eq:uv_9/2_ir}
\mathcal{L}_{\Delta_N=9/2} \supset \frac{\lambda_2 y_\mathrm{DS} y}{\Lambda_\mathrm{UV}^3} (\tilde{H} \bar{L})(\chi^3) + \frac{y_\mathrm{DS}^2 \tilde{y}^2}{16\pi^2 \Lambda_\mathrm{UV}} (H^\dagger H)(\bar{\chi} \chi),
\end{equation}
with $\tilde{y} = \max[\lambda_{H\phi}, y^2, \lambda_2^2 y^2 / 16\pi^2]$. Again, the non-renormalizable Higgs portal generally dominates due to the lower dimension of $\bar{\chi} \chi$. 

Because of the vector-like QCD-like UV completion, the scalar operator $\mathcal{O}_S \equiv \bar{\chi} \chi$ ($\Delta_S = 3$) is always present in the IR, along with its Higgs mixing, which typically dominates phenomenology compared to the composite HNL portal.

Beyond production of light fields, indirect constraints arise from the higher dimensional operators obtained integrating out the heavy particles. For masses above a few TeV, virtual $N$ and $\phi$ exchange generate operators such as
\begin{equation}
\frac{\alpha_{lH}}{\Lambda_\mathrm{UV}^2} (H L)^\dagger \gamma^\mu p_\mu (H L),\quad
\frac{\alpha_{\square H}}{\Lambda_\mathrm{UV}^2} (H^\dagger H) \square (H^\dagger H),\quad
\frac{\alpha_3}{\Lambda_\mathrm{UV}^2} (H^\dagger H)^3,
\end{equation}
where we assume the various Wilson coefficients to be $\alpha \sim \mathcal{O}(1)$. Current electroweak precision data constrain $\Lambda_\mathrm{UV} \gtrsim 10~\mathrm{TeV}$~\cite{Ahmed:2024hpg} (this is close in spirit to the non-unitarity constraints mentioned previously). A similar outcome is obtained if $\alpha_{lH}$ is the only non-zero coefficient. Both this bound and the limits from Higgs mixing suppress the expected number of events for a composite HNL portal in QCD-like UV completions, unless the Wilson coefficient of the operators are carefully chosen to make the HNL portal dominant. 

In the examples above, we identified theories where the relevant fermionic operator has dimension $\Delta_N = 5/2$ (when both the scalar $\phi$ and fermion $\chi$ are light) or $\Delta_N = 9/2$ (when $\chi$ is the only light field and dark-baryon number is broken).

To construct a theory in which the most relevant fermionic operator has dimension $\Delta_N = 7/2$, additional structure is required. For illustration, consider a theory with $N_c = 3$, similar to the one discussed previously, but extended by introducing an additional heavy fermion $\Psi$, transforming in the fundamental of $SU(3)$, alongside the heavy vector-like singlet fermion $N$. The light fields remain the fermion $\chi$ and the scalar $\phi$, both in the fundamental. We impose a global $U(1)$ symmetry with charge assignments $q_\chi = -2 q_\phi$, $q_\Psi = q_\phi$, and $q_N = 0$. The allowed interactions are
\begin{equation}\label{eq:uv_7/2}
\mathcal{L}_\mathrm{UV} = \tilde{y} \tilde{H} \bar{L} N + y \bar{\Psi} \phi N + \lambda_2 \chi \Psi \phi - V(\phi^\dagger \phi, H^\dagger H),
\end{equation}
where in the $\lambda_2$ term the gauge indices are contracted with an antisymmetric tensor.

Below the scale $\Lambda_{\mathrm{UV}}$, the only gauge-invariant fermionic operator neutral under the global symmetry is $\mathcal{O}_N = \chi \phi^2$, with dimension $\Delta_N = 7/2$, where gauge indices are again contracted antisymmetrically. The role of the additional heavy fermion $\Psi$ is to provide an extra link between the Standard Model and the light dark-sector fields, effectively raising the dimension of the leading fermionic operator. The imposed global $U(1)$ symmetry ensures that this additional link is necessary to form a singlet operator.

Given the full field content of the UV completion, one can compute the normalization of the two-point function of the fermionic operator $\mathcal{O}_N$.
We illustrate the calculation for a generalization of the model introduced in \cref{eq:uv_7/2}, in which $\mathcal{O}_N \propto \chi \phi^n$, with all fields transforming in the fundamental representation of the dark-color group $SU(N_c)$ and with their gauge indices contracted antisymmetrically. This condition fixes $N_c = n+1$. However, in order to highlight the distinct roles played by the gauge structure and by the field multiplicity in $\mathcal{O}_N$, we will keep the two symbols separate.
The computation proceeds by matching the high-energy behavior of $HL \to HL$ scattering in the UV-complete model with that in the CFT description (see also Ref.~\cite{Borrello:2025hal}):
\begin{equation}\label{eq:cft_uv_match}
N_f^n \, N_c! \, \Gamma(n)\, (16\pi^2)^n \sim A_N
\quad \Rightarrow \quad
c_N \sim \frac{2}{(4\pi^2)^n} (n+1)! \, N_f^n \, N_c! ,
\end{equation}
where $N_f$ denotes the number of scalar flavors appearing in the operator. We can now identify the multiplicity factor as $C\equiv N_f^n \, N_c!\, \Gamma(n)$.

\paragraph{Strongly coupled CFT} Weakly coupled UV completions generically induce lower dimensional scalar operators such as $\bar{\chi}\chi$ or $\phi^\dagger \phi$, particularly in models with vector-like matter. Constraints on the associated scalar portals are typically stronger than those arising from higher-dimensional fermionic portals, unless either (i) some of the heavy states reside at intermediate scales between $\Lambda_{\mathrm{IR}}$ and $\Lambda_{\mathrm{UV}}$, as discussed in \cref{sec:breakingCFT}, or (ii) the UV couplings are arranged to suppress the Higgs portal relative to the fermionic one.

A structural way to avoid low-dimensional scalar operators $\mathcal{O}_S$ is to consider chiral matter content, thereby forbidding such operators by symmetry. This generally requires extending the gauge sector and ensuring anomaly cancellation.

Alternatively, a dynamical mechanism can suppress scalar operators by generating large anomalous dimensions. In a holographic framework based on the AdS/CFT correspondence \cite{Maldacena:1997re,Gubser:1998bc,Witten:1998qj}, this corresponds to a hierarchy in the bulk masses of scalar and fermion fields in $\mathrm{AdS}_5$. At leading order in the $1/N$ expansion, weakly coupled bulk fields map to generalized free operators on the boundary, with scaling dimensions fixed by their bulk masses \cite{Witten:1998qj}.

In this setup, one can construct theories in which a fermionic operator $\mathcal{O}_N$ is the only single-trace primary operator at low dimension. Scalar operators then arise only as double-trace composites in the operator product expansion, $\mathcal{O}_S \sim \bar{\mathcal{O}}_N \mathcal{O}_N$, with scaling dimension $\Delta_S = 2\Delta_N + \mathcal{O}(1/N^2)$, as expected in large-$N$ conformal field theories \cite{Heemskerk:2009pn,Fitzpatrick:2010zm}.

As a result, the scalar portal operator $H^\dagger H \mathcal{O}_S$ has dimension $\Delta(H^\dagger H \mathcal{O}_S) \simeq 2\Delta_N + 2$, which is parametrically larger than that of the fermionic portal $H \ell \mathcal{O}_N$, with dimension $\Delta_N + \tfrac{5}{2}$. For sufficiently large $\Delta_N$, constraints from the fermionic portal can therefore dominate over those from the scalar portal. A similar argument applies to vector operators, provided the theory does not contain conserved currents with protected dimension $\Delta = 3$.

\section{Results}\label{sec:results}
In this section we describe experimental probes for all the scenarios described in \cref{sec:UVcompletions}: i) conformal all the way to the EW scale, ii) conformal breaking before the EW scale, iii) conformal all the way to the EW scale with light pNGBs, iv) conformal breaking before the EW scale with light pNGBs. 

Assuming a flavor-democratic scenario with $y_e \approx y_\mu \approx y_\tau$, we focus primarily on probes controlled by $y_\mu$, and to a lesser extent on those sensitive specifically to $y_e$. The coupling $y_\tau$ becomes relevant only in scenarios where the first two couplings are suppressed, and will not be considered further here.

\begin{figure}[t!] 
    \centering

    \begin{subfigure}[b]{0.47\linewidth}
        \centering
        \includegraphics[width=\linewidth]{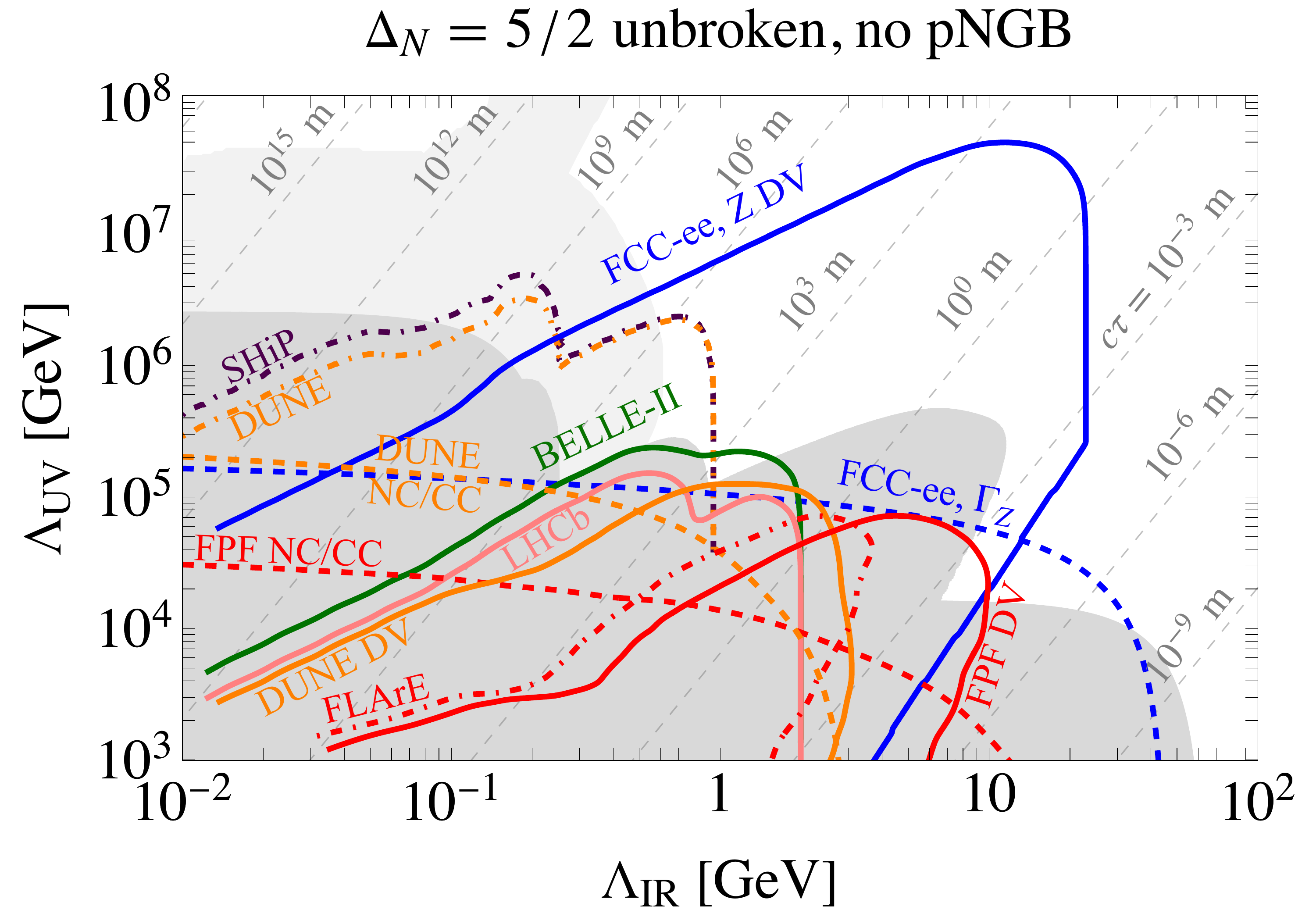}
    \end{subfigure}
    \hfill
    \begin{subfigure}[b]{0.47\linewidth}
        \centering

        \includegraphics[width=\linewidth]{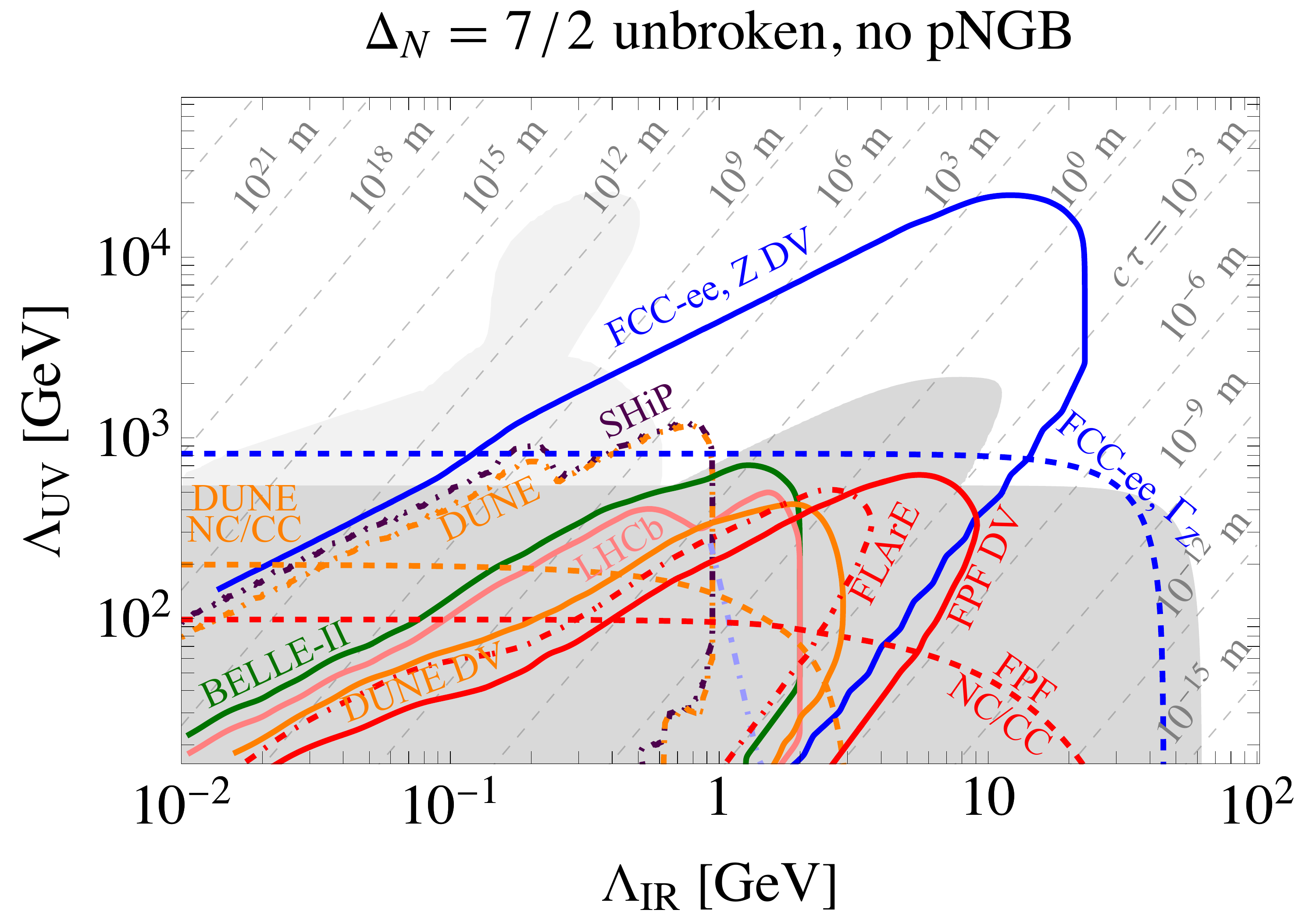}
    \end{subfigure}
    
    \vspace{0.1cm} 

    \begin{subfigure}[b]{0.47\linewidth}
        \centering
      
        \includegraphics[width=\linewidth]{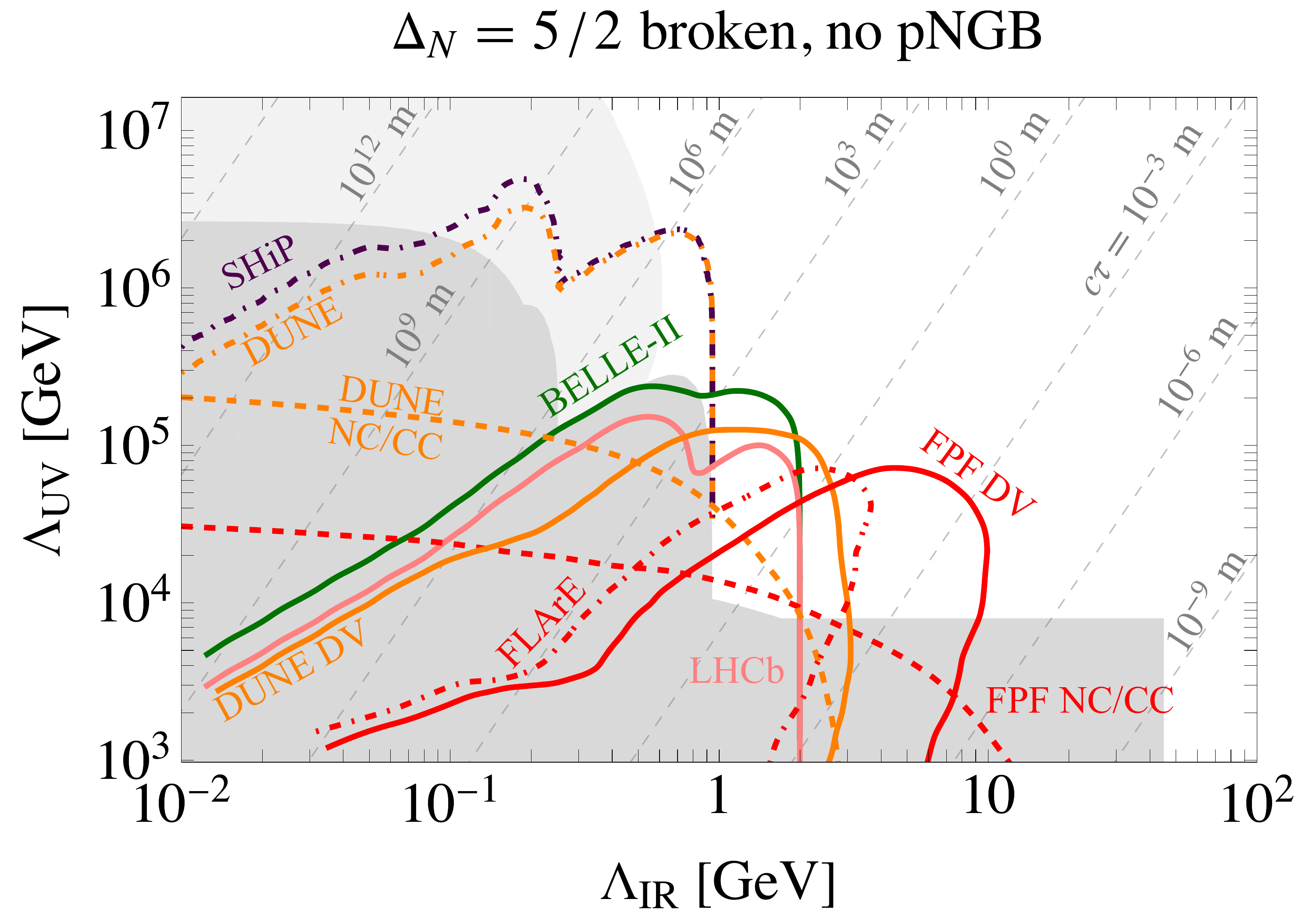}
    \end{subfigure}
    \hfill
    \begin{subfigure}[b]{0.47\linewidth}
        \centering
        \includegraphics[width=\linewidth]{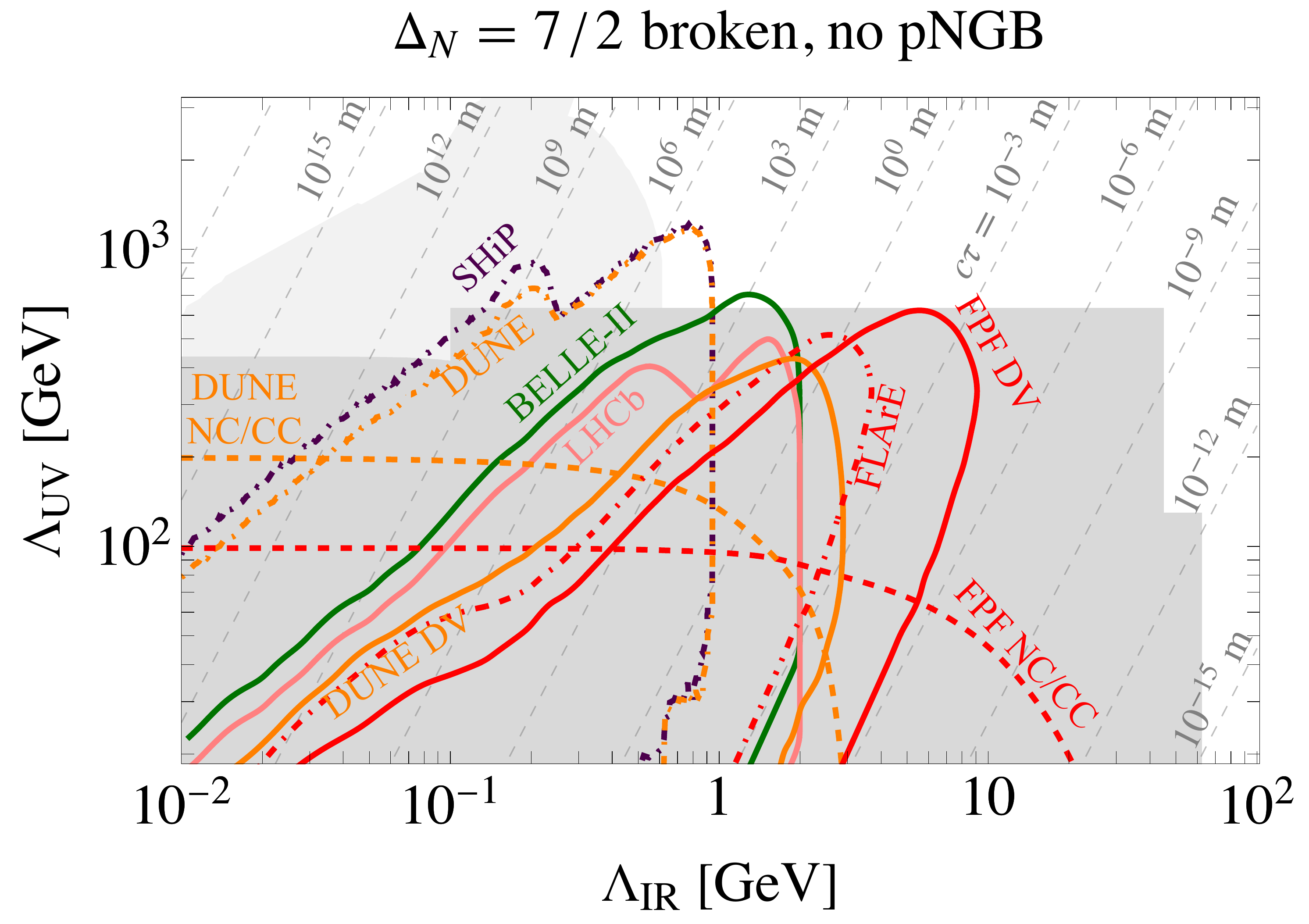}
    \end{subfigure}

    \caption{Experimental limits in the $\Lir$–$\Luv$ plane, for $\Delta_N = 5/2$ (left) and $\Delta_N = 7/2$ (right), considering the first two UV scenarios of \cref{sec:CompositenessModel}: unbroken CFT up to the EW scale (top), and CFT broken below the EW scale (bottom). The dark-gray shaded region denotes the envelope of present constraints discussed in \cref{sec:present}, excluding supernova limits, which are shown in light gray. Dashed gray lines indicate isocontours of the fragment decay length. Solid lines show projected sensitivities from displaced-vertex searches with dark-jet production inside the detector. Dashed lines correspond to the enhancement of the neutral-current to charged-current ratio, while dot-dashed lines represent limits from displaced decays of dark states produced in the target. The experiments considered, described in \cref{sec:results}, are DUNE (orange), FPF (red), SHiP (purple), Belle~II (green), LHCb (pink) and FCC-ee (blue).}
    \label{fig:delta_combined}
\end{figure}

\begin{figure}[t!] 
    \centering

    \begin{subfigure}[b]{0.47\linewidth}
        \centering
        \includegraphics[width=\linewidth]{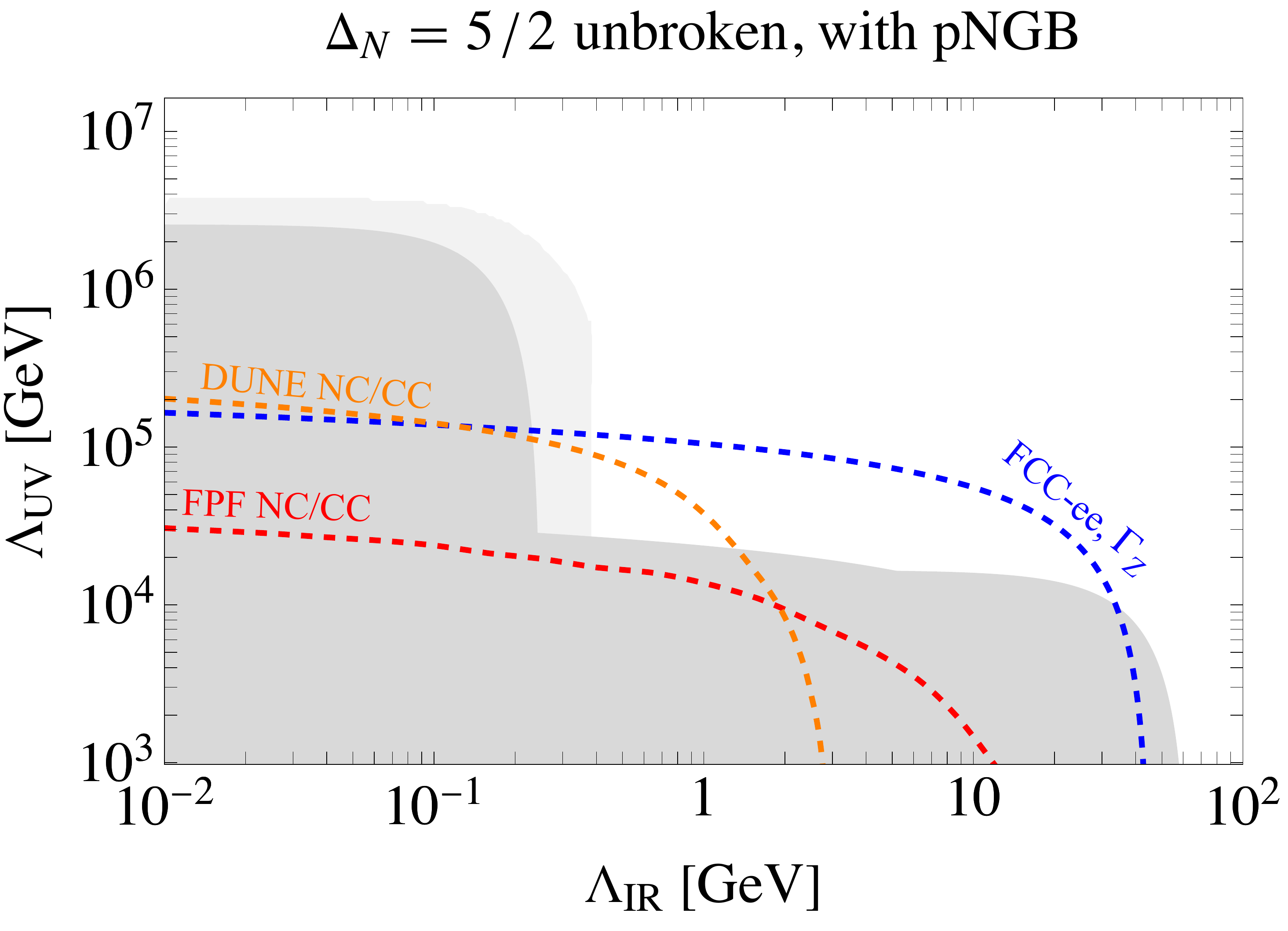}
    \end{subfigure}
    \hfill
    \begin{subfigure}[b]{0.47\linewidth}
        \centering

        \includegraphics[width=\linewidth]{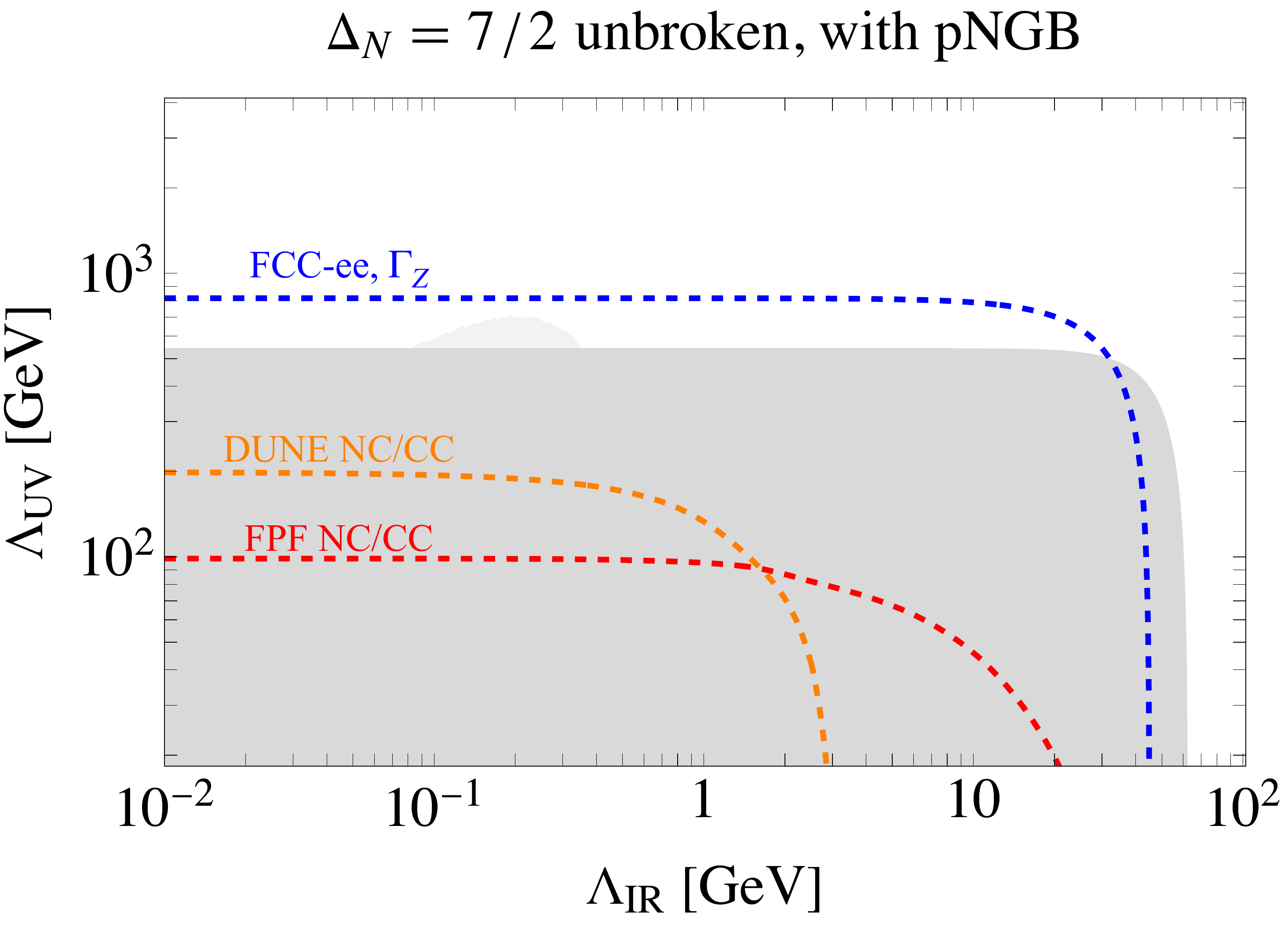}
    \end{subfigure}
    
    \vspace{0.1cm} 

    \begin{subfigure}[b]{0.47\linewidth}
        \centering
      
        \includegraphics[width=\linewidth]{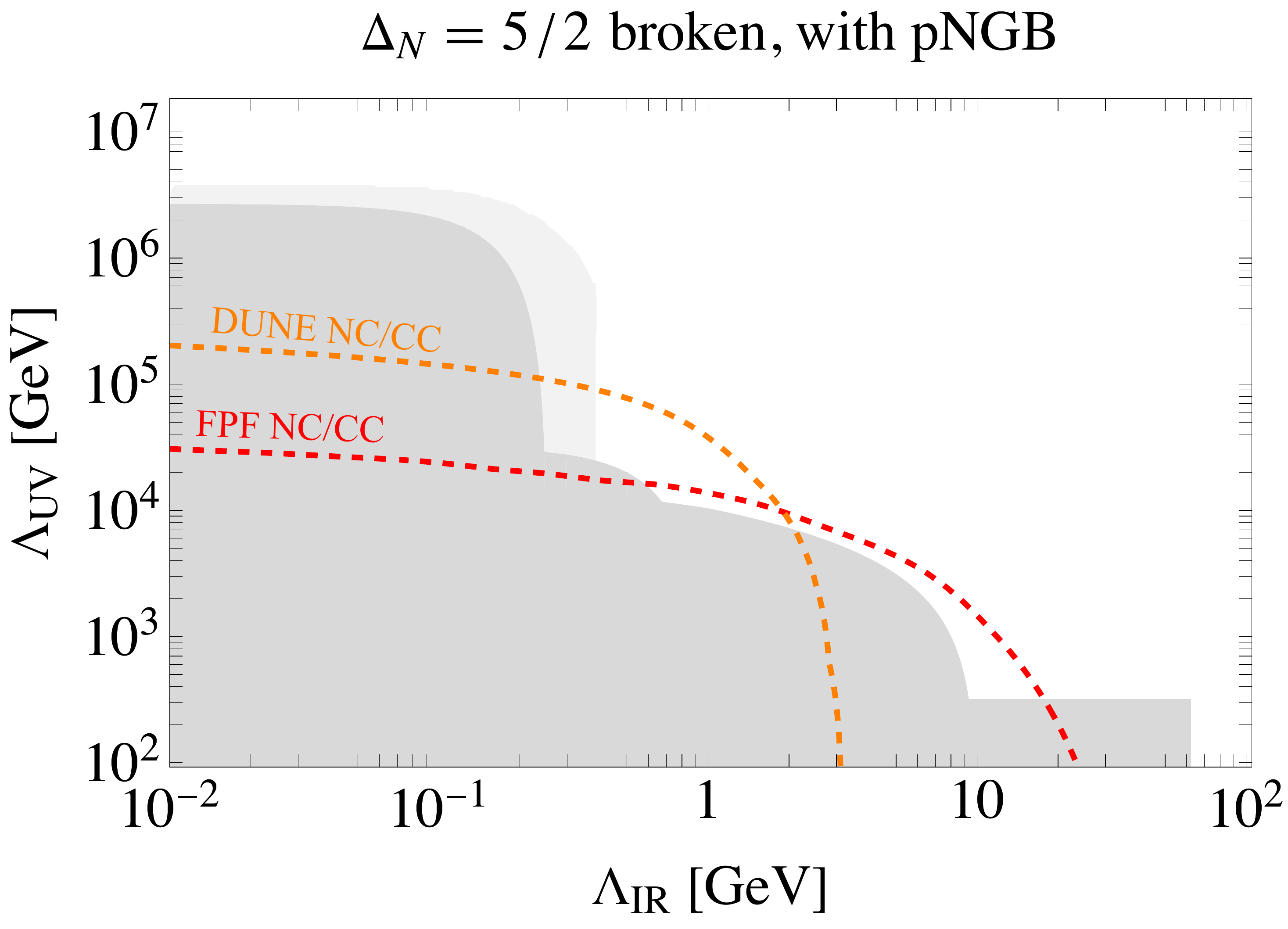}
    \end{subfigure}
    \hfill
    \begin{subfigure}[b]{0.47\linewidth}
        \centering
        \includegraphics[width=\linewidth]{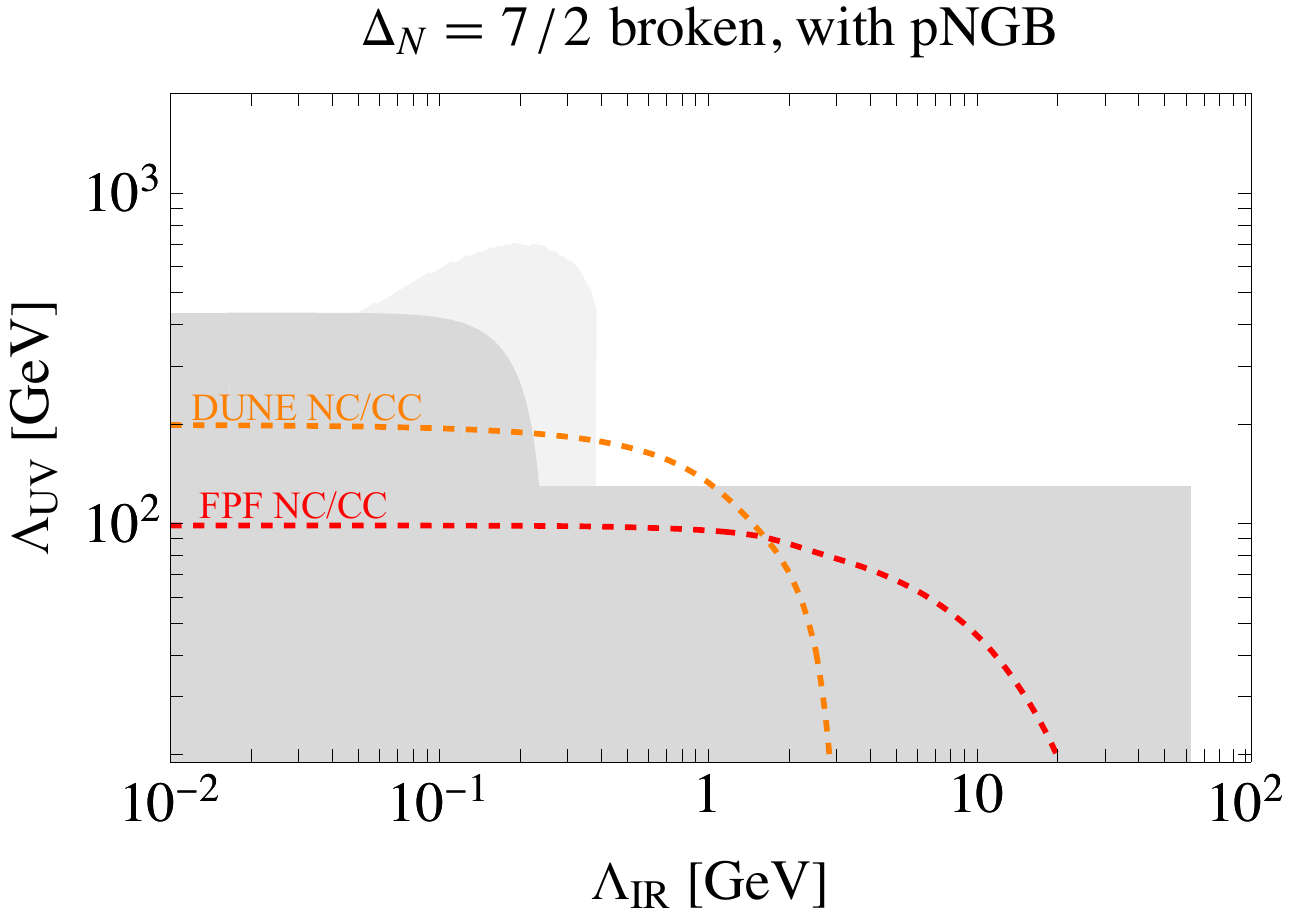}
    \end{subfigure}

    \caption{As in \cref{fig:delta_combined}, but for the two additional scenarios with light pNGBs introduced in \cref{sec:goldstone}. The gray shaded regions indicate parameter space excluded by current constraints, with supernova limits shown in light gray (see \cref{sec:present} for further details). In scenarios with light pNGBs, only constraints from invisible signals survive; these are shown as dashed lines for DUNE (orange), FPF (red), and FCC-ee (blue).}

    \label{fig:delta_combined_pNGB}
\end{figure}

We start in \cref{sec:present} by listing present probes, and then discuss in \cref{sec:future} the projected sensitivities of future searches. For each production channel, we consider both invisible and displaced decay searches where relevant. We collect in \cref{tab:collexp,tab:dumpexp,tab:nuexp} the main features and numerical values of each experiment that is used in our analysis. For simplicity and easiness of comparison, we group them as neutrino experiments (\cref{tab:nuexp}), beam dumps (\cref{tab:dumpexp}) and collider experiments (\cref{tab:collexp}). Finally, we show the results in \cref{fig:delta_combined,fig:delta_combined_pNGB}, where the dark and light gray shaded regions are probed by present searches, while colored lines indicate the sensitivity of future experiments. Similar plots projected in the $\Lir-|V_{\psi\nu_i}|^2$ plane are shown in \cref{app:BoundsMixing}.

\subsection{Present probes}\label{sec:present}

\paragraph{Light Meson decays} 
When produced in meson decays, composite HNL resonances are typically long-lived, as the kinematically accessible values of $\Lambda_{\rm IR}$ are small, leading to a suppressed width of the resonances.

Constraints can arise from light meson decays with missing energy (${\slashed E}$); namely, measurements of $\pi^+ \to \ell^+ +{\slashed E}$ at PIENU~\cite{PIENU:2011aa,PiENu:2018lsf} and of $K^+ \to \ell^+ + {\slashed E}$ at NA62~\cite{NA62:2020mcv,NA62:2021bji}. In general, for light mesons, constraints from $e^+ + {\slashed E}$ decays are more stringent than those from $\mu^+ + {\slashed E}$, and can probe a wider kinematical window, resulting in a broader accessible range for $\Lambda_{\mathrm{IR}}$. 

In the well-studied case of weakly coupled HNLs, one can exploit the full reconstruction of the two-body decay kinematics to perform a bump hunt in the missing mass, $m_{\rm miss}^2 = (P_{\pi,K} - P_\ell)^2$, over the background arising from the radiative tail with a missed photon in leptonic decays ($\pi^+ \to \ell^+ \nu (\gamma)$ or $K^+ \to \ell^+ \nu (\gamma)$). This should be contrasted with the composite case, where the missing mass distribution instead reflects the conformal scaling of the decay width. 
For this reason, we adopt a conservative approach and use the current uncertainties on relevant meson branching ratios, reported in \cref{tab:SMdecay}, to derive constraints on the composite HNL.

\paragraph{Gauge boson and Higgs decays} 
SM boson decays provide a powerful probe both through invisible and displaced signatures. For invisible decays, the presence of the DS modifies the invisible branching ratios of SM bosons. The current experimental measurements, including their uncertainties, are summarized in \cref{tab:SMdecay}. In our scenario, the most stringent constraint arises from the Higgs boson invisible decay. The strongest current bound is set by ATLAS~\cite{ATLAS:2023tkt}, ${\rm Br}(h \to {\rm inv}) \leq 10.7\%$.

For displaced signatures, searches for long-lived particles produced in $Z$ and $W$ boson decays have been performed at LEP~\cite{DELPHI:1996qcc} and at the LHC~\cite{ATLAS:2025uah}, respectively. These analyses typically interpret the results within a simplified model containing a single HNL. We recast these limits for the DS jet scenario by translating them into constraints on the branching ratios of gauge bosons into displaced final states.

In the case of the $Z$ boson, we consider the full LEP dataset, corresponding to approximately $N_Z = 10^6$ produced bosons. For long-lived HNLs yielding displaced signatures, the DELPHI analysis achieved essentially background-free conditions, leading to a limit ${\rm Br}(Z \to \nu N) < 10^{-6}$. At shorter lifetimes, the sensitivity decreases as the signal becomes prompt and is overwhelmed by hadronic $Z$ decay backgrounds (e.g.\ monojet-like topologies). Conversely, for very long lifetimes, detection relies on energy deposits or track clusters in the outer detector regions, which significantly reduces the efficiency.

For the $W$ boson, LHC searches focus on displaced vertices with either two leptons or a lepton plus a pion in the final state. Despite the much larger production yield, $N_W \simeq 10^9$, the sensitivity is currently limited by sizable backgrounds. Consequently, present $W$-based searches do not yet provide competitive constraints on the composite DS scenario.

In addition, at large $\Lambda_{\rm IR}$ the DS fragments decay promptly. We conservatively constrain this region using existing bounds on untagged branching ratios of the Higgs bosons. However, we stress that dedicated searches targeting such prompt, possibly non-standard final states with multileptons plus jet could in principle lead to significantly stronger constraints than those adopted here.

\begin{table}[t!]

    \centering
    
    \begin{tabular}{c|ccccccc }
                Exp. & $N_{\nu OT}$ &  $\langle E_\nu\rangle $ [GeV]&$L_{\rm det}$ [m]&$\sigma_{\rm t}^{-1}$ [GeV$^2$]&Technology\\
        \hline
      DUNE~\cite{DUNE:2020lwj,DUNE:2020ypp,DUNE:2021tad}   & $10^{20}$&   3&5&0.2&LArTPC \\
    NuTeV~\cite{NuTeV:2001whx,Zeller:2002he}  & $ 10^{15}$&   60&20&1&Fe-Scintillator\\
       FPF~\cite{MammenAbraham:2024gun,Feng:2022inv}  & $ 10^{15}$&   300&1&0.2&Emulsion/LArTPC \\
       SHiP~\cite{SHiP:2015vad}&$10^{16}$&50&2&0.7&Emulsion, Scintillator
    \end{tabular}
    \caption{Main features of the neutrino beam experiments considered in this work. The first column indicates the number of neutrino produced in one year of running, while the second shows the average neutrino energy. The last three column refer to the detector length, the target material density and the detection technology used, respectively.}
    \label{tab:nuexp}
    \end{table}

\paragraph{Neutral to charge current ratio}
The NuTeV \cite{NuTeV:2001whx} experiment provided the highest-energy neutrino beam to date, with $E_\nu \sim 300~\mathrm{GeV}$ ($\sqrt{s} \sim \mathcal{O}(10)~\mathrm{GeV}$) and a total integrated flux of $N_\nu \simeq 5\times10^{16}$. The wide energy spread of the neutrino beam does not allow to reconstruct the event kinematics  making NuTeV mostly sensitive to inclusive enhancement of neutral-current interactions such as the ones discussed in \cref{sec:invisible}. The NC-to-CC ratio was extracted using a sign-selected $\nu_\mu$ and $\bar{\nu}_\mu$ beam scattering on an iron-scintillator calorimeter, where CC events were identified by a long muon track and NC events as short hadronic showers. The observable was thus determined from the ratio of short to long events, with corrections for misidentification and detector effects. The statistical uncertainty reached the $\mathcal{O}(10^{-3})$ level, while dominant systematics—primarily CC/NC misclassification, charm production, and parton distribution functions—were of comparable size. An enhancement of the NC rate therefore appears as an excess of short events, rendering the measurement sensitive to per-mille-level deviations in the NC/CC ratio. 

\paragraph{Proton beam dumps} High energy proton beams impinging on fixed target produce a large flux of forward mesons, which in turn can produce DS jets, as discussed in \cref{sec:CrossSecScaling}. If the DS fragments are long-lived enough, they can pass the shielding lenght, $L_{\rm shield}$, and decay inside the detector volume. The leading experiment that probes our parameter space is CHARM~\cite{CHARM:1983ayi,Marocco:2020dqu,Barouki:2022bkt}, where a 400 GeV proton beam impinges on a copper target, with a total luminosity of $N_{POT}=2\times10^{18}$. We consider the meson flux parametrizations in \cref{eq:Athertoneq,eq:dmesonflux}, and the $D$ meson fragmentation constants shown in Ref.~\cite{VanDijk:2641500}.

We find that CHARM constrains the low-$\Lambda_{\rm IR}$ region for $\Delta_N = 5/2$, both in scenarios where conformal symmetry is approximately preserved up to the electroweak scale and where it is broken below it. When light pNGBs are introduced, the branching ratio of DS jets into visible final states is reduced and thus the expected rate. For $\Delta_N = 7/2$, the stronger scaling with $\Lambda_{\rm UV}^{-2}$ cannot be compensated by the luminosity, leading to generically weaker bounds. 

\begin{table}[t!]

    \centering
     \begin{tabular}{c|cccccc }
                Exp. & Particle $(x)$ & $N_{xOT}$  &$ E_x$ [GeV]&$L_{\rm shield}$ [m] &$L_{\rm dec}$ [m]&$L_{\rm det}$ [m]\\
        \hline
       CHARM~\cite{CHARM:1983ayi,Davier:1989wz}& $p$& $10^{18}$&400&480&35&10\\
       SHiP~\cite{SHiP:2015vad}&$p$&$ 10^{20}$&400&50&14&50\\
      DUNE ND~\cite{DUNE:2021tad,DUNE:2020fgq} &$p$& $10^{21}$ & 120 & 574 & 5 & 5 \\
    DUNE FD~\cite{DUNE:2020ypp,DUNE:2020lwj} & $p$& $10^{21}$ & 120 & $1.3\times 10^6$ & 58 & 58 \\
\hline
   ORSAY~\cite{Davier:1989wz}&$e^-$&$10^{16}$&2&2&2&1\\
       E137~\cite{Bjorken:1988as,Batell:2014mga}&$e^-$&$10^{20}$&20&179&204&3\\
        NA64~\cite{NA64:2024klw}&$e,\mu$&$10^{13}$&150&0&10&0.5\\
    \end{tabular}
    \caption{Main features of the beam dump experiments considered in this work. The first three column show the particle beam type, the number of particles on target and their energy, respectively, while the last three show the shield, decay volume and detector lengths.}
    \label{tab:dumpexp}
    \end{table}

\paragraph{Electron beam dumps}
DS jets can be produced also at electron beam dumps~\cite{Andreas:2012mt}, such as ORSAY~\cite{Davier:1989wz} and E137~\cite{Bjorken:1988as}. Similarly to the proton beam dump case, the DS jet is produced by electron scattering on target and can travel downstream through the shielding, where scintillators detectors are positioned. However, the region of parameter space probed by these searches is already excluded by other probes, and we therefore do not show any limit from them.

\paragraph{NA64$\mu$} 
The NA64 experiment at CERN searches for DS particles using an intense electron beam dump from the SPS. The collaboration has recently extended its physics program by exploiting an intense muon beam, thereby providing sensitivity to DS scenarios coupled to the second lepton generation. This configuration, referred to as NA64$\mu$, is described in~\cite{NA64:2024klw}.

The expected number of signal events can be estimated using the inputs from \cref{tab:dumpexp} as
\beq
N_{\rm events} \sim 2 \times 10^{6} \left( \frac{17~{\rm GeV}}{\Lambda_{\rm UV}} \right)^{2\Delta_N - 3} \, ,
\eeq
where we have assumed a geometric acceptance $\epsilon_{\rm geo} \sim 1$, a decay probability $P_{\rm dec} \sim 1$ and a target density of $\mathcal{O}(1)$ GeV$^2$. This estimate corresponds to $\mathcal{O}(1)$ events for $\Lambda_{\rm UV} \lesssim 10^{4}~{\rm GeV}$ when $\Delta_N = 5/2$, and $\Lambda_{\rm UV} \lesssim 5 \times 10^{2}~{\rm GeV}$ for $\Delta_N = 7/2$.

We emphasize that this estimate is obtained assuming the projected number of muons on target expected at the HL-LHC, whereas the current dataset corresponds to $\sim 10^{11}$ muons on target. These results therefore indicate that NA64$\mu$ at full luminosity could probe signatures of strongly coupled dynamics in the DS, potentially leading to emerging SM particles within the detector volume.

\paragraph{Supernova bounds}
Light states can be efficiently produced in astrophysical environments, most notably in supernovae (SN). Bounds on heavy neutral leptons (HNLs) mixing with muon or tau neutrinos were derived in~\cite{Carenza:2023old} using the standard luminosity argument: particles that are produced and escape the SN contribute to anomalous cooling, while decays inside or outside the star can modify the explosion dynamics or yield observable signals.

The typical SN temperature distribution peaks at $T_{\rm SN} \sim 30~\mathrm{MeV}$~\cite{Fischer:2018kdt}, with a tail up to $\sim 100~\mathrm{MeV}$, allowing us to distinguish two regimes. For $\Lambda_{\rm IR} > T_{\rm SN}$, production occurs in the confined regime, and bounds can be recast by identifying the lightest fermionic bound state $\psi$ with an HNL with effective mixing angle given in \cref{eq:mixingangle}. If $\psi$ predominantly decays into invisible hidden states, constraints based on visible decays are weakened, and cooling bounds dominate.

For $\Lambda_{\rm IR} < T_{\rm SN}$, production is dominated by DIS processes and is enhanced relative to the confined case by $(2 m_p T_{\rm SN}/\Lambda_{\rm IR}^2)^{2\Delta_N - 3}$. Decay-based bounds are correspondingly strengthened, while cooling bounds, which depend only on production, lead to a rescaling of the $\Lambda_{\rm UV}$ limit by $(2 m_p T_{\rm SN}/\Lambda_{\rm IR}^2)^{(2\Delta_N - 5)/(2\Delta_N - 3)}$. This results in an $\mathcal{O}(10)$ strengthening for $\Delta_N = 7/2$, while the $\Delta_N = 5/2$ case remains approximately unchanged.

Overall, SN bounds can dominate over terrestrial constraints for $\Lambda_{\rm IR} \lesssim 500~\mathrm{MeV}$, although they are subject to larger theoretical and astrophysical uncertainties. We thus show these limits in \cref{fig:delta_combined} as a light gray shaded region, in order to distinguish them from accelerator based constraints.

\begin{table}[t!]

    \centering
    
    \begin{tabular}{c|ccccccc}
                Exp.  &$ \sqrt{s}$&beam&${\cal L}$ &Particle $(P)$&$N_P$&$L_{\rm det}$ [m]\\
        \hline
        LHCb~\cite{LHCb:2018roe}&14 TeV&$pp$&300 fb$^{-1}$&$B$&$10^{13}$&10\\
        
        \hline
        BELLE-II~\cite{Belle-II:2018jsg}&10.58 GeV&$e^+e^-$&50 ab$^{-1}$&$B$&$5\times 10^{10}$&8\\
        FCC-ee~\cite{FCC:2025lpp,Blondel:2022qqo}&91 GeV&$e^+e^-$&150 ab$^{-1}$&$Z$&$5\times10^{12}$&2\\
    \end{tabular}
    \caption{Main features of $B$ and $Z$ factories considered in this work. The first three columns show the center of mass energy, the colliding beams and the integrated luminosity of the experiments, respectively. The particle $P$ produced and its total number are indicated in the fourth and fifth column, while the last column indicates the typical detector length. For FCC-ee we assumed also an improvement in the missing energy measurement of a factor of 20 compared to the one in \cref{tab:SMdecay}.}
    \label{tab:collexp}
    \end{table}

\paragraph{Astrophysical neutrinos}

Astrophysical sources provide a broad neutrino flux spanning many orders of magnitude in energy~\cite{Vitagliano:2019yzm}. Above the GeV scale the flux is dominated by atmospheric neutrinos, while at ultra-high energies additional sources produce neutrinos with energies $E_\nu \sim 100~\mathrm{TeV}$--$1~\mathrm{PeV}$. The fluxes can be approximated as
\beq
\phi_\nu^{\rm atmo} \sim \frac{1.5\times10^{17}}{{\rm km^2~GeV~y}} \left( \frac{E_\nu}{{\rm GeV}} \right)^{-2.7}, 
\qquad
\phi_\nu^{\rm astro} \sim \frac{7.7\times10^{-2}}{{\rm km^2~GeV~y}} \left( \frac{E_\nu}{{\rm PeV}} \right)^{-2.37}.
\eeq
High energy astrophysical fluxes can in principle be probed at neutrino telescopes such as IceCube~\cite{IceCube:2016zyt} and KM3NeT~\cite{Margiotta:2014gza}. However, their sensitivity is limited by the relatively low event yield compared to beam experiments, the rapid fall of the flux with energy, and the difficulty of background rejection. In addition, the typical partonic center-of-mass energy remains limited by small Bjorken-$x$~\cite{Palmisano:2025abd}, so that the accessible kinematic range is not significantly extended compared to accelerators.

DS production inside the detector can lead to displaced ``double-bang'' signatures~\cite{Coloma:2017ppo, Airen:2025uhy}, where a NC event is followed by displaced hadronic and leptonic tracks from the decay of DS fragments. The expected number of events scales as
\beq
S_{\rm DB} \sim \Delta T \, V_{\rm det} \, n_{\rm det} \int \! \di E_\nu \, \frac{\di \Phi_\nu}{\di E_\nu} \, \sigma(E_\nu) \, P_{\rm dec},
\eeq
where $\Delta T$ is the time exposure, $V_{\rm det}$ is the detector volume and $n_{\rm det}$ is the nuclear number density of the detector material. Even neglecting backgrounds, this channel yields sensitivities roughly a factor of $2$ better than current bounds for $\Lambda_{\rm IR} \lesssim \rm{few}~\mathrm{GeV}$ for $\Delta_N = 5/2$, while for $\Delta_N = 7/2$ it can improve the $\Luv$ bound by up to a factor of $4$ in the same $\Lir$ range. This gain, however, does not survive once more realistic detector efficiencies are taken into account.

By contrast, the enhanced NC signal from escaping states is overwhelmed by the large SM background~\cite{IceCube:2022kff}, rendering this channel ineffective. A more promising possibility arises from DS states produced in the Earth and decaying inside the detector~\cite{Plestid:2020ssy}. This may yield competitive sensitivity for $\Delta_N = 7/2$ and $\Lambda_{\rm IR} \gtrsim \mathcal{O}(100~\mathrm{MeV})$ if the background can be suppressed. A more detailed study of this possibility is left for future work.

\paragraph{Indirect probes} 

Independently of the details of the UV completion, the portal in \cref{eq:compositeN} predicts that, at energies well below the confinement scale, the spectrum contains a pseudo-Dirac heavy neutral lepton with mixing angle given by \cref{eq:mixingangle}. Standard low-energy probes of HNLs therefore apply. In the parameter region of interest, however, these constraints are subleading, as high-energy probes are enhanced by the UV scaling of composite production and dominate the sensitivity.

We next consider indirect constraints from dimension-six operators built out of SM fields. The dominant contribution to precision observables arises from
\begin{equation}
\frac{1}{\Lambda_{\rm UV}^2}(HL)^\dagger \sigma_\mu p^\mu (HL)\,,    
\end{equation}
which modifies neutrino couplings to electroweak gauge bosons and hence affects precisely measured observables such as decay rates.

Generic UV completions are expected to generate this operator with an \(\mathcal{O}(1)\) Wilson coefficient, implying bounds on \(\Lambda_{\rm UV}\) at the \(\mathcal{O}(10)\,\mathrm{TeV}\) scale~\cite{Antusch:2006vwa,Antusch:2014woa,Blennow:2023mqx,Ahmed:2024hpg}. If the operator arises only at one loop, these bounds are relaxed to \(\mathcal{O}(1)\,\mathrm{TeV}\).  

Importantly, this constraint is independent of \(\Delta_N\), in contrast to bounds from the mixing operator, which weaken for increasing \(\Delta_N\). A tuned UV completion could suppress the Wilson coefficient, rendering indirect bounds arbitrarily weak. As shown in \cref{fig:delta_combined}, precision constraints dominate for \(\Delta_N \geq 7/2\). Therefore, indirect probes provide the leading sensitivity in the regime of a highly composite neutrino partner, unless additional tuning is invoked.

We also consider IR contributions to SMEFT operators from virtual exchange of light states excited by $\mathcal{O}_N$, which induce momentum-dependent neutrino form factors. For semi-integer $\Delta_N$, the operator in momentum space reads
\begin{equation}\label{eq:nu_FF}
    (HL)^\dagger \langle T \{ O_N \bar{O}_N \} \rangle (HL)
    = \frac{1}{\Lambda_{\rm UV}^{2\Delta_N-3}}
    (HL)^\dagger (p^2)^{\Delta_N-5/2}\sigma_\mu p^\mu
    \log\!\left(\frac{p^2}{\Lambda_{\rm UV}^2}\right) (HL)\,.
\end{equation}
The logarithm signals a non-local form factor associated with running. These contributions affect processes mediated by higher-dimensional contact operators. At high energies, if the conformal regime extends to the electroweak scale, they yield constraints comparable to LEP $Z$-pole measurements. At low energies, they match onto operators such as $\bar{\nu}\nu\,\Box\,\bar{q}q$, inducing model-dependent corrections to neutrino scattering. Rather than producing new signal events, they modify the neutral-current background due to their distinct kinematics.

\subsection{Future searches}\label{sec:future}

\paragraph{DUNE -- early 2030}
The DUNE experiment~\cite{DUNE:2020lwj,DUNE:2020ypp} will deliver an intense neutrino flux of approximately $N_\nu \sim 3\times10^{20}/{\rm year}$ in the energy range $0.4$--$10~\mathrm{GeV}$~\cite{DUNE:2021cuw}; this is produced at Fermilab by a high-intensity proton beam, with $1.1\times10^{21}$ POT/year. The detector complex consists of a Near Detector (ND), located $574~\mathrm{m}$ downstream, and a Far Detector (FD) at a baseline of $\sim1300~\mathrm{km}$. Both detectors employ liquid argon time projection chamber (LArTPC) technology, enabling precise energy and track reconstruction. In the following, we focus on the ND, as the FD receives a significantly attenuated flux and has a geometrical acceptance for dark states produced on target suppressed by a factor $\sim 10^3$.

Despite the relatively low center-of-mass energy, DUNE will collect a very large sample of neutrino interactions. Dark sector jets can be produced through the interaction in \cref{eq:compositeN}. The resulting constraints from modifications of the NC/CC ratio are shown as orange dashed lines in \cref{fig:delta_combined,fig:delta_combined_pNGB}. We note that this estimate assumes deep-inelastic scattering (DIS) scaling over the full energy range; although the flux peaks at a few GeV, where DIS is relevant, quasi-elastic and coherent processes may contribute and a more refined treatment would be required for a dedicated analysis.

For shorter lifetimes, the DUNE ND can probe displaced and emerging signatures, thanks to a vertex resolution of $L_{\rm min} \sim 1~\mathrm{cm}$~\cite{DUNE:2021tad} and timing capabilities; we assume a detector length $L_{\rm det} = 5~\mathrm{m}$. Given the relatively low $\sqrt{s}$, the multiplicity of DS fragments is small (see \cref{fig:jets_LE}), favoring displaced vertices over fully developed emerging jet signatures. The resulting projection is shown as a solid orange line in \cref{fig:delta_combined}.

Finally, DUNE can also operate as a proton beam dump experiment~\cite{Brdar:2022vum}, producing dark states with $\Lambda_{\rm IR} < m_D/2$ via secondary meson decays~\cite{Batell:2009di,deNiverville:2011it}, constraining the low $\Lir$ region; this is shown by the dotdashed orange line in \cref{fig:delta_combined}.

\paragraph{SHiP -- early 2030}

SHiP~\cite{SHiP:2015vad,SHiP:2021nfo} is a proposed proton beam dump experiment at CERN SPS, with a projected integrated flux of $2\times10^{20}$ protons on target over five years. The experimental setup includes a decay volume located $\sim 90~\mathrm{m}$ downstream of the target, instrumented with tracking and calorimetric detectors, as well as a compact $\nu_\tau$ detector, enabling the study of both neutrino interactions and decays of long-lived dark sector states. In the following, we focus on the latter. SHiP is expected to observe a relatively small sample of neutrino interactions, $\sim 10^6$, and, as discussed in~\cite{Borrello:2025hal}, the lower center-of-mass energy compared to FPF, together with the reduced neutrino flux relative to DUNE, leads to weaker projected sensitivities.

In beam-dump mode, however, the large luminosity and compact detector geometry, compared to previous experiments such as CHARM and to DUNE, allow for an improvement over existing limits, as illustrated in \cref{fig:delta_combined} as a dotdashed purple line.

\paragraph{Belle II and LHCb -- present to early 2040} 
Decays of heavy charm ($D$) and bottom ($B$) mesons can produce DS jets in a larger kinematical window than the light meson case, reaching $\Lir\sim$GeV scales and leading to both invisible and displaced signatures. Flavor factories as Belle II~\cite{Belle-II:2018jsg} and LHCb~\cite{LHCb:2018roe} are already actively taking data, although these do not yield relevant constraints in the scenarios considered here yet.

Invisible final states can be probed through missing-energy measurements at Belle II in two-body~\cite{CLEO:2008ffk,BESIII:2013iro,BaBar:2008fhj,Belle-II:2026flt} and semileptonic decays~\cite{Belle:2000cnh,BaBar:2010efp,Belle:2013ytx,Belle:2013hlo,Belle:2015pkj,Belle-II:2018jsg,Belle-II:2025pye}, as well as at LHCb~\cite{LHCb:2014osd,LHCb:2016inz,LHCb:2017xxn,LHCb:2025ymr,LHCb:2018roe}; however, their present sensitivity is limited. We summarize the limits on these invisible branching ratios in \cref{tab:SMdecay}.

Displaced signatures from DS jets, on the other hand, provide a promising probe. We focus on the clean topology where the displaced decay produces a dilepton pair. This signature can be efficiently targeted and probed with the full luminosity of Belle~II and LHCb, see \cref{tab:collexp}.

In the displaced regime, a large fraction of the signal arises from semileptonic processes, \({\bf m} \to {\bf m}' \ell \nu\), where an off-shell neutrino can mediate the production of a DS jet. We focus on inclusive semileptonic \(B\) decays, \(B \to X_{c,u} \ell + n\rm{DV}(2\ell)\), which provide the dominant sensitivity. The search requiring at least 1 DV with a di-lepton pair would be applicable also to weakly coupled HNL while asking for at least 2 DVs would identify unambiguously our composite sterile scenario. To our knowledge, no dedicated displaced search has been proposed in these channels, despite existing studies of dark showers and displaced signatures at Belle~II~\cite{Duerr:2019dmv,Bernreuther:2022jlj}. We strongly encourage the collaborations to study these channels.

Regarding the computation of the signal rate we highlight few complications. In contrast to the SM case, where the massless leptonic current fixes the hadronic structure, the presence of a massive DS jet with \(p_D^2 > 0\) requires two independent form factors, \(F_+\) and \(F_0\), and leads to dynamically varying Dalitz boundaries. Both \(X_c\) and \(X_u\) final states contribute: although \(X_u\) is Cabibbo-suppressed, its larger phase space enhances sensitivity to heavier DS states. We provide more details on the computation in \cref{sec:SMdecay}.
A fully inclusive treatment would require summing over all hadronic final states. In practice, we set all form factors to one and verify that this reproduces the SM inclusive rates. We require a single prompt charged lepton to tag the \(B\) decay, together with at least a displaced vertex containing a dilepton pair. The projected sensitivity at Belle II and LHCb are shown in \cref{fig:delta_combined} as a solid green line and pink, respectively. 

Together with heavy mesons, at Belle~II a large number of $\tau^\pm$ pairs are produced. The ratio of $\tau^\pm$ pairs produced with respect to the  \(B\) meson ones is approximately $N_{B\bar{B}}/N_{\tau\bar{\tau}}\approx1.2$. Since all tau decays contain at least a \(\nu_\tau\), the full sample is effectively available for neutrino-portal signatures, whereas only semileptonic \(B\) decays contribute. As a result, tau decays can provide a slightly larger effective signal yield for \(\Lambda_{\rm IR} \lesssim 1~\mathrm{GeV}\), where the available phase space is maximal.

In contrast, at LHCb $\tau$ production is strongly suppressed relative to \(B\) production, due both to the \(\mathcal{O}(10^5)\) smaller production cross section and to the forward acceptance of the detector, which further reduces sensitivity to centrally produced processes, such as Drell--Yan \(\tau^+\tau^-\). In this sense, Belle~II uniquely offers the opportunity to probe the neutrino portal using both \(\tau\) and \(B\) decays, potentially providing insight into the flavor structure of the underlying dynamics.

\paragraph{FPF -- early 2030} The FPF~\cite{Feng:2022inv} will consist of a suite of detectors located along the HL-LHC beam axis, approximately $600~\mathrm{m}$ downstream of the ATLAS interaction point. These detectors are designed to capture the intense forward flux of particles produced in $pp$ collisions, including neutrinos and displaced decays of DS states.

The proposed detectors rely on different technologies and therefore require distinct assumptions on event topology and displaced track multiplicities for background suppression. For instance, FLArE, based on LArTPC technology, provides timing information that enables the association of a primary vertex—identified through hadronic activity from neutrino-induced nucleon disintegration—with secondary vertices, without requiring the decay products to point back to the interaction point. In contrast, emulsion detectors rely on topological reconstruction and typically require multiple displaced vertices to geometrically associate them with the corresponding primary interaction.

For simplicity, we consider a single FASER-like detector with timing capabilities, and show the corresponding sensitivity for DV in \cref{fig:delta_combined} as a solid red line. We assume a detector length $L_{\rm det}=1~\mathrm{m}$ and a minimum displacement $L_{\rm min}=1~\mathrm{cm}$. This requirement is conservative, as FASER~\cite{FASER:2019dxq,FASER:2024ref,Hayakawa:2025xra} can achieve a spatial resolution at the level of a few~$\mu\mathrm{m}$~\cite{FASER:2025qaf}.

The red dotdashed lines in \cref{fig:delta_combined} correspond to processes in which neutrino disintegration occurs in the upstream FLArE detector, providing a tagged primary vertex, while the displaced decays take place in the downstream FASER-like detector, separated by a decay volume of length $L_{\rm dec}\sim 10$ m. The reach of invisible signals is shown instead as a red dashed line in \cref{fig:delta_combined,fig:delta_combined_pNGB}

Finally, we also note that the FPF can probe displaced vertices from long-lived particles produced directly in LHC collisions~\cite{FASER:2018eoc,Kose:2025cff}. A detailed modeling of this contribution is left for future work.

\paragraph{Future $e^+e^-$ colliders -- mid 2040}
Future electron-positron colliders, such as FCC-ee~\cite{FCC:2025lpp,Blondel:2022qqo} and CEPC~\cite{CEPCStudyGroup:2018ghi,CEPCStudyGroup:2025kmw}, will significantly enhance the sensitivity to boson decays by producing large event samples in a clean experimental environment. Operating at the $Z$ pole, $\sqrt{s}\simeq91~\mathrm{GeV}$, these machines are expected to produce up to $N_Z \simeq 6\times10^{12}$ $Z$ bosons over their full integrated luminosity.

We focus on the FCC-ee proposal and consider the IDEA detector concept~\cite{IDEAStudyGroup:2025gbt}. We require a displaced vertex within the drift chamber, with a characteristic size $L_{\rm det}\sim2~\mathrm{m}$, and assume a minimum displacement set by the projected vertex resolution near the interaction point, $\Delta L \simeq 5~\mu\mathrm{m}$. The reach for a DV is shown as a solid blue line in \cref{fig:delta_combined}; the reach for multiple DV is in the bottom row of \cref{fig:jets_HE} (for the general scenario of unbroken CFT). Backgrounds can be efficiently suppressed by vetoing events with back-to-back visible tracks compatible with a two-body $Z$ decay. In addition, the large statistics will allow for an improvement of the $Z$ invisible branching ratio measurement by approximately a factor of $20$. This limit is shown as dashed blue lines in \cref{fig:delta_combined,fig:delta_combined_pNGB}.  Analogously to the case of constraints from Higgs decays, we extend the invisible reach into the region where the dark resonance becomes prompt. This regime approximately corresponds to the sensitivity of untagged \(Z\) decays, although dedicated searches targeting prompt non-standard final states, such as multileptons accompanied by jets, are expected to yield significantly stronger constraints. We leave a careful study of these channels for further studies.

\section{The composite $\nu$-SMEFT}
\label{sec:nusmeft}
In this section we briefly discuss higher-dimensional portals connecting the Standard Model (SM) to the dark sector (DS). As discussed in \cref{sec:HLO}, the operator $HL\mathcal{O}_N$ is the unique portal at dimension $\Delta_N + 5/2$. At the next order in the SMEFT expansion, operators of dimension $\Delta_N + 9/2$ arise, providing a natural generalization of the SMEFT extended with heavy neutral leptons~\cite{delAguila:2008ir,Liao:2016qyd} (see also~\cite{Alcaide:2019pnf,Fernandez-Martinez:2023phj} for phenomenological studies).

A first class of operators involves two additional insertions of the Higgs field,
\beq
\Q_{2H,i} = \frac{(\bar L_i H \mathcal{O}_N)(H^\dagger H)}{\Lambda^{\Delta_N + 1/2}}, 
\qquad 
\Q_{2HD,i} = \frac{(\tilde{H}^{\dagger} \overset{\longleftrightarrow}{D_\mu} H)(\bar{\mathcal{O}}_N\gamma^{\mu} e_i)}{\Lambda_{\rm UV}^{\Delta_N +1/2}},
\eeq
where $e_i$ denotes the right-handed charged lepton and $D_\mu$ the SM covariant derivative. 

Dipole operators can be constructed by including gauge field strengths,
\beq
\Q_{HB,i} = \frac{(\bar{L}_i\sigma^{\mu\nu}\mathcal{O}_N)(\tilde{H}B_{\mu\nu})}{\Lambda_{\rm UV}^{\Delta_N +1/2}}, 
\qquad 
\Q_{HW,i} = \frac{(\bar{L}_i\sigma^{\mu\nu}\mathcal{O}_N)(\tau^a\tilde{H}W^a_{\mu\nu})}{\Lambda_{\rm UV}^{\Delta_N +1/2}},
\eeq
where $B_{\mu\nu}$ and $W_{\mu\nu}$ are the SM field strengths. At low energies, these operators map onto dipole interactions of the fermionic states.

A second class consists of four-fermion operators, which can be written schematically as
\beq
\Q_{4F,ijk}= \frac{1}{\Lambda_{\rm UV}^{\Delta_N +1/2}} \left[ (\bar L_i \Gamma \mathcal{O}_N)(\bar F_j \Gamma F_k) + (\bar e_i \Gamma \mathcal{O}_N)(\bar F_j \Gamma F_k) \right],
\eeq
where $F=Q,L,u,d,e$ and $\Gamma$ denotes a generic Dirac structure. Neglecting operators with two insertions of $\mathcal{O}_N$, one obtains in particular~\cite{Li:2021tsq}
\begin{align}
    \Q_{duNe} &= (\bar d\gamma_\mu u)(\bar{\mathcal{O}}_N \gamma^\mu e), \\
    \Q_{LeLN}^{(1)} &= (\bar L^a e)\,\epsilon_{ab}\,(\bar L^b \mathcal{O}_N), \\
    \Q_{QuNL}^{(1)} &= (\bar Q u)(\bar{\mathcal{O}}_N L), \\
    \Q_{QdLN}^{(1)} &= (\bar Q^a d)\,\epsilon_{ab}\,(\bar L^b \mathcal{O}_N).
\end{align}
For scalar operators $\Q^{(1)}$, tensor counterparts with $\sigma_{\mu\nu}$ structures, denoted $\Q^{(3)}$, can also be constructed, e.g. $\Q_{LeLN}^{(3)} = \lp \bar L^a \sigma_{\mu\nu} e \rp \epsilon_{ab} \lp \bar L^b \sigma^{\mu\nu} {\cal O}_N \rp$.

Most four-fermion operators contain both neutral- and charged-current components, with the exception of $\Q_{duNe}$, which is purely charged-current. These features are summarized in \cref{tab:portal}. We also note that, after electroweak symmetry breaking, operators contributing dominantly to one current can induce the other through off-shell neutrino exchange.

\begin{table}[t!]

    \centering
    
    \begin{tabular}{c|c c c c c c c c c}
        $\Q$ & $\Q_{HL}$ & $\Q_{2HD}$ & $\Q_{HB}$ & $\Q_{HW}$& $\Q_{2H}$ & $\Q_{duNe}$ & $\Q_{LeLN}$ & $\Q_{QuNL}$ & $\Q_{QdNL}$ \\
        \hline
        $\nu$ & \checkmark &  & \checkmark & \checkmark &\checkmark &  & \checkmark & \checkmark & \checkmark  \\
        $\ell$ & & \checkmark & & \checkmark &  & \checkmark & \checkmark & \checkmark & \checkmark
    \end{tabular}
    \caption{Summary table of the portals considered in this work. The checkmarks indicate which operators connect to neutrinos and charged leptons.}
    \label{tab:portal}
    \end{table}

\paragraph{Cross sections}
The production cross sections of higher-dimensional operators can be computed following the procedure outlined in \cref{sec:CrossSecScaling}. The parametric scaling in \cref{eq:totcatt} acquires additional powers of $s/\Lambda_{\rm UV}^2$ due to the extra field insertions.

\begin{figure}[t!]
        \centering

        \includegraphics[width=0.45\linewidth]{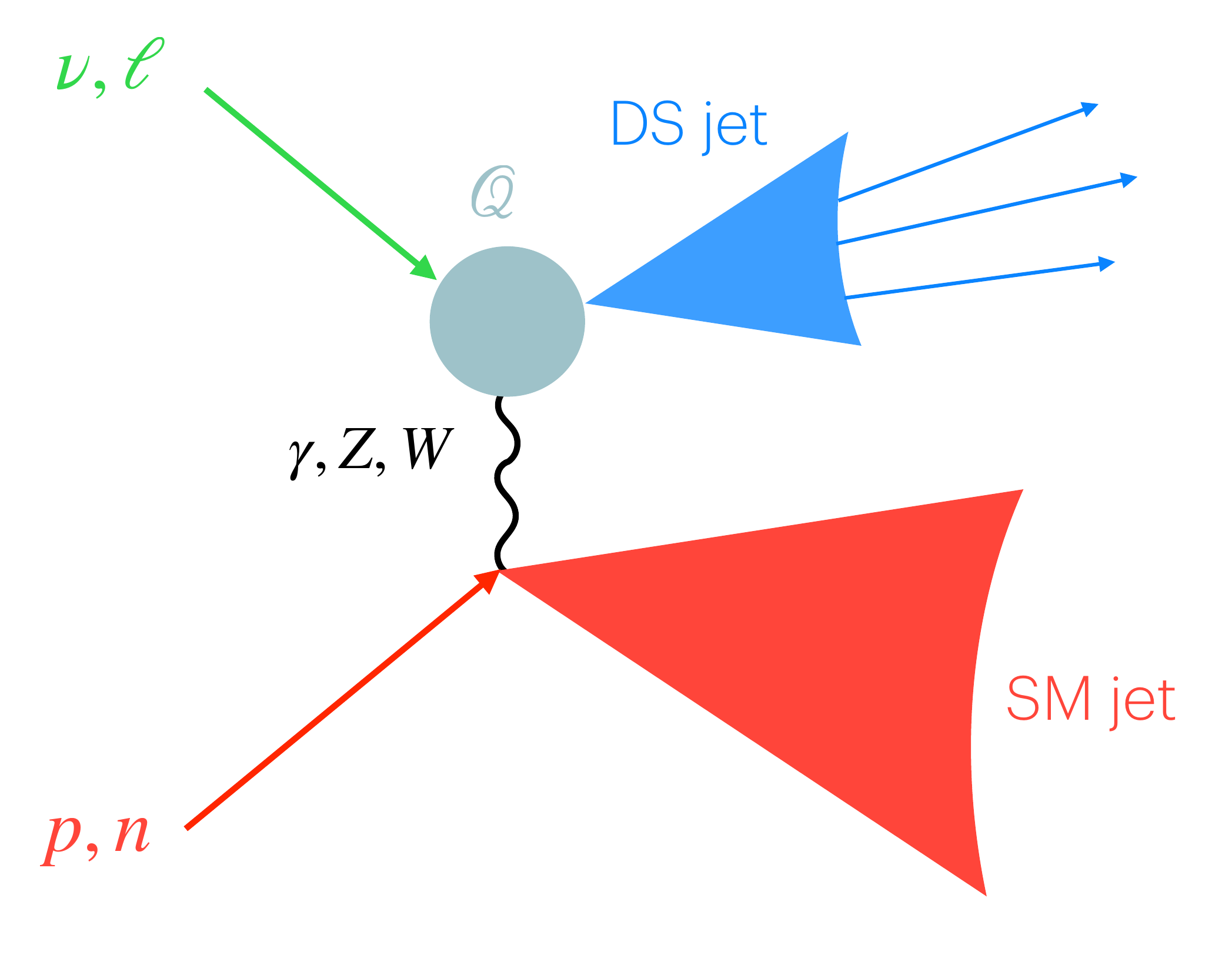}
        \includegraphics[width=0.49\linewidth]{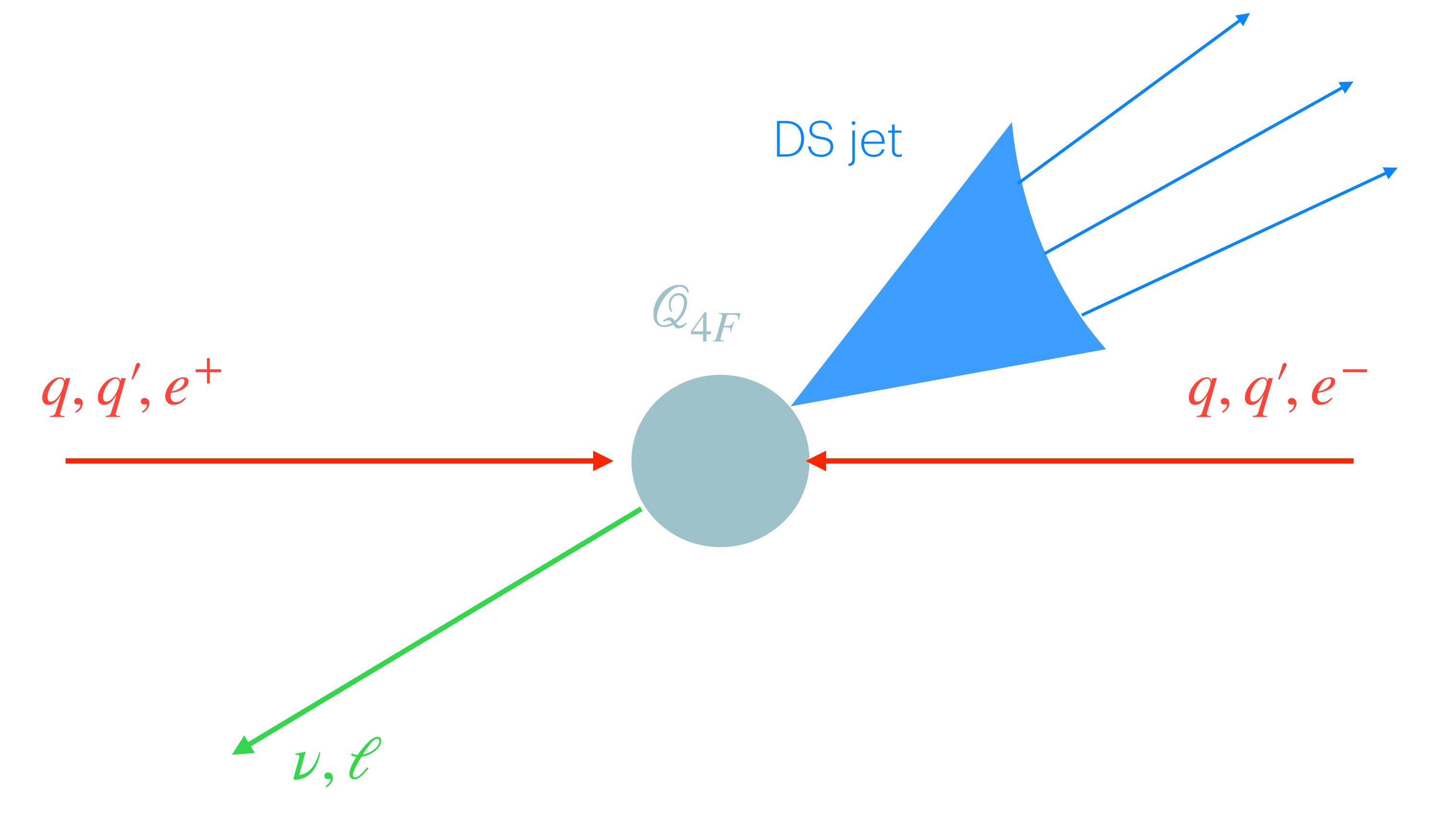}
    
    \caption{Sketch diagrams for DS jet production from higher order portals. {\bf Left:} production via SM boson exchange. {\bf Right:} production via contact interaction. }
    \label{fig:HighDimPortals}
\end{figure}

Starting from the operator $\Q_{2H}$, its dynamics are analogous to $\Q_{HL}$, with two additional Higgs insertions. One finds
\beq
\sigma_{2H} \sim  \C_{2H}^2 \frac{v^2}{s^2} \left( \frac{s}{\Lambda_{\rm UV}^2} \right)^{\Delta_N + 1/2}\,.
\eeq
The operators involving a single Higgs and a derivative or field strength, $\Q_{2HD}$ and $\Q_{HB,W}$, induce dark jet production via vector boson exchange (see the left panel of \cref{fig:HighDimPortals}), and scale as
\begin{align}
\sigma_{2HD} &\sim  \frac{\C_{2HD}^2}{s}  \left( \frac{s}{\Lambda_{\rm UV}^2} \right)^{\Delta_N + 1/2}\,,\\
\sigma_{HB,W} &\sim \left( \frac{\C_{HB} v}{s} \right)^2 \left( \frac{s}{\Lambda_{\rm UV}^2} \right)^{\Delta_N + 1/2}\,.
\end{align}
Finally, four-fermion operators correspond to contact interactions (see the right panel of \cref{fig:HighDimPortals}), with cross section
\beq\label{eq:4Fxsec:scaling}
\sigma_{4F} \sim  \frac{\C_{4F}^2}{s} \left( \frac{s}{\Lambda_{\rm UV}^2} \right)^{\Delta_N + 1/2}\,.
\eeq
For fixed $\Delta_N$, the $\Q_{4F}$ and $\Q_{2HD}$ contributions exhibit the strongest growth with $s$ among the higher-dimensional operators.
The two classes of operators exhibit the same power counting because $\mathcal{Q}_{2HD} \propto v^2\, W_\mu\, \bar{\mathcal{O}}_N \gamma^\mu e$. After integrating out the $W$ boson below the electroweak scale, additional four-fermion operators are generated; however, their Wilson coefficients are further suppressed by electroweak couplings.

\paragraph{Lifetimes}
Similarly to the cross section computation, the resonance lifetime induced by higher dimensional portals can be estimated following \cref{sec:WidthScaling}, or by using the approximate scaling in \cref{eq:GammaPsi}. 

For the $\Q_{2H}$ and $\Q_{2HD}$ operators one finds
\begin{align}
\Gamma_{2H} &\sim \frac{(\C_{2H} v)^2}{64\pi^3} \left( \frac{\Lambda_{\rm IR}^2}{\Lambda_{\rm UV}^2} \right)^{\Delta_N + 1/2} \Lambda_{\rm IR}^{-1}\,, \\
\Gamma_{2HD} &\sim \frac{\C_{2HD}^2}{64\pi^3} \left( \frac{\Lambda_{\rm IR}^2}{\Lambda_{\rm UV}^2} \right)^{\Delta_N + 1/2} \Lambda_{\rm IR}\,,
\end{align}
where the scaling of $\Gamma_{2HD}$ with $\Lambda_{\rm IR}$ reflects the fact that this operator induces purely charged-current processes mediated by $W$ exchange. Similarly to the operator in \cref{eq:compositeN}, when $\Lambda_{\rm IR} > m_\ell + m_\pi$ the resonance can undergo two-body decays via an off-shell $W$, up to additional CKM and electroweak coupling factors.

Dipole operators, $\Q_{HB}$ and $\Q_{HW}$, lead to two-body decays into $\gamma\nu$, with width
\beq
\Gamma_{HB,W} \sim \frac{(\C_{HB,W} v)^2}{8\pi} \left( \frac{\Lambda_{\rm IR}^2}{\Lambda_{\rm UV}^2} \right)^{\Delta_N + 1/2} \Lambda_{\rm IR}^{-1}\,.
\eeq
For four-fermion operators, it is useful to distinguish hadronic operators, denoted $4F,q$, from purely leptonic ones such as $\Q_{LeLN}$. 
When kinematically allowed, the former can lead to a decay into a lepton and a pion without extra couplings insertions, with width:
\beq\label{eq:Width:4Fq}
\Gamma_{4F,q} \sim \frac{\C_{4F,q}^2}{8\pi} \left( \frac{\Lambda_{\rm IR}^2}{\Lambda_{\rm UV}^2} \right)^{\Delta_N + 1/2} \Lambda_{\rm IR}\,,
\eeq
while the latter proceeds through three-body decays, $\Lambda_{\rm IR} > 3 m_\ell$, and is phase-space suppressed,
\beq\label{eq:Width:4Fl}
\Gamma_{LeLN} \sim \frac{\C_{LeLN}^2}{64\pi^3} \left( \frac{\Lambda_{\rm IR}^2}{\Lambda_{\rm UV}^2} \right)^{\Delta_N + 1/2} \Lambda_{\rm IR}\,.
\eeq
These decay channels are generically suppressed with respect to the standard HNL ones discussed in \cref{sec:WidthScaling} because their widths require extra powers of $(\Lambda_{\rm IR}/\Lambda_{\rm UV})^2$. The least suppressed contributions arise from dipole and $\Q_{2H}$ operators, scaling as $\Gamma_{HB,W}/\Gamma_\psi \sim 16\pi^2  v^4 / \Lambda_{\rm UV}^4$ and $\Gamma_{2H}/\Gamma_\psi \sim  v^4 / \Lambda_{\rm UV}^4$, respectively, while all other channels are suppressed by $(\Lambda_{\rm IR}/\Lambda_{\rm UV})^4$.

We note that the apparent $16\pi^2$ enhancement of dipole decays is typically not realized in ultraviolet completions, as the corresponding Wilson coefficients are loop suppressed, $\C_{HB} \sim (16\pi^2)^{-1} y_\ell$. Although scenarios in which dipole operators dominate can be constructed~\cite{Voloshin:1987qy,Brdar:2020quo}, we do not pursue this possibility further here.

\subsubsection*{Signal yield at colliders from $\Q_{4F}$}

The phenomenology of higher-dimensional portals can largely be understood by analogy with the $\Q_{HL}$ operator discussed in this work, once the modified scaling behaviors and possible phase-space suppressions from additional final-state particles are taken into account (e.g.\ dipole operators can induce extra photons in meson decays). A detailed analysis is left for future work.

A key difference, however, is that the processes considered in this paper (with the exception of Higgs decays) rely on the propagation of an off-shell neutrino from a SM weak interaction, which subsequently fragments into a dark jet via $\Q_{HL}$. Higher-dimensional portals can instead couple directly to the SM through contact interactions, without requiring an intermediate neutrino propagator, as illustrated in \cref{fig:HighDimPortals}.

The left panel of \cref{fig:HighDimPortals} corresponds to processes mediated by $\Q_{2HD}$, $\Q_{HB}$, and $\Q_{HW}$, where a gauge boson is emitted and acts as a mediator. These contributions are typically suppressed by electroweak couplings. While an enhancement may arise from the photon propagator in the $q^2 \to 0$ regime, relevant for coherent neutrino--nucleus scattering, such processes occur at low $\sqrt{s}$ and are not sufficient to produce states with masses in the $\Lambda_{\rm IR}$ range of interest.

We therefore focus on direct production of $\mathcal{O}_N$ through four-fermion portals, shown in the right panel of \cref{fig:HighDimPortals}. Using the scaling in \cref{eq:4Fxsec:scaling}, the number of produced DS jets can be estimated as
\beq\label{eq:N_4F:Scaling}
N_{\rm DS} \sim \mathcal{L}_{\rm int}\,\sigma_{4F} \sim \mathcal{L}_{\rm int}\,\frac{1}{s} \left( \frac{s}{\Lambda_{\rm UV}^2} \right)^{\Delta_N + 1/2}\,,
\eeq
where we have set $\C_{4F}=1$ and assumed the conformal regime $\Lambda_{\rm IR} \ll \sqrt{s} \ll \Lambda_{\rm UV}$. The strong energy dependence implies a rapid growth of the production rate with $s$. However, realistic sensitivities depend on the full event topology: inclusive searches suffer from large SM backgrounds, while displaced searches are limited by the long lifetimes associated with $\Q_{4F}$-induced decays (see \cref{eq:Width:4Fq,eq:Width:4Fl}).

To estimate the experimental reach, we rescale results obtained for analogous interactions with elementary HNLs~\cite{Fernandez-Martinez:2023phj}, using the parametrics of \cref{eq:4Fxsec:scaling} and evaluating $s$ at the relevant energy scale of each experiment. Assuming $\mathcal{O}(1)$ Wilson coefficients, the resulting bounds should be regarded as optimistic and model dependent.

Existing constraints are dominated by collider searches, which benefit from the largest center-of-mass energies. For the purely leptonic operator $\Q_{LeLN}$, LEP monophoton searches exclude $\Lambda_{\rm UV} \sim 200~\mathrm{GeV}$ for $\Delta_N=5/2$. For quark-lepton operators, mono-lepton searches at ATLAS~\cite{ATLAS:2017jbq,ATLAS:2019lsy} constrain $\Lambda_{\rm UV} \sim \mathcal{O}(\mathrm{TeV})$, while meson decay experiments such as NA62 can reach similar scales below the kaon threshold. These results confirm that collider searches provide the leading sensitivity, especially for larger $\Delta_N$ where the cross section grows more rapidly with energy.

At lepton colliders, such as Belle~II in the near term and FCC-ee in the longer term, production through $\Q_{LeLN}$ leads to invisible final states when the dark fragments are long-lived. Although the production rates can be sizable, the lack of visible activity significantly reduces the experimental sensitivity. A rough projection suggests that mono-photon searches at FCC-ee can constrain the operator $\Q_{LeLN}$ up to scales of order $\Luv \lesssim 500~\mathrm{GeV}$ for $\Delta_N = 5/2$. Displaced signatures are also suppressed, as the decay probability within the detector volume is typically $P_{\rm dec} \sim 10^{-6}$, resulting in negligible event yields.

In contrast, hadron colliders provide a more favorable environment. At the HL-LHC, with $\mathcal{L}_{\rm int}=3~\mathrm{ab}^{-1}$ and  assuming an energy for the partonic subprocess $\sqrt{\hat s}=500~\mathrm{GeV}$, one expects from \cref{eq:N_4F:Scaling} 
\beq
N_{\rm DS,  HL-LHC} \sim 10^{7} \left( \frac{\mathrm{TeV}}{\Lambda_{\rm UV}} \right)^6
\eeq
for $\Delta_N=5/2$. In this case, the signal consists of a high-$p_T$ charged lepton recoiling against a dark jet, leading to distinctive mono-lepton plus missing energy signatures. With appropriate kinematic cuts, the signal can populate regions with suppressed SM backgrounds, allowing sensitivity to $\Lambda_{\rm UV}$ at the TeV scale.

Requiring in addition a displaced vertex from the decay of dark-sector fragments can further enhance the sensitivity. Accounting for the typical boosts at hadron colliders, $\gamma\beta \sim \mathcal{O}(10\text{--}100)$, for $\Lir=1$ GeV and a detector length of 2 m, one expects 
\beq
N_{\rm HL-LHC}^{\rm displaced} \sim 300 \left( \frac{\mathrm{TeV}}{\Lambda_{\rm UV}} \right)^{12} \,,
\eeq
yielding a nearly background-free signature. 

Finally, we note that for $\Delta_N=5/2$ collider sensitivities are relevant primarily in scenarios where the four-fermion portal dominates. In models where it is generated together with the $HL\mathcal{O}_N$ operator at a common scale, existing constraints on $\Lambda_{\rm UV}$ significantly reduce the expected signal yield (see e.g. \cref{fig:delta_combined}). Overall, four-fermion portals favor high-energy hadron colliders, where large production rates can partially compensate for the suppression from long lifetimes; however, when they coexist with the composite HNL portal, they typically lead to subdominant signals and weaker overall sensitivity.

\section{Conclusions}
\label{sec:Conclusions}

Composite sterile sectors provide a compelling and testable extension of the SM, directly linked to neutrino mass generation and naturally connected to dark sectors (DS) that interact predominantly through neutrinos. Their phenomenology is both rich and distinctive, yielding experimental signatures that are not only novel but, in several cases, essentially unique to this framework. The results presented here identify concrete and experimentally accessible targets across multiple facilities.

A useful organizing principle is the scaling dimension \(\Delta_N\) of the composite fermionic operator in the DS. For \(\Delta_N \gtrsim 7/2\), the phenomenology is typically dominated by precision constraints on the dimension-6 operator \((HL)^\dagger \partial (HL)\), barring fine tuning of the UV parameters. In this regime, high-energy machines provide the leading probes only if such tuning is present. In contrast, the window \(3/2 < \Delta_N \lesssim 7/2\) exhibits a particularly rich phenomenology, where both invisible and visible signatures can be explored across a wide range of experiments.

In particular, we have identified distinct UV scenarios in which the relative importance of electroweak-scale observables and lower-energy probes is reversed, thereby motivating a comprehensive exploration of these signals across complementary experimental frontiers. For clarity, we summarize below the main experimental targets.

Below the electroweak scale, neutrino experiments emerge as a particularly powerful probe. We have identified two qualitatively new signatures: a measurable enhancement of the neutral-current to charged-current ratio (NC/CC) driven by long-lived sterile-sector resonances, and DS production in neutrino disintegration giving rise to a hierarchy of signatures, ranging from single displaced vertices (DVs) to multiple DVs and, in extreme cases, high-multiplicity events such as emerging jets. The observation of the latter two would constitute a smoking-gun signature of the composite sterile scenario. High-energy beams such as those at the FPF are especially well suited to probe neutrino-induced DS showers with high multiplicity, while the high luminosity of DUNE ensures complementary sensitivity. A decisive step forward to improve the experimental reach will require surpassing the precision of existing NC/CC measurements, such as those from NuTeV~\cite{NuTeV:2001whx}, and fully characterizing backgrounds to displaced signatures.

Similarly, at high-luminosity $B$ factories, including Belle~II and LHCb, rare decays of the form $B \to \ell X_{u,s} + n\,\mathrm{DV}(2\ell)$ provide a clean and powerful discovery channel. At Belle~II DVs produced in inclusive tau decays provide an interesting complementary probe. While single displaced dilepton vertices can also be a consequence of weakly coupled HNLs, the observation of multiple DVs would constitute a smoking-gun signature of a composite sterile sector.

Beam-dump experiments such as SHiP and DUNE can further access this parameter space, although predominantly through events with a single DV. In this context, requiring two DVs has the potential to dramatically suppress backgrounds, opening qualitatively new search strategies beyond those previously employed, such as by CHARM~\cite{CHARM:1983ayi,CHARM:1985nku}.

At the energy frontier, higher-dimensional four-fermion operators at the HL-LHC can induce exotic and striking signatures, such as mono-leptons accompanied by one or more displaced vertices, motivating dedicated searches. 
Looking ahead, FCC-ee provides an ideal environment to probe these scenarios, with displaced and invisible \(Z\) decays offering unprecedented sensitivity in a clean experimental setting.

Several directions merit further investigation. In particular, a more detailed assessment of astrophysical neutrino fluxes, beyond the preliminary discussion in \cref{sec:results}, and their interplay with terrestrial searches could open additional discovery channels. Finally, it will be important to generalize the well-known renormalizable portals to the composite case and to study systematically their phenomenology and interplay within QCD-like UV completions. We leave this program to future work.


\subsection*{Acknowledgments}
This work would not have been possible without the hospitality of the Galileo Galilei Institute for Theoretical Physics, to which all the authors are deeply grateful. We also acknowledge the valuable feedback received during the many presentations on this subject from both theoretical and experimental colleagues, whom we prefer not to list individually to avoid unintentionally omitting anyone. MT acknowledge hospitality and support during the final parts of this work by the Munich Institute for Astro-, Particle and BioPhysics (MIAPbP), which is funded by the Deutsche Forschungsgemeinschaft (DFG, German Research Foundation) under Germany´s Excellence Strategy - EXC-2094 - 390783311.
MC is supported in part by Perimeter Institute for Theoretical Physics. Research at Perimeter Institute is supported by the Government of Canada through the Department of Innovation, Science and Economic Development Canada and by the Province of Ontario through the Ministry of Research, Innovation and Science. MT acknowledges support by Next Generation EU, as part of Piano Nazionale di Ripresa e Resilienza (PNRR), Missione 4, Componente 2, Investimento 1.2 - CUP I13C25000150006.  The work of DR is supported in part by the European Union - Next Generation EU through the PRIN2022 Grant n. 202289JEW4.


\appendix

\section{Details on computations for DS production}\label{app:xsdetail}
In this Appendix we review the main steps and provide details to compute the production rates of DS jets induced by the portal in \cref{eq:compositeN}. In \cref{sec:NeutrinoNucleonScattering} we compute the production via neutrino-nucleon scattering, while in \cref{sec:SMdecay} we compute the relevant decay widths of SM particles into final states that include $\On$.

\subsection{Neutrino-Nucleon scattering}
\label{sec:NeutrinoNucleonScattering}
In high--energy neutrino scattering off an hadronic target, the initial--state nucleon can be approximated as a collection of free partons. For momentum transfers larger than the nucleon mass, $Q^2 \gtrsim 2 m_P^2$ with $P=p,n$\footnote{Here we assume isospin symmetry, so that we can take the masses of the nucleons to be equal, $m_p=m_n\simeq939$ MeV, and the parton PDFs to be equal under $u\to d$ replacement: $f_u\equiv f_{u/p}=f_{d/n}$ and $f_d\equiv f_{d/p}=f_{u/n}$. }, the interaction probes distances short enough that neutrino scattering effectively occurs on individual quarks inside the nucleon. In this regime, QCD is perturbative and a partonic description is justified.  At lower values of \(Q^2\), the strong coupling increases, partons are no longer weakly interacting, and non--perturbative QCD effects become dominant. To exclude events near the QCD resonance region, we impose a cut on the invariant mass of the hadronic final state, $m_{\rm j}^2 > 2 m_P^2$~\cite{Borrello:2025hal}.

The dark jet production cross--section is obtained by considering the scattering of a neutrino off a quark \(q\) carrying a fraction \(x\) of the proton momentum. The relevant neutral--current couplings of the quark are \(\ell_q\) and \(r_q\), corresponding to left-- and right--handed interactions, respectively. They are given by
\[
\ell_q = I_{3,q} - Q_q \sin^2\theta_W , 
\qquad
r_q = - Q_q \sin^2\theta_W ,
\]
where \(I_{3,q}\) is the third component of the quark weak isospin and \(Q_q\) its electric charge in units of \(e\), while $\theta_W$ is the weak mixing angle.

The matrix element squared, averaged on the initial spins, can be split into $|\bar M|^2=|\bar M_L|^2+|\bar M_R|^2$, where $M_L$ $(M_R)$ involves the left-handed (right-handed) quark current. We have
\begin{equation}
|\bar M_L|^2=y^2_{\ell}2\ell_q^2A_{N}\frac{\hat s(\hat s-p_D^2)}{v^2\Lambda_{\rm UV}^4}\left(\frac{p_D}{\Lambda_{\rm UV}}\right)^{2\Delta_N-7}\,,        
\end{equation}
and
\beq
|\bar M_R|^2=y^2_\ell2r_q^2A_{N}\frac{\hat u(\hat u-p_D^2)}{v^2\Lambda_{\rm UV}^4}\left(\frac{p_D}{\Lambda_{\rm UV}}\right)^{2\Delta_N-7}\,,
\eeq
where $\hat s=xs=2xE_\nu m_p$ and $\hat u$ are the partonic Mandelstam variables. We define \(x\) and the inelasticity \(y\) as
\begin{equation}
x = \frac{Q^2}{2p_P \cdot q}, \quad y = \frac{2p_P \cdot q}{s} \implies Q^2 = x y s \,.   
\end{equation}
Here $p_P$ is the momentum of the incoming nucleon , $p_j$ is the momentum of the hadronic final state, $p_\nu$ is the momentum of the incoming neutrino, $p_D$ is the momentum of the outgoing dark excitation, $\hat p_2$ and $\hat p_4$ are respectively the ingoing and outgoing momenta of the quark involved in the interaction, and $q=p_\nu-p_D=p_j-p_P$ is the momentum exchanged.
The fully differential cross-section for the process \(N + \nu \to \text{SM  jet} + \text{DS}\), obtained as a sum of partonic subprocesses \(q + \nu \to q + \text{DS}\), is 
\begin{align}
\di\sigma_{\rm 2DIS} =& y_\ell^2\frac{2A_N}{\Lambda_{\rm UV}^4 v^2} \left(\frac{p_D^2}{\Lambda_{\rm UV}^2}\right)^{\Delta_N - 7/2} \sum_{q, \bar{q}} f_q(Q^2, x) \left[\ell_q^2(\hat{s} - p_D^2) + r_q^2(1 - y)(\hat{s} - \hat{s}y - p_D^2)\right]\notag\\&\times \frac{\di p_D^2 \di^3 p_D \di^3\hat{p}_j}{(2\pi)^3(2\pi)^4 2 E_D 2 \hat{E}_4}(2\pi)^4 \delta^4(p_\nu + \hat{p}_2 - p_D - \hat{p}_4)\,,
\end{align}
where each subprocess is weighted by the respective quark PDFs \(f_q(Q^2, x)\), which describe the probability of finding a quark $q$ carrying a fraction $x$ of the proton momentum.
Integrating over the final state phase space at fixed $p_D^2$, the phase space simplifies to
\[
\frac{\di p_D^2}{16\pi^2} \frac{\hat{E}_4 \di \hat{E}_4 \di \cos\theta}{E_D} \delta(\hat{s} - E_D - \hat{E}_4)\,,
\]
where in the center-of-mass we have
\[
\hat{E}_4 = \frac{\hat{s} - p_D^2}{2\sqrt{\hat{s}}}, \quad E_D = \frac{\hat{s} + p_D^2}{2\sqrt{\hat{s}}}\,.\label{eq:kinreg}
\]
Defining \(\hat{t} = -Q^2 = (p_P - p_j)^2\) and using the relation \(\di \hat{t} = \di \cos\theta \frac{\hat{s} - p_D^2}{2}\), we find 
$$\di  x \di \hat{t} = x s \di x \di y\,.$$
Combining all together, the differential cross-section as a function of \(x, y, p_D^2\) is
\begin{align}\label{eq:diffxs}
    \frac{\di \sigma_{\rm 2DIS}}{\di x \di y \di p_D^2} &= y_\ell^2\frac{A_N }{8\pi^2 \Lambda_{\rm UV}^4 v^2} \left(\frac{p_D^2}{\Lambda_{\rm UV}^2}\right)^{\Delta_N - 7/2} \sum_{q, \bar{q}} f_q(Q^2, x) \notag\\&\times\left[\ell_q^2(xs - p_D^2) + r_q^2(1 - y)(xs - xsy - p_D^2)\right]\,.
\end{align}
The SM DIS differential cross-section is 
\begin{equation}\label{eq:calcdisbkg}
  \frac{d\sigma_B^{\rm DIS}}{dx dy} = \frac{s}{2\pi v^4} \sum_{q, \bar{q}} x f_q(Q^2, x) \left[\ell_q^2 + r_q^2(1 - y)^2\right]  \,.
\end{equation}
The kinematically allowed region in the \((x,y)\) plane for BSM dark–jet production is smaller than the one of SM DIS. This reduction originates from the kinematic constraints imposed on the final--state hadronic jet,
\begin{equation}
E_j \geq m_j \,, \qquad \cos^2\theta_j \leq 1 \,,
\end{equation}
which in the \((x,y)\) plane translate into the conditions
\begin{equation}
1 - \sqrt{y(1-x)} \geq \frac{p_D}{\sqrt s} \,, 
\qquad
x(1-y) \geq \frac{p_D^2}{s} \,.
\end{equation}

\subsection{Details on the decay widths}
\label{sec:SMdecay}
We provide explicit expressions for the two-body decay widths, shown in \cref{eq:meson,eq:boson,eq:higgs}, and for the three-body meson decay widths, used for computing the constraints from $D$ and $B$ meson decays in \cref{sec:results}.

\paragraph{Meson decays: two-body}
We consider the decay ${\bf m}^\pm\to \ell^\pm + \On$, where ${\bf m}^\pm$ is a charged pseudoscalar meson. To define the hadronic charged current, we introduce the form factor $H^\mu_{\bf m}$. By symmetry arguments, we can write
\begin{equation}
    H^\mu_{\bf m}=f_{\bf m} p^\mu|V_{qq'}|
\end{equation}
where $f_{\bf m}$ is a momentum-independent constant (see for instance Ref.~\cite{Rosner:2015wva}), since the meson is on-shell, $p^\mu$ is the meson momentum and $V_{qq'}$ is the CKM matrix element involved in the transition. The squared and spin averaged matrix element in the meson rest frame then reads 
\begin{equation}
    |\bar M|^2=y_\ell^2\frac{A_Nf_{\bf m}^2|V_{qq'}|^2}{v^2}\frac{ m_{\bf m}^2(m_\ell^2+p_D^2)-(m_\ell^2-p_D^2)^2}{p_D^4}\left(\frac{p_D^2}{\Luv^2}\right)^{\Delta_N-3/2}\,.
\end{equation}
The fully differential decay width is
\begin{align}
    \di \Gamma&=\frac{2y^2_\ell}{m_{\bf m}}\frac{A_Nf_{\bf m}^2|V_{qq'}|^2}{v^2}\frac{ m_{\bf m}^2(m_\ell^2+p_D^2)-(m_\ell^2-p_D^2)^2}{p_D^4}\left(\frac{p_D^2}{\Luv^2}\right)^{\Delta_N-3/2}\notag\\&\times\frac{\di p_D^2 \di^3 p_D \di^3 p_\ell}{(2\pi)^3(2\pi)^4 2 E_D 2 E_\ell} (2\pi)^4\delta^4(p_{\bf m} -p_\ell - p_D )
\end{align}
We can now integrate over the lepton momentum and over the final state angles and get
\begin{align}
   \frac{\di \Gamma(\mathbf{m}^\pm \to \ell^\pm + \On)}{\di p_D^2} &= \frac{y^2_\ell A_N}{32\pi^2} \frac{|V_{qq'}|^2 f_\mathbf{m}^2 m_\mathbf{m}^3}{v^2 \Lambda_{\rm UV}^4} \left[ \frac{p_D^2}{\Lambda_{\rm UV}^2} \right]^{\Delta_N - 7/2} \notag\\&\times\left[ \frac{p_D^2}{m_\mathbf{m}^2} \left(1 - \frac{p_D^2}{m_\mathbf{m}^2} \right) + \frac{m_\ell^2}{m_\mathbf{m}^2} \left( 1 + 2\frac{p_D^2}{m_\mathbf{m}^2} - \frac{m_\ell^2}{m_\mathbf{m}^2} \right) \right] |\mathbf{p}_\ell|\,,\label{eq:mesondecayw}
\end{align}
where
\begin{equation}
   |\mathbf{p}_\ell| \equiv \sqrt{ \left(1 - \frac{p_D^2}{m_\mathbf{m}^2} \right)^2 + \frac{m_\ell^2}{m_\mathbf{m}^2} \left( \frac{m_\ell^2}{m_\mathbf{m}^2} - 2\frac{p_D^2}{m_\mathbf{m}^2} - 2 \right) }
\end{equation}
is the magnitude of the lepton’s three-momentum in the meson rest frame.
It is noteworthy that for \( m_\ell \ll p_D \lesssim m_{\mathbf{m}} \), the decay width is dominated by the first term in the square brackets of \cref{eq:mesondecayw}. Indeed, we can write 
\beq
\Gamma(\mathbf{m}^\pm \to \ell^\pm + \On) \sim \frac{f_\mathbf{m}^2m_\mathbf{m}}{8\pi v^2}\left[\frac{m_\mathbf{m}^2}{\Lambda_{\rm UV}^2}\right]^{\Delta_N-3/2}\,.
\eeq
This contribution becomes UV-dominated for scaling dimensions \( \Delta_N \geq \tfrac{5}{2} \), as first shown in Ref.~\cite{Chacko:2020zze}. By contrast, for heavier leptons such as the muon, kinematics restricts the decay to the production of only the lightest dark-sector resonances. In this regime, the decay width is instead dominated by the second term in the brackets, which becomes UV-dominated for \( \Delta_N \geq \tfrac{7}{2} \). 

Finally, we can arrive at \cref{eq:meson} by defining the phase space factor in \cref{eq:mesondecayw} as
\begin{equation}
    \Pi_I(p_D^2/m_I^2)=\left[ \frac{p_D^2}{m_\mathbf{m}^2} \left(1 - \frac{p_D^2}{m_\mathbf{m}^2} \right) + \frac{m_\ell^2}{m_\mathbf{m}^2} \left( 1 + 2\frac{p_D^2}{m_\mathbf{m}^2} - \frac{m_\ell^2}{m_\mathbf{m}^2} \right) \right] |\mathbf{p}_\ell|\,.
\end{equation}
and by computing the SM respective decay width of the meson, namely
\begin{equation}
    \Gamma_{\rm SM,I}=\frac{1}{8\pi v^4}f_\mathbf{m}^2m_{\mathbf{m} }m_\ell^2V_{qq'}^2\,.
\end{equation}

\paragraph{Meson decays: three body}
We consider the semi-leptonic decay ${\bf m}\to {\bf m}' + \ell^\pm +\On$, where ${\bf m}'$ is a different meson, with $m_{\bf m}>m_{{\bf m}'}$. 
The squared and spin averaged matrix element in the meson rest frame then reads
\begin{equation}
    |\bar M({\bf m}\to {\bf m}'+\ell+\On)|^2=\frac{y_\ell^2\vert V_{qq'}\vert ^2}{4v^2}\frac{(p_D^2)^{\Delta_N-7/2}}{\Luv^{2\Delta_N-3}}H^{\alpha\beta} L_{\alpha \beta}\,,
\end{equation}
where we define the hadronic and leptonic tensor as
\begin{align}
    H^{\alpha\beta}&\equiv\langle {\bf m}'\vert \bar q'\gamma^\alpha q\vert {\bf m}\rangle \langle  {\bf m}\vert \bar q\gamma^\beta q'\vert {\bf m}'\rangle\,,\\
     L^{\alpha\beta}&\equiv8(p_2^{\alpha}p_3^{\beta}+p_2^{\beta}p_3^{\alpha}-g^{\alpha\beta} p_2\cdot p_3+i\epsilon{\alpha\beta\rho\sigma}p_2^{\rho}p_3^{\sigma})\,.
\end{align}
Here, $p_1$, $p_2$ and $p_3$ are the momenta of the final meson, lepton and dark jet, respectively, while $P$ is the momentum of the initial meson. The structure of the hadronic tensor can be again decomposed by use of form factors as~\cite{Becirevic:1999kt}
\begin{equation}
\begin{split}
    H^{\alpha\beta} &= F_+^2(q^2)(P^{\alpha}+p_1^\alpha)(P^\beta+p_1^{\beta})+F_0^2(q^2)q^{\alpha}q^\beta \\
    &+ F_+(q^2)F_0(q^2)[(P^{\alpha}+p_1^\alpha)q^\beta+(P^{\beta}+p_1^\beta)q^\alpha]\,.
    \end{split}
\end{equation}
We introduce the Mandelstam variables $m_{ij}^2=(p_i+p_j)^2$; note that in this notation we have $q^2=m_{23}^2$. 
The differential decay width can be reduced to
\begin{equation}
    \frac{\di \Gamma({\bf m}\to {\bf m}'+\ell+\On)}{\di p_D^2}=\frac{y_\ell^2\vert V_{qq'}\vert ^2}{v^2m_{\bf m}^3(512\pi^4)}\frac{(p_D^2)^{\Delta_N-7/2}}{\Luv^{2\Delta_N-3}}C(m_{12}^2,m_{23}^2,m_3^2;F_+,F_0)\di m_{12}^2dm_{23}^2\,,
\end{equation}
where $C$ is defined as
\begin{align}
   & C(m_{12}^2,m_{23}^2,m_3^2;F_+,F_0)\equiv F_0^2 (q^2)(m_{23}^2 - p_D^2) m_3^2\notag\\&+2 F_+(q^2) F_0 (q^2)(2 m_{12}^2 + m_{23}^{2} - m_3^2-2 m_1^2) m_3^2\\&+F_+^2 (q^2)\left[(m_{23}^2 - m_3^2) m_3^2 + 4 m_1^2 (m_{12}^2 - m_{\bf m}^2) + 
  4 m_{12}^2 (m_3^2 + m_{\bf m}^2-m_{23}^2 )-4 (m_{12}^2)^2\right]\notag\,.
\end{align}
The structure of the final result is richer than in the SM case due to the presence of massive states created by the operator $\On$. In particular, this requires the introduction of two form factors; unlike \cref{eq:meson}, the final result for the three-body width cannot be reduced to a compact formula.

It is also worth stressing that the form factors $F_+$ and $F_0$ differ significantly in character depending on the type of transition under consideration. For heavy-to-heavy transitions, such as $B \to D$, the form factors are well described within Heavy Quark Effective Theory (HQET)~\cite{Bernlochner:2022ywh,Veseli:1995fr} and exhibit a mild, smooth dependence on $q^2$ over the entire kinematic range. This behavior reflects the fact that the initial and final hadrons are kinematically and structurally similar, so that the overlap of their wavefunctions remains large and varies only weakly with $q^2$.

For heavy-to-light transitions, such as $B \to \pi$, the form factors exhibit a much more pronounced dependence on $q^2$~\cite{Wu:2006rd,Bordone:2019vic}. This reflects the fact that the final-state meson is structurally very different from the initial one and is often produced with large recoil, so that the overlap of the corresponding wavefunctions becomes strongly sensitive to the kinematics, leading to a steep and non-trivial $q^2$ behavior.

For simplicity, we fix the form factors to 1 to get the result of \cref{fig:delta_combined}.

\paragraph{EW bosons decay} 
We consider the leptonic two-body decays of the $Z,W$ and $h$ bosons. We can procede with the computation in a similar fashion as the two-body meson decays, obtaining the amplitudes
\begin{align}
    |\bar M(Z\to \nu_\ell+\On)|^2&=y_\ell^2\frac{A_Nm_Z^2}{3p_D^2}\frac{3E_\ell m_Z-m_\ell^2-2E_\ell^2}{m_Z^2+m_\ell^2-2m_ZE_\ell}\lp\frac{p_D^2}{\Luv^2}\rp^{\Delta_N - 3/2}\,,\\
    |\bar M(W^\pm\to \ell^\pm +\On)|^2&=y_\ell^2\frac{2A_Nm_W^2}{3p_D^2}\frac{3E_\ell m_W-m_\ell^2-2E_\ell^2}{m_W^2+m_\ell^2-2m_WE_\ell}\lp\frac{p_D^2}{\Luv^2}\rp^{\Delta_N - 3/2}\,,\\
    |\bar M(h\to \nu_\ell+\On)|^2&=y_\ell^2A_N\frac{m_h^2-p_D^2}{p_D^2}\lp\frac{p_D^2}{\Luv^2}\rp^{\Delta_N-3/2}\,,
\end{align}
and the differential decay widths
\begin{align}
   \frac{ \di \Gamma(Z\to \nu_\ell+\On)}{\di p_D^2}&=y_\ell^2\frac{A_N}{48\pi^2}\frac{m_Z^3}{\Lambda_{\rm UV}^4}\left[\frac{p_D^2}{\Lambda_{\rm UV}^2}\right]^{\Delta_N-7/2}\left[1-\frac{p_D^2}{m_Z^2}\right]^2\left[2+\frac{p_D^2}{m_Z^2}\right]\,,\\
     \frac{\di \Gamma(W^\pm\to \ell^\pm+\On)}{\di p_D^2}&=y_\ell^2\frac{A_N}{24\pi^2}\frac{m_W^3}{\Lambda_{\rm UV}^4}\left[\frac{p_D^2}{\Lambda_{\rm UV}^2}\right]^{\Delta_N-7/2}\left[1-\frac{p_D^2}{m_W^2}\right]^2\left[2+\frac{p_D^2}{m_W^2}\right]\,,\\
     \frac{\di \Gamma(h\to \nu_\ell+\On)}{\di p_D^2}&=y_\ell^2\frac{A_N}{32\pi^2}\frac{m_h}{\Lambda_{\rm UV}^2}\left[\frac{p_D^2}{\Lambda_{\rm UV}^2}\right]^{\Delta_N-5/2}\left[1-\frac{p_D^2}{m_h^2}\right]^2\, .\label{eq:hwidth}
\end{align}
In the UV dominated limit, we can approximate the decay widths as
\begin{align}
     \Gamma(h\to \nu + \On)&\sim \frac{m_h}{8\pi}\left[\frac{m_h^2}{\Lambda_{\rm UV}^2}\right]^{\Delta_N-3/2}\,,\\
      \Gamma(Z(W^\pm)\to \nu(\ell^\pm) + \On)&\sim \frac{m_{Z,W}}{8\pi}\left[\frac{m_{z,W}^2}{\Lambda_{\rm UV}^2}\right]^{\Delta_N-3/2}\,.
\end{align}
Notice that for $m_\ell>0$ UV dominated production from $W,Z$ $(h)$ requires $\Delta_N\geq7/2$ (5/2).

From the expressions above, we can define the phase space factors
\begin{equation}
    \Pi_b=\left[1-\frac{p_D^2}{m_b^2}\right]^2\left[2+\frac{p_D^2}{m_b^2}\right]\,,\quad
      \Pi_h=\frac{1}{4\pi}\left[1-\frac{p_D^2}{m_h^2}\right]^2
\end{equation}
which appear in \cref{eq:boson,eq:higgs}.
The respective SM decay widths are
\begin{equation}
    \Gamma_{{\rm SM},Z}=\frac{1}{v^2}\frac{m_Z^3}{12\pi}=(g^2+g'^2)\frac{m_Z}{48\pi}\,,\quad\Gamma_{{\rm SM},W}=\frac{1}{v^2}\frac{m_W^3}{6\pi}=g^2\frac{m_W}{24\pi}\,,\quad \Gamma_{{\rm SM},h}=\frac{3y_b^2m_h}{4\pi}\,.
\end{equation}

\subsection{Meson fluxes at proton beam dumps}\label{app:mesonfluxes}
The production rate of $\On$ at proton beam dumps strongly depend on the fraction of mesons produced by the proton scattering on target. Following Ref.~\cite{SHiP:2018xqw}, we can write for heavy mesons ($D$ and $B$)
\begin{equation}
N_{\text{prod}} = \sum_{q\in(c,b)} N_q \times \sum_{{\bf m}} f(q \to {\bf m}) \times \text{BR}({\bf m} \to \On + X)\,,
\end{equation}
where $N_q$ is the total number of quark $q$ produced, $f(q \to {\bf m})$ is the quark fragmentation fraction into the meson ${\bf m}$, and $\text{BR}({\bf m} \to \On + X)$ is the meson inclusive branching ratios into $\On$. The number $N_q$ can be computed as
\begin{equation}
N_q \sim 2 \times X_{\bar{q}q} \times N_{\text{POT}}\,,
\end{equation}
where the production fractions $X_{qq}$ have been computed to be $X_{cc}\sim 10^{-3},~X_{bb}\sim 10^{-7}$ for proton beams of 400 GeV~\cite{HERA-B:2007rfd,Lourenco:2006vw}. Hence, in our analysis we consider production on $\On$ from $D$ mesons, but we neglect production from $B$ meson (see Ref.~\cite{Schubert:2024hpm} for more details).

Light mesons, $\pi$ and $K$, are instead produced with larger fractions. Their fluxes can be described using the empirical Atherton parametrization~\cite{Atherton:1980vj,VanDijk:2641500}
\begin{equation}\label{eq:Athertoneq}
\frac{\di ^2N}{\di p\,\di \Omega} = N_{\rm int} A_{\bf m} \left(\frac{B_{\bf m}}{p_0} e^{-B_Mp/p_0}\right)\left(\frac{2C_{\bf m} p^2}{2\pi} e^{-C_M p^2 \theta^2}\right)\,,
\end{equation}
where $A_{\bf m},B_{\bf m},C_{\bf m}$ are meson-dependent constants, which we summarize in \cref{tab:mesparam}, $p_0$ is the beam energy, and $N_{\rm int}$ is the number of proton-target interactions. For a target of length $L$ and interaction length $\lambda$, we have
\begin{equation}
N_{\rm int} = N_{\rm POT}\,\frac{L}{\lambda}\,e^{-L/\lambda}\,.
\end{equation}
This parametrization captures forward production of light mesons in the soft-QCD regime, where $Q^2 \sim \mathcal{O}(\Lambda_{\rm QCD}^2)$, while large transverse momenta are exponentially suppressed. For $D$-meson production, we instead adopt the parametrization of Ref.~\cite{Lourenco:2006vw} 
\begin{equation}\label{eq:dmesonflux}
\frac{\di ^2 N}{\di \Omega \, \di p} = N_{\rm int} A_D\,\frac{(1 - p/p_0)^n}{(1 + p^2 \theta^2/p_D^2)^m}\,,
\end{equation}
with $n=4$ and $m=5$. In this case, the production is governed by a hard scale $Q^2 \sim m_Q^2 \gtrsim \Lambda_{\rm QCD}^2$, and the transverse-momentum spectrum follows a power-law behavior characteristic of perturbative QCD. $A_D$ is a phenomenological constant which we fix to reproduce the existing bound from CHARM on HNL.
\begin{table}[t!]
    \centering
    \begin{tabular}{c|c c c}
      $M$ & $A_M$ &  $B_M$& $C_M$\\
      \hline
     $\pi$&1.2&9.5&5    \\
     $K$&0.16&8.5&3
    \end{tabular}
    \caption{Atherton parameters for light meson fluxes at beam dump.}
    \label{tab:mesparam}
\end{table}
We have verified that using these parameterizations  allows us to reproduce existing bounds on standard HNLs with reasonable accuracy~\cite{Bondarenko:2018ptm}. 
\subsection{Mixing angle plots}\label{app:BoundsMixing}
We add for completeness the parameter space of present and future searches in the $\Lir-|V_{\psi\nu_i}|^2$ plane.
\begin{figure}[t!] 
    \centering

    \begin{subfigure}[b]{0.47\linewidth}
        \centering
\includegraphics[width=\linewidth]{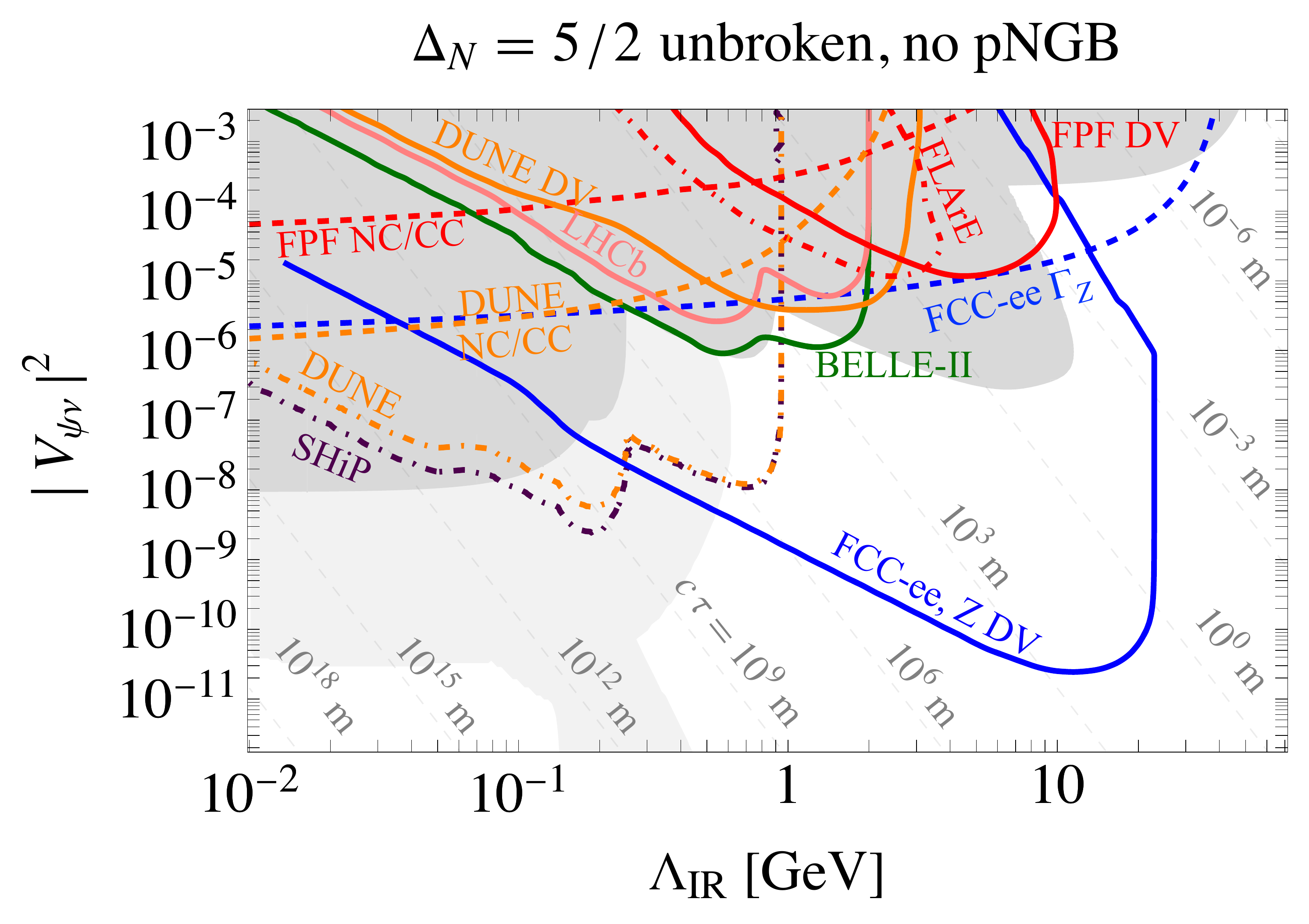}
    \end{subfigure}
    \hfill
    \begin{subfigure}[b]{0.47\linewidth}
        \centering

        \includegraphics[width=\linewidth]{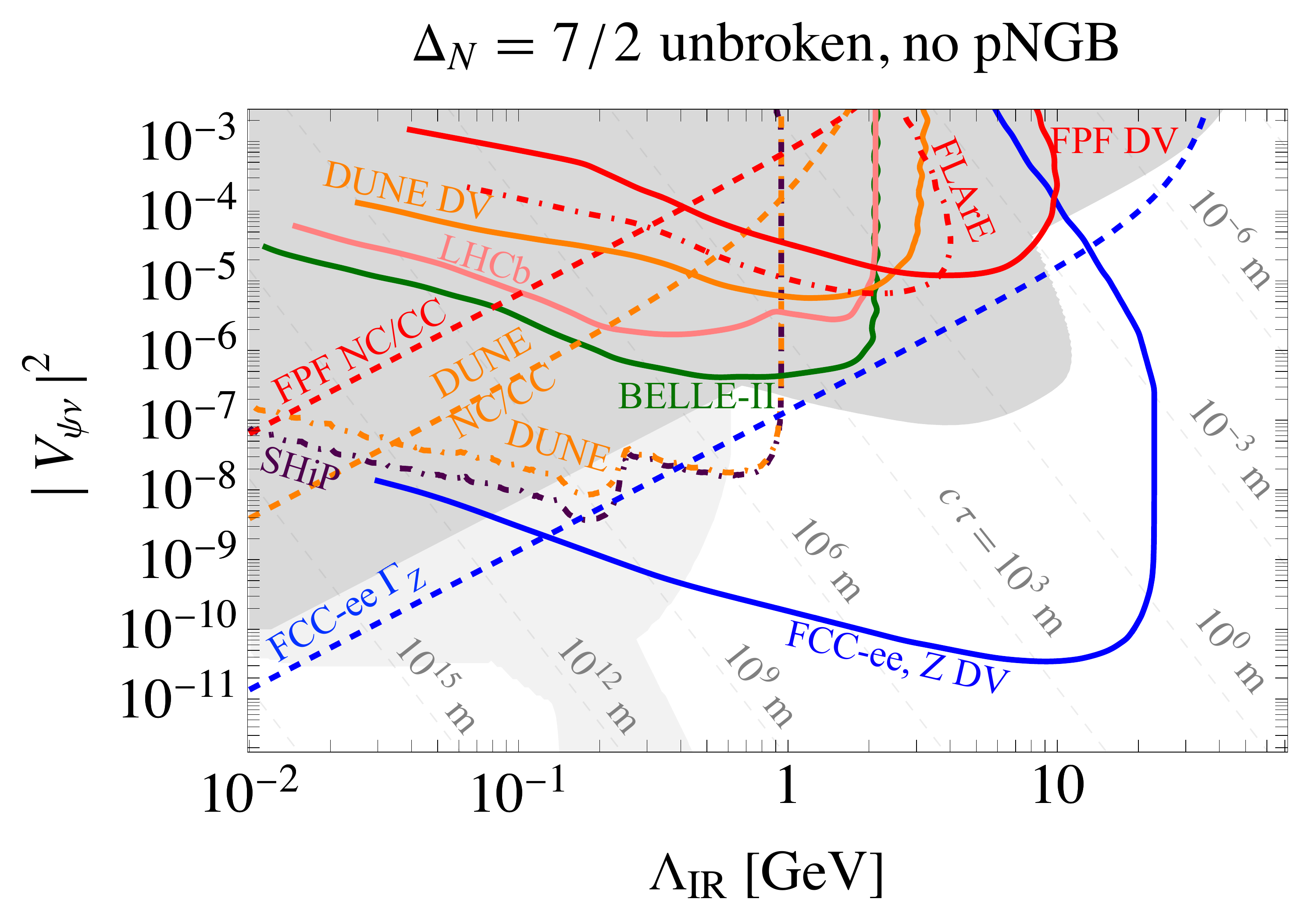}
    \end{subfigure}
    
    \vspace{0.1cm} 

    \begin{subfigure}[b]{0.47\linewidth}
        \centering
      
        \includegraphics[width=\linewidth]{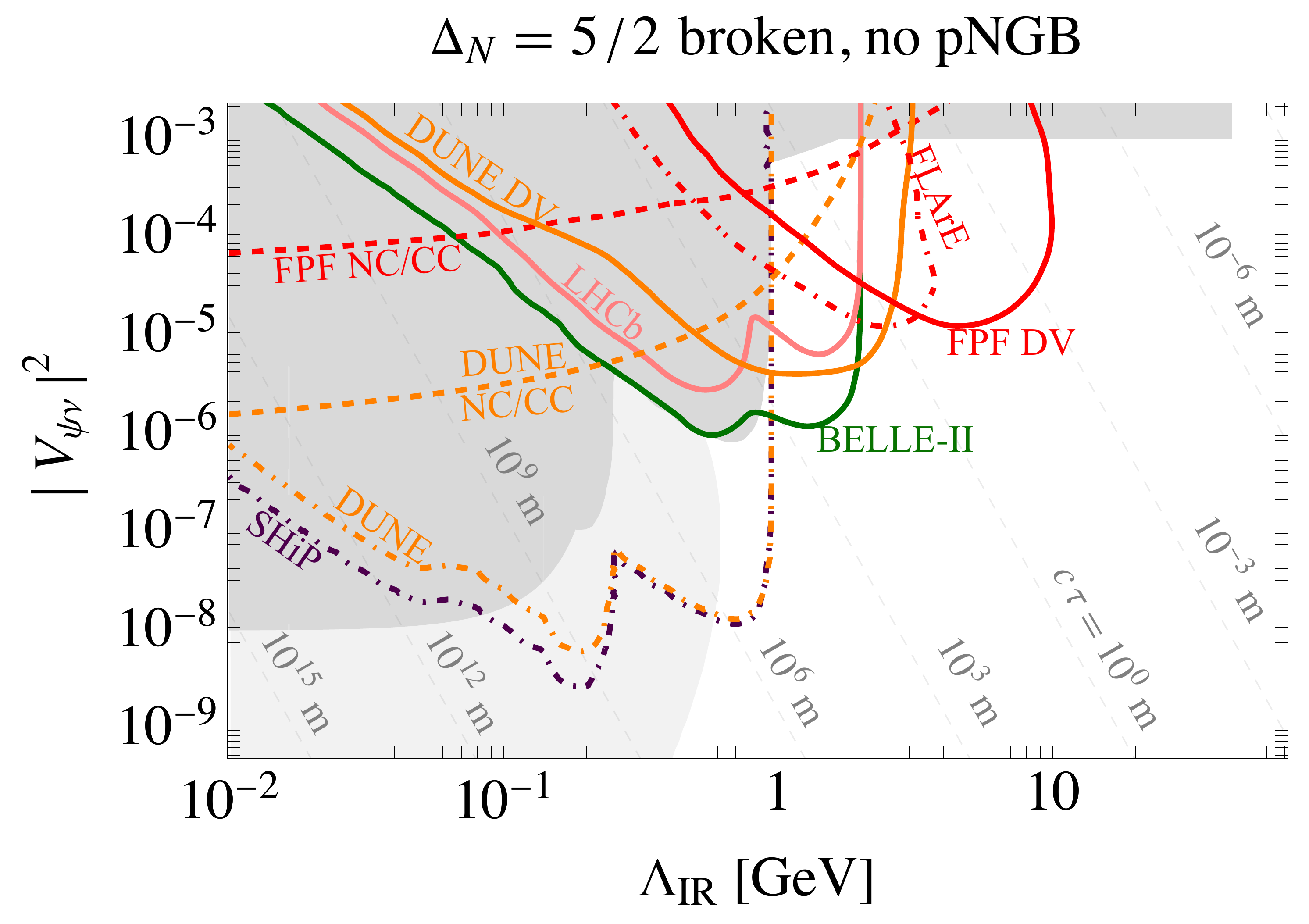}
    \end{subfigure}
    \hfill
    \begin{subfigure}[b]{0.47\linewidth}
        \centering
        \includegraphics[width=\linewidth]{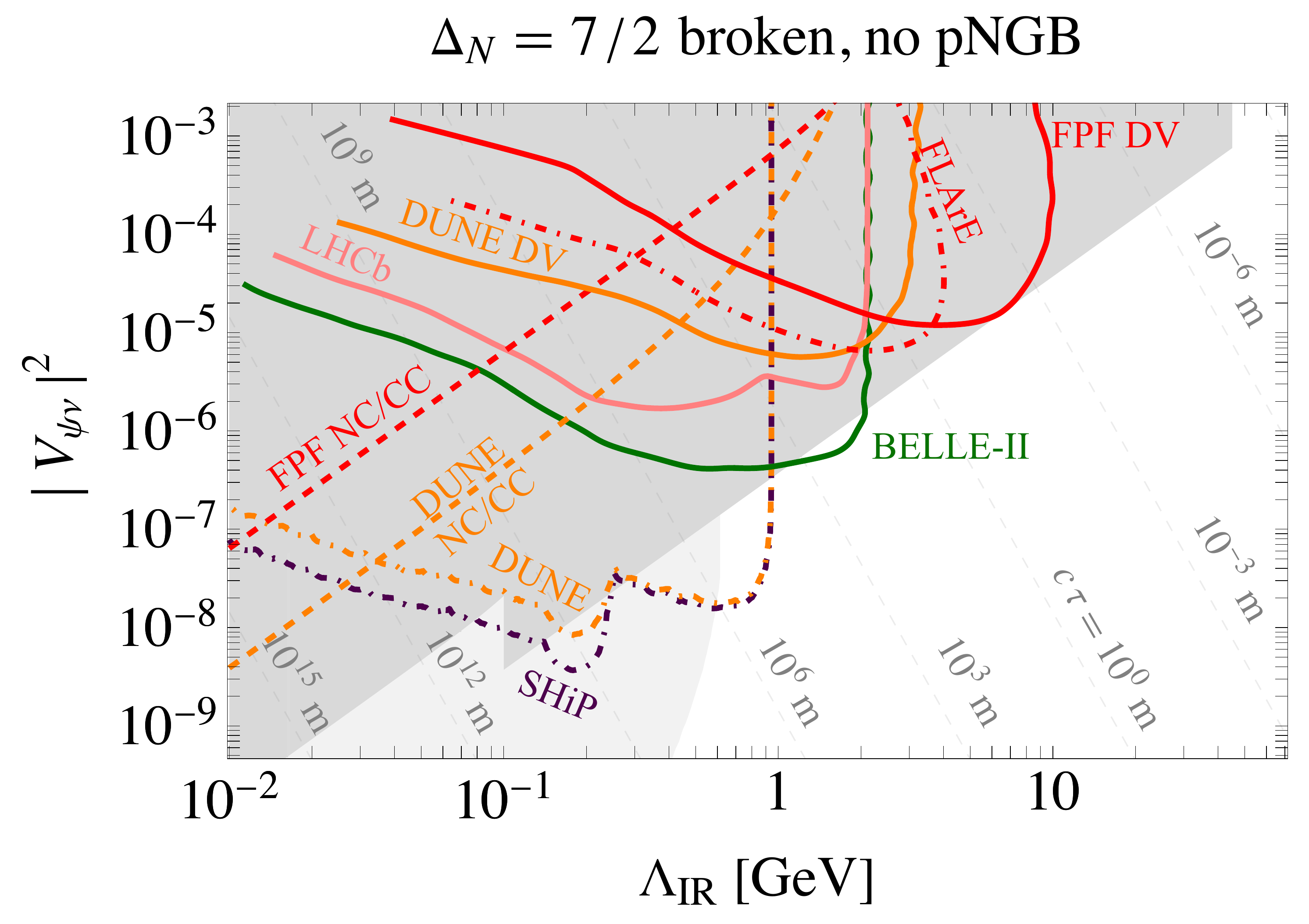}
    \end{subfigure}

    \caption{Experimental limits in $\Lir$-$|V_{\psi\nu}|^2$ plane, where the mixing angle is defined in \cref{eq:mixingangle}. The color code is the same as \cref{fig:delta_combined}. }
    \label{fig:delta_combinedu2}
\end{figure}

\section{Composite vs Fundamental Heavy Neutral Leptons}\label{app:strongISS}

In this section we discuss in detail the main differences between weakly and strongly coupled heavy neutral lepton scenarios. First, we show that the enhancement of the NC to CC event ratio is a distinctive feature of strong coupling or more precisely of scenarios in which the neutrino interacts through an operator with scaling dimension $\Delta_N > 3/2$ that can excite a multiparticle continuum. Second, we discuss how neutrino mass generation mechanisms, such as the Seesaw and Inverse Seesaw, are changed in presence of the strong dynamics.

\subsection{Cross-sections enhancement and mixing}
The effect of a weakly coupled HNL portal is to mix mass and interaction eigenstates without modifying inclusive cross-sections, up to phase-space corrections that reduce the total rate. 

To illustrate this, consider three scalar fields $(\phi,\psi,\eta)$, where $\phi,\psi$ mix (analogous to SM neutrinos and an HNL), while $\phi$ interacts with $\eta$ (analogous to SM neutrinos with quarks or leptons). For simplicity, we take $\eta$ to be massless. The Lagrangian is
\begin{equation}\label{eq:scalarmixing}
    2\mathcal{L}=(\partial\phi)^2+(\partial\psi)^2+(\partial\eta)^2+m^2\phi^2+\Lambda^2\psi^2+\frac{\lambda}{2}\phi^2\eta^2+\delta\phi\psi\,,
\end{equation}
with mass eigenstates $(\phi',\psi')$ defined by
\begin{equation}
    \phi=\phi'\cos\theta+\psi'\sin\theta\,,\qquad 
    \psi=-\phi'\sin\theta+\psi'\cos\theta\,,
\end{equation}
and masses $(m',\Lambda')$. In the limit $m\ll\Lambda$, the mixing angle is $\theta\sim\delta/(2\Lambda^2)$.

We consider $2\to2$ scattering with $\eta$ in the ultra-relativistic limit. The squared amplitudes are
\begin{align}
   | M(\phi'\eta\to\phi'\eta)|^2&=\lambda^2\cos^4\theta\,,\\
   | M(\psi'\eta\to\phi'\eta)|^2=| M(\phi'\eta\to\psi'\eta)|^2&=\lambda^2\cos^2\theta\sin^2\theta\,,\\
   | M(\psi'\eta\to\psi'\eta)|^2&=\lambda^2\sin^4\theta\,.
\end{align}
In the limit $s\gg m_i^2$, all phase-space factors coincide, $\mathrm{d}\Phi \;\to\; 1/(8\pi)\,,$ up to corrections $\mathcal{O}(m_i^2/s)$. The inclusive cross-section then reads
\begin{equation}
    \sigma \simeq \frac{\lambda^2}{16\pi s}
    \left[(\cos^2\theta+\sin^2\theta)^2 - \mathcal{O}\!\left(\frac{m^2}{s},\frac{\Lambda^2}{s}\right)\right]
    = \frac{\lambda^2}{16\pi s}\left[1 - \mathcal{O}\!\left(\frac{m^2}{s},\frac{\Lambda^2}{s}\right)\right].
\end{equation}
This coincides with the result in the absence of mixing, showing that mixing effects cancel in the inclusive rate. This is a direct consequence of the unitarity of the rotation matrix.

The same conclusion holds more generally. Consider a light state $\phi$ mixing with a set of heavier states $\psi_i$, described by a mass matrix $\mathfrak{M}$ diagonalized by a unitary matrix $U$. Writing the interaction eigenstates as $\Psi_i$, with $i=1$ corresponding to $\phi$, matrix elements take the form
\begin{equation}
    |M_{ij}|^2 = \lambda^2 |U^\dagger_{i1}U_{1j}|^2\,.
\end{equation}
The total cross-section is
\begin{equation}
\sigma_{\mathrm{tot}} = \sum_{ij} \int \mathrm{d}\Phi_{ij}\, |M_{ij}|^2\,.
\end{equation}
In the high-energy limit $\sqrt{s}\gg \Lambda_i$, all phase-space factors coincide, $\mathrm{d}\Phi_{ij}\simeq \mathrm{d}\Phi_0$, yielding
\begin{equation}
\sigma_{\mathrm{tot}} = \lambda^2 \int \mathrm{d}\Phi_0 
\sum_{ij} |U^\dagger_{i1}U_{1j}|^2 
= \lambda^2 \int \mathrm{d}\Phi_0 
= \sigma_0 \,,
\end{equation}
where unitarity of $U$ has been used. Deviations arise only from mass effects $\mathcal{O}(\Lambda_i^2/s)$.

This argument extends to fermions and more general interactions: although individual matrix elements depend non-trivially on kinematics, the cancellation persists in the high-energy limit where phase space becomes universal. We conclude that weakly coupled mixing does not enhance inclusive cross-sections. Instead, it leads to a suppression due to reduced phase space from heavier states.

\subsection{Seesaw models from a strongly coupled Dark Sector}
\label{sec:SSvsISS}

Seesaw mechanisms can be implemented in strongly coupled theories in multiple ways. Broadly, strong dynamics allows for two (non-exhaustive) possibilities to generate IR mass terms: (i) via higher-dimensional operators that interpolate masses below the confinement scale $\Lir$ (as in Nambu-Jona-Lasinio-like scenarios), and (ii) via dimensional transmutation (as in QCD). 

As an example of the first case, a standard Seesaw can be realized with composite vectorlike fermions~\cite{Arkani-Hamed:1998wff}:
\begin{equation}
    {\cal L}_{\rm SS}=\frac{HL{\cal O}_N}{\Luv^{\Delta_N-3/2}}+\frac{\bar{\mathcal{O}}_N {\cal O}_N}{\Luv^{2\Delta_N-4}}\,,
\end{equation}
where ${\cal O}_N$ is a composite operator (see \cref{sec:HLO}), e.g. a dark baryon. We assume $\Delta_N\leq 2$ to avoid UV sensitivity in the two-point function and focus on a single flavor. Near $\Lir$, we match ${\cal O}_N$ onto a fermionic resonance $\psi$,
${\cal O}_N\sim \Lir^{\Delta_N-3/2}\psi\,,$ up to $\mathcal{O}(1)$ coefficients. After EWSB, the IR Lagrangian becomes
\begin{equation}\label{eq:SS:LowEn}
{\cal L}_{\rm IR}=v\epsilon^{\Delta_N-3/2}\psi \nu + \Luv\epsilon^{2\Delta_N-3}\bar\psi\psi\,,
\end{equation}
with $\epsilon\equiv \Lir/\Luv$, leading to
\begin{equation}
M\sim
\begin{bmatrix}
0 & v \epsilon^{\Delta_N-3/2} \\
v \epsilon^{\Delta_N-3/2} & \Luv\epsilon^{2\Delta_N-3}
\end{bmatrix}\ ,\qquad m_{\nu}\sim \frac{v^2}{\Luv}\,,\quad 
m_{\psi}\sim \Luv\epsilon^{2\Delta_N-3}\,.
\end{equation}
For $\Delta_N\leq 2$, $m_\psi$ is controlled by the UV scale, whereas we are interested in $m_\psi\sim\Lir$. This motivates CFT deformations such as the ones of Ref.~\cite{Chacko:2020zze}, where the fermion mass is generated dynamically:
\begin{equation}
M\sim
\begin{bmatrix}
0 & v \epsilon^{\Delta_N-3/2} \\
v \epsilon^{\Delta_N-3/2} & \Lir
\end{bmatrix},
\qquad
m_{\nu}\sim \frac{v^2}{\Lir}\epsilon^{\Delta_N-3/2},\quad 
m_{\psi}\sim \Lir.
\end{equation}
In both cases, for $m_\psi\gg m_\nu$,
\begin{equation}
\theta_{\nu\psi}\sim \sqrt{\frac{m_\nu}{m_\psi}}\,,
\end{equation}
as in standard Seesaw models, implying suppressed experimental signatures.

The Inverse Seesaw (ISS) avoids this suppression. In the weakly coupled case,
\begin{equation}
{\cal L}_{\rm ISS}=HL\Psi+M\Psi\bar\Psi+\frac{\mu}{2}\bar\Psi^2\,,
\end{equation}
yielding $m_\nu=\frac{\mu v^2}{M^2}\sim \mu\,\theta_{\nu\psi}^2$. In this setup, the smallness of neutrino masses originates from the hierarchy $\mu\ll v$, allowing for large active–sterile mixing. While this hierarchy is technically natural, it is not dynamically explained, making the ISS less compelling than the standard Seesaw in accounting for the origin of neutrino masses.

In a strongly coupled realization of ISS, both Dirac and Majorana masses may arise from higher-dimensional operators:
\begin{equation}
{\cal L}_{\rm UV}=\frac{HL\On}{\Luv^{\Delta_N-3/2}}+\frac{\bar{\mathcal{O}}_N\On}{\Luv^{2\Delta_N-4}}+\frac{\tilde O \On^2}{\Luv^{\tilde\Delta+2\Delta_N-4}}.
\end{equation}
Assuming that at low energy $\tilde O\sim \mu^{\tilde\Delta}$, the IR Lagrangian becomes
\begin{equation}
{\cal L}_{\rm IR}\sim \epsilon^{\Delta_N-3/2}HL\psi+\Luv\epsilon^{2\Delta_N-3}\bar\psi\psi+\Luv\epsilon^{2\Delta_N-3}\left(\frac{\mu}{\Luv}\right)^{\tilde\Delta}\psi^2,
\end{equation}
leading to
\begin{equation}
m_\nu\sim \frac{v^2}{\Luv}\left(\frac{\mu}{\Luv}\right)^{\tilde\Delta},\qquad
\theta_{\nu\psi}\sim \frac{\sqrt{m_\nu}}{\sqrt{\Luv}}\left(\frac{\Luv}{\mu}\right)^{\tilde\Delta/2}\left(\frac{\Luv}{\Lir}\right)^{\Delta_N-3/2}.
\end{equation}
Compositeness thus controls both neutrino mass suppression and mixing enhancement. However, both $m_\psi$ and $\mu$ remain controlled by the UV scale. UV convergence of $\langle \On\On\tilde O\rangle$ requires $2\Delta_N+\tilde\Delta\leq 8.$

An alternative implementation (as first proposed in Ref.~\cite{Chacko:2020zze}) assumes $\tilde O$ directly interpolates $\psi^2$:
\begin{equation}
{\cal L}_{\rm UV}=\frac{HL\On}{\Luv^{\Delta_N-3/2}}+\hat \mu\frac{\tilde O}{\Luv^{\tilde\Delta-4}},
\end{equation}
leading to
\begin{equation}
{\cal L}_{\rm IR}\sim \epsilon^{\Delta_N-3/2}HL\psi+\Lir\bar\psi\psi+\hat\mu\,\Lir\,\epsilon^{\tilde\Delta-4}\psi^2.
\end{equation}
The resulting neutrino mass and mixing are
\begin{equation}
m_\nu\sim \hat\mu\,\Lir\left(\frac{\Lir}{\Luv}\right)^{\tilde\Delta-4}\left(\frac{v}{\Lir}\right)^2\left(\frac{\Lir}{\Luv}\right)^{2\Delta_N-3},\qquad
\theta_{\nu\psi}\sim \frac{v}{\Lir}\left(\frac{\Lir}{\Luv}\right)^{\Delta_N-3/2}.
\end{equation}
In this setup, compositeness controls whether neutrino masses are suppressed by small mixing or by small lepton-number violation.

For $\hat\mu=1$ and $2\Delta_N+\tilde\Delta=8$,
\begin{equation}
m_\nu\sim \frac{v^2}{\Luv}\sim \text{eV}
\qquad\Rightarrow\qquad
\Luv\sim 10^{14}\,\text{GeV}.
\end{equation}
An alternative that avoids such a large $\Luv$ is to generate the Majorana mass through the more irrelevant operator
\begin{equation}
\mathcal{L}_{\rm{UV}}\supset\frac{\tilde O \On^2}{\Lambda_{\rm UV}^{2\Delta_N+\tilde\Delta-4}}\,,  
\end{equation}
so that the lepton-number–carrying operator $\tilde O$ does not directly interpolate the IR Majorana term $\psi^2$. In this case the Majorana mass is more suppressed, and correspondingly smaller neutrino masses can be obtained for lower $\Luv$. For instance, taking $\Delta_N=2$ and $\tilde\Delta=4$, one finds
\begin{equation}
m_\nu\sim \frac{v^2}{\Luv}\left(\frac{\Lir}{\Luv}\right)^4\sim \text{eV}
\qquad\Rightarrow\qquad
\Luv\sim 600\,\text{GeV}\left(\frac{\Lir}{1\,\text{GeV}}\right)^{4/5}.
\end{equation}
The model in \cref{eq:uv_5/2}, supplemented by the symmetry-breaking terms in \cref{eq:su_3_break}, is of this type: the Dirac mass is generated dynamically by confinement, while the Majorana mass arises from an analogous higher-dimensional operator obtained after integrating out the heavy scalar $\phi$.

\newpage

\bibliography{bibliography}
\bibliographystyle{JHEP}

\end{document}